\documentclass[fleqn,usenatbib]{mnras}

\usepackage{graphicx}
\usepackage{amsmath}
\usepackage[dvipsnames]{xcolor}
\usepackage{amssymb}
\usepackage{newtxtext}
\usepackage[varvw]{newtxmath}
\usepackage{ulem}
\usepackage{soul}
\usepackage{siunitx}
\usepackage{macros_vdb}
\usepackage{tikz}
\usetikzlibrary{shapes,arrows}

\defcitealias{Navarro.etal.97}{NFW}
\defcitealias{Green.vdBosch.19}{GB19}

\graphicspath{{./figures/}}

\title[The impact of artificial disruption]{The tidal evolution of dark matter substructure -- II. The impact of artificial disruption on subhalo mass functions and radial profiles}

\author[S. B. Green, F. C. van den Bosch, and  F. Jiang]{%
Sheridan~B.~Green,$^{1}$\thanks{E-mail: \href{sheridan.green@yale.edu}{sheridan.green@yale.edu} (SBG)}\thanks{NSF Graduate Research Fellow} Frank~C.~van den Bosch,$^{1,2}$ and Fangzhou Jiang$^{3,4}$\thanks{Troesh Scholar}
\vspace*{8pt}
\\
$^{1}$Department of Physics, Yale University, P.O. Box 208120, New Haven, CT 06520-8120\\
$^{2}$Department of Astronomy, Yale University, P.O. Box 208101, New Haven, CT 06520-8101\\
$^{3}$TAPIR, California Institute of Technology, Pasadena, CA 91125\\
$^{4}$Carnegie Observatories, 813 Santa Barbara Street, Pasadena, CA 91101\\
}

\date{}

\pubyear{2021}

\begin{document}
\label{firstpage}
\pagerange{\pageref{firstpage}--\pageref{lastpage}}
\maketitle

\begin{abstract}
Several recent studies have indicated that artificial subhalo disruption (the spontaneous, non-physical disintegration of a subhalo) remains prevalent in state-of-the-art dark matter-only cosmological simulations. In order to quantify the impact of disruption on the inferred subhalo demographics, we augment the semi-analytical \texttt{SatGen} dynamical subhalo evolution model with an improved treatment of tidal stripping that is calibrated using the \textit{DASH} database of idealized high-resolution simulations of subhalo evolution, which are free from artificial disruption. We also develop a model of artificial disruption that reproduces the statistical properties of disruption in the \textit{Bolshoi} simulation. Using this framework, we predict subhalo mass functions (SHMFs), number density profiles, and substructure mass fractions and study how these quantities are impacted by artificial disruption and mass resolution limits. We find that artificial disruption affects these quantities at the $10-20\%$ level, ameliorating previous concerns that it may suppress the SHMF by as much as a factor of two. We demonstrate that semi-analytical substructure modeling must include orbit integration in order to properly account for splashback haloes, which make up roughly half of the subhalo population. We show that the resolution limit of $N$-body simulations, rather than artificial disruption, is the primary cause of the radial bias in subhalo number density found in dark matter-only simulations. Hence, we conclude that the mass resolution remains the primary limitation of using such simulations to study subhaloes. Our model provides a fast, flexible, and accurate alternative to studying substructure statistics in the absence of both numerical resolution limits and artificial disruption.
\end{abstract}


\section{Introduction}\label{sec:intro}

The standard $\Lambda$ cold dark matter ($\Lambda$CDM) cosmological model predicts that structure forms as the consequence of primordial dark matter overdensities that collapse to form self-bound haloes. Smaller perturbations collapse earlier and merge to form larger haloes, resulting in a hierarchical halo assembly process that spans all mass scales. By studying halo evolution via cosmological $N$-body simulations, it is clear that the tightly bound central regions of smaller haloes survive the merger process, persisting as orbiting subhaloes within the treacherous environment of their host halo, where they are subjected to dynamical friction and disruptive tidal forces \citep[e.g.,][]{MBW10}. Neglecting the impact of baryonic physics, this merger process is roughly self-similar due to the scale-free nature of gravitational collapse, ultimately resulting in an entire hierarchy of substructure, where subhaloes themselves host sub-subhaloes, and so on all the way down \citep{Tormen.etal.97, Gao.etal.04, Kravtsov.etal.04, Giocoli.etal.10}.

The population statistics of dark matter (DM) substructure are most often summarized in terms of subhalo mass functions (SHMFs) and radial profiles; these summary statistics depend heavily on the underlying particle nature of DM. For example, the free-streaming cutoff scale, set by the DM thermal velocity, impacts the low-mass end of the SHMF \citep[e.g.,][]{Knebe.etal.08, Lovell.etal.14, Colin.etal.15, Bose.etal.17}, while non-negligible DM self-interactions result in cored inner halo density profiles \citep[e.g.,][]{Burkert2000, Vogelsberger.etal.12, Rocha.etal.13}, which impacts the survivability of substructure in the presence of tides \citep[e.g.,][]{Penarrubia.etal.10}. The predictions of substructure demographics made by these various dark matter models differ primarily at the low mass end. Consequently, many observational searches are underway in the attempt to constrain the abundance of low mass substructure, leveraging gravitational lensing \citep[e.g.,][]{Dalal.Kochanek.02, Keeton.Moustakas.09, Vegetti.etal.14, Shu2015, Hezaveh2016, Gilman2020, Vattis2020}, indirect detection via DM annihilation to $\gamma$-rays or decay signals \citep[e.g.,][]{Strigari.etal.07, Pieri.etal.08, Hayashi2016, Hiroshima2018, Delos2019, Facchinetti2020, Rico2020, Somalwar2021}, and gaps in stellar streams \citep[e.g.,][]{Carlberg2012, Ngan2014, Erkal.etal.16, Bonaca2020, Necib2020}, among other approaches.  Since satellite galaxies are inferred to live within subhaloes, with their respective properties related via the galaxy-halo connection, DM substructure statistics are intimately connected to satellite galaxy abundances \citep[e.g.,][]{Vale.Ostriker.06, Hearin.etal.13, Behroozi.etal.13c, Newton.etal.18, Nadler2019, Nadler2020b, Nadler2020} and thus can be used to constrain cosmology through their impact on small-scale clustering statistics \citep[e.g.,][]{Benson.etal.01, Berlind.etal.03, vdbosch.etal.05b, Lange2019, vdBosch2019}. Clearly, accurately modeling the evolution of DM subhalo populations is a prerequisite for their use as a cosmological probe and as a tool to study the particle nature of dark matter. Unfortunately, since the evolution of DM substructure is highly non-linear, modeling all but the most idealized circumstances has proven analytically intractable. Thus, to date, cosmological $N$-body simulations have been the most common avenue used for studying the demographics of DM substructure.

In recent years, cosmological simulations have successfully and repeatedly passed an important convergence test: as resolution is varied, the SHMFs remain in agreement above the 50--100 particle limit \citep[e.g.,][]{Springel.etal.08, Onions.etal.12, Knebe.etal.13, vdBosch.Jiang.16, Griffen.etal.16, Ludlow2019}. While this is promising, the physical correctness of cosmological simulations is not guaranteed by the convergence of mass functions alone. Using the state-of-the-art \textit{Bolshoi} simulation \citep{Klypin.etal.11}, \citet{vdBosch.17} found that the evolved SHMF of surviving subhaloes looks identical to the SHMF of disintegrated subhaloes, noting that total subhalo disruption is prevalent. The inferred disruption rates from various studies are extremely high, with roughly 55-65\% (90\%) of subhaloes accreted at $z=1$ ($2$) being disrupted by the present day \citep{Han.etal.16, vdBosch.17, Jiang.vdBosch.17}. \citet{Hayashi.etal.03} has shown that the total binding energy of a halo that is instantaneously stripped down to a sufficiently small radius (encompassing roughly 5--10\% of the original mass) can be positive; hence, the authors suggested that such systems could disrupt spontaneously. Motivated by this analysis, subsequent works have incorporated physical disruption via tidal stripping and heating into their models or used such an argument as a justification for their results \citep{Zentner.Bullock.03, Taylor.Babul.04, Klypin.etal.15, Garrison-Kimmel.etal.17}. Recently, however, \citet{vdBosch.etal.2018a} demonstrated that the boundedness of a subhalo remnant does not depend solely on the total binding energy, but rather on the radial distribution of the binding energies of the constituent particles. In fact, by using idealized simulations with sufficiently high resolution, \citet{vdBosch.etal.2018a} showed that it is possible for a self-bound remnant to survive even after 99.9\% of the original mass has been stripped. More broadly, the study used analytical arguments to show that neither tidal heating nor tidal stripping alone are capable of causing complete physical disruption of cuspy CDM subhaloes \citep[consistent with earlier work by][]{Penarrubia.etal.10}. As a follow-up, \citet{vdBosch.etal.2018b} ran a suite of idealized numerical simulations, concluding that disruption of $N$-body subhaloes in cosmological simulations is largely numerical in nature and can be primarily attributed to (i) discreteness noise caused by insufficient particle resolution and (ii) inadequate softening of gravitational forces (see \citealt{Mansfield2020} for a recent analysis of the impact of the force softening scale on various halo properties). In agreement with these findings, \citet{vdBosch.17} assessed that approximately 80\% of subhalo disruption in the \textit{Bolshoi} simulation is most likely numerical in nature (see Section~\ref{ssec:artdisr} below for details).

If the majority of subhalo disruption in cosmological simulations is indeed numerical, the implications for small-scale cosmology and astrophysics are profound. For example, a disruption-driven reduction in subhalo statistics would result in systematic biases in predictions from subhalo abundance matching \citep[e.g.,][]{Conroy.etal.06, Vale.Ostriker.06, Guo.etal.10, Hearin.etal.13, Chaves2016}. Semi-analytical models of galaxy and dark matter substructure evolution \citep[e.g.,][]{Taylor.Babul.01, Penarrubia.Benson.05, Zentner.etal.05, Diemand.etal.07a, Kampakoglou.Benson.07, Gan2010, Pullen.etal.14, Jiang.vdBosch.16, Benson2020, Jiang2021, Yang2020} have historically been calibrated to reproduce the results of cosmological simulations and thus end up having inherited any systematic issues present in such simulations. As a specific example, \citet{Jiang.vdBosch.16} constructed a semi-analytical model that accurately matches the statistics of subhaloes in the \textit{Bolshoi} simulation by simply tuning an orbit-averaged mass-loss rate and including an empirical model of subhalo disruption that, by construction, reproduces the disruption demographics in the simulation. As shown in \citet[][hereafter \citetalias{Green.vdBosch.19}]{Green.vdBosch.19}, in the absence of such disruption, the normalization of the evolved SHMF predictions from \citet{Jiang.vdBosch.16} is boosted by a factor of two. Hence, depending on the fraction of subhalo disruption in cosmological simulations that is indeed artificial, it remains possible that such simulations (and derivative semi-analytical models) may be underestimating subhalo abundances by up to a factor of two. Such a systematic bias would have serious implications for dark matter indirect detection searches and could help explain the `galaxy clustering crisis' in subhalo abundance matching \citep{Campbell.etal.18}, since both of these applications, among others, depend heavily on evolved SHMFs from simulations. As long as the effects of artificial disruption remain as an unknown variable in the analysis of cosmological simulations, we will be unable to extract the maximum amount of cosmological and astrophysical information content that will soon be made available in various large upcoming surveys, including DESI, LSST, EUCLID, and WFIRST. Clearly, there is still work to be done towards better understanding the tidal evolution of DM substructure, hence the motivation of the present study.

Recently, we released \texttt{SatGen} \citep{Jiang2021}, a semi-analytical modeling framework for studying galaxy and DM substructure evolution. The core components of the dark matter-only side of the framework include prescriptions for (i) analytical merger trees \citep{Cole2000, Parkinson.etal.08, Benson2017}, from which the internal properties of subhaloes at accretion are derived, (ii) orbital parameter distributions for infalling subhaloes \citep[][]{Zentner.etal.05, Wetzel.11, Jiang.etal.15, Li2020}, (iii) the integration of subhalo orbits, including dynamical friction \citep{Chandrasekhar.43}, (iv) the evolved subhalo density profile, which captures how the internal structure of the subhalo responds to tidal heating and stripping \citep[e.g.,][]{Hayashi.etal.03, Penarrubia.etal.10, Drakos.etal.17, Green.vdBosch.19, Errani2020b}, and (v) the instantaneous mass-loss rate, which depends on the structure of both the host- and subhalo in addition to the orbit \citep[e.g.,][this work]{Drakos2020}. In contrast to \citet{Jiang.vdBosch.16}, which followed \citet{vdBosch.etal.05} by only considering orbit-averaged subhalo evolution, \texttt{SatGen} integrates individual subhalo orbits, thereby allowing for a proper treatment of splashback haloes \citep[e.g.,][]{Ludlow2009, Aung2021, Diemer2020b, Diemer2020a, Fong2020}. As we will show, this treatment of splashback haloes is crucial for properly comparing model predictions with simulation results.

The goal of this work is to build a semi-analytical model of substructure evolution that is independent of any tidal evolution-related numerical artifacts that may be present in cosmological simulations. Thus, in \citet{Ogiya2019}, we introduced the Dynamical Aspects of SubHaloes (\textit{DASH}) database, a large library of idealized, high-resolution $N$-body simulations of the tidal evolution of individual subhaloes. This simulation library has two key strengths: (i) the simulations span a wide range of parameter space, varying the initial orbital parameters and host- and subhalo concentrations and (ii) the live $N$-body subhaloes satisfy the strict set of convergence criteria laid out in \citet{vdBosch.etal.2018b}, suppressing numerical artifacts caused by discreteness noise and inadequate force softening. In \citetalias{Green.vdBosch.19}, we used \textit{DASH} to calibrate a highly accurate, simply parametrized empirical model of the evolved subhalo density profile (ESHDP), which is unimpeded by numerical artifacts and is applicable to a far wider range of subhalo parameter space than that of previous works \citep{Hayashi.etal.03, Penarrubia.etal.10, Drakos.etal.17}. In this work, we use the results of \citetalias{Green.vdBosch.19} as a component in a simple, physically motivated model of the instantaneous mass-loss rate. After calibrating this model to faithfully reproduce the subhalo mass trajectories across the range of \textit{DASH} simulations, we incorporate it into \texttt{SatGen}, yielding the aforementioned artifact-free semi-analytical model. We use this tool to make predictions for evolved subhalo mass functions, radial profiles, and substructure mass fractions and compare these findings to \textit{Bolshoi} as an independent attempt to quantify the impact of artificial disruption on the abundance of dark matter subhaloes in cosmological simulations.

This paper is organized as follows. In Section~\ref{sec:methods}, we describe our methods, giving an overview of \texttt{SatGen} and our modifications, which include the incorporation of the \citet{Li2020} orbital parameter model (summarized in Appendix~\ref{app:initorbits}), the ESHDP model of \citetalias{Green.vdBosch.19}, and an improved, \textit{DASH}-calibrated mass-loss rate. We also detail our procedures for modeling the impact of artificial disruption and calibrating the dynamical friction strength. In Section~\ref{sec:results}, we present the results of our augmented \texttt{SatGen} model, focusing on SHMFs, radial profiles, substructure mass fractions, and the numerical disruption rate in simulations. We conclude in Section~\ref{sec:summary} by summarizing our research program, highlighting the updates made to \texttt{SatGen}, and discussing our findings and their implications. 

The cosmology used throughout this work is consistent with that of the \textit{Bolshoi} simulation \citep{Klypin.etal.11}: $\Omega_\rmm = 0.270$, $\Omega_\Lambda = 0.730$, $\Omega_\rmb = 0.0469$, $h=0.7$, $\sigma_8 = 0.82$, and $n_\rms = 0.985$. The halo mass is defined as the mass enclosed within the virial radius, $r_\mathrm{vir}$, inside of which the mean density is equal to  $\Delta_\mathrm{vir}(z)$ times the critical density. For the $\Lambda$CDM cosmology adopted in this work, $\Delta_\mathrm{vir}(z = 0) \approx 100$ and is otherwise well-described by the fitting formula presented by \citet{Bryan.Norman.98}. Throughout, we use $m$ and $M$ to denote subhalo and host halo masses, respectively. We use $l$ and $r$ to reference subhalo- and host halo-centric radii, respectively. Projected radii are indicated by upper-case letters. The base-10 logarithm is denoted by $\log$ and the natural logarithm is denoted by $\ln$.

\section{Methods}\label{sec:methods}

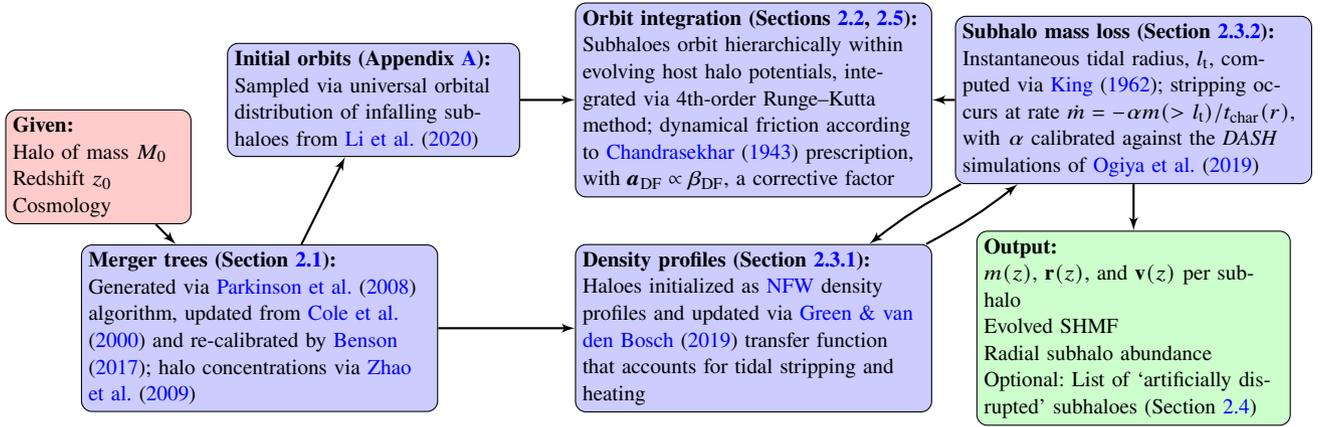
\begin{figure*}
    \centering
    
    \tikzstyle{block} = [rectangle, draw, fill=blue!20,
    text width=16em, rounded corners, minimum height=4em]
    \tikzstyle{narrowblock} = [rectangle, draw, fill=blue!20,
    text width=13em, rounded corners, minimum height=4em]
    \tikzstyle{blockgreen} = [rectangle, draw, fill=green!20,
    text width=12em, rounded corners, minimum height=4em]
    \tikzstyle{blockred} = [rectangle, draw, fill=red!20,
    text width=12em, rounded corners, minimum height=4em]
    \tikzstyle{line} = [draw, thick, -latex']
    
    \begin{tikzpicture} [node distance = 4cm, auto]
        \node [blockred, text width=8em] (input) {
        \textbf{Given:}\\
        Halo of mass $M_0$\\
        Redshift $z_0$\\
        Cosmology};
        \node [block, below right of=input, node distance=3cm] (trees) {
        \textbf{Merger trees (Section~\ref{ssec:trees}):}\\
        Generated via \citet{Parkinson.etal.08} algorithm, updated from \citet{Cole2000} and re-calibrated by \citet{Benson2017}; halo concentrations via \citet{Zhao2009}};
        \node [narrowblock, above of=trees, node distance=3cm, xshift=1.5cm] (orbits) {
        \textbf{Initial orbits (Appendix~\ref{app:initorbits}):}\\
        Sampled via universal orbital distribution of infalling subhaloes from \citet{Li2020}};
        \node [block, right of=trees, node distance=6.5cm] (profiles) {
        \textbf{Density profiles (Section~\ref{sssec:eshdp}):}\\
        Haloes initialized as \citetalias{Navarro.etal.97} density profiles and updated via \citet{Green.vdBosch.19} transfer function that accounts for tidal stripping and heating};
        \node [block, right of=orbits, node distance=5cm] (integrate) {
        \textbf{Orbit integration (Sections~\ref{ssec:orbitint}, \ref{ssec:dfstrength}):}\\
        Subhaloes orbit hierarchically within evolving host halo potentials, integrated via 4th-order Runge--Kutta method; dynamical friction according to \citet{Chandrasekhar.43} prescription, with $\boldsymbol{a}_\mathrm{DF} \propto \beta_\mathrm{DF}$, a corrective factor};
        \node [block, right of=integrate, node distance=5cm] (massloss) {
        \textbf{Subhalo mass loss (Section~\ref{sssec:massloss}):}\\
        Instantaneous tidal radius, $l_\rmt$, computed via \citet{King.62}; stripping occurs at rate $\dot{m} = -\alpha m(>l_\rmt) / t_\mathrm{char}(r)$, with $\alpha$ calibrated against the \textit{DASH} simulations of \citet{Ogiya2019}};
        \node [blockgreen, text width=14em, below of=massloss, node distance=3cm] (out) {
        \textbf{Output:}\\
        $m(z)$, $\mathbf{r}(z)$, and $\mathbf{v}(z)$ per subhalo\\
        Evolved SHMF\\
        Radial subhalo abundance\\
        Optional: List of `artificially disrupted' subhaloes (Section~\ref{ssec:artdisr})};
        \path [line] (input) -- (trees);
        \path [line] (trees) -- (orbits);
        \path [line] (trees) -- (profiles);
        \path [line] (orbits) -- (integrate);
        \path [line] (massloss) -- (integrate);
        \draw[-latex', thick] (massloss) to[bend right=5] node[above,rotate=60] {} (profiles);
        \draw[-latex',thick] (profiles) to[bend right=5] node[below,rotate=60] {} (massloss);
        \path [line] (massloss) -- (out);
    \end{tikzpicture}
    \caption{A flowchart that summarizes the \texttt{SatGen} framework employed in this study.}
    \label{fig:schematic}
\end{figure*}

Our work builds on the original \texttt{SatGen} model that is presented in \citet{Jiang2021}; we refer the reader to that paper for any additional model details that are omitted below. In what follows, we highlight the salient features of \texttt{SatGen} and discuss in greater detail the new modifications that we make as part of this study. Fig.~\ref{fig:schematic} presents a schematic flowchart that summarizes all of the individual components of our framework. 

\subsection{Merger trees}\label{ssec:trees}

Given an input that includes host halo virial mass, $M_0$, redshift of observation, $z_0$, and underlying cosmology, \texttt{SatGen} generates a user-defined number of halo merger trees that specify the subhalo masses and redshifts at which they are accreted by the main progenitor of each halo. Merger trees are constructed using the method of \citet{Parkinson.etal.08}, which is a modified version of the \texttt{GALFORM} `binary method with accretion' introduced by \citet{Cole2000}. As demonstrated in \citet{Jiang.vdBosch.14} and \citet{vdBosch.etal.14}, this method yields results that are in excellent agreement with numerical simulations.\footnote{As an aside, we acknowledge that several components of the model, including the analytical merger tree algorithm and the orbital parameter distribution model, are still calibrated to agree with cosmological simulations. However, the calibration of these components only depend on properties of \textit{unevolved} subhaloes (i.e., prior to accretion) and hence are not adversely impacted by any artifacts that may manifest in their subsequent tidal evolution.} As detailed in \citet{Jiang2021}, we use the \citet{Parkinson.etal.08} method with the updated set of parameters advocated for by \citet{Benson2017} that are applicable to the \citet{Bryan.Norman.98} virial halo mass definition. Each merger tree is characterized by a minimum progenitor mass, $M_{\rm res}$, which we set to be a fixed fraction, $\psi_{\rm res}$, of the final host halo mass, i.e., $M_\mathrm{res} = \psi_\mathrm{res} M_0$. The value of $\psi_\mathrm{res}$ used varies depending on the application and is specified accordingly. Following \citet{Parkinson.etal.08}, the merger tree is sampled using small time steps of $\Delta z \approx 10^{-3}$; however, in order to reduce memory usage, the tree is subsequently down-sampled to a temporal resolution of $\Delta t = \min[0.1t_\mathrm{dyn}(z), 0.06 \, \mathrm{Gyr}]$, where $t_\mathrm{dyn}(z)=\sqrt{3 \pi/ [16G \Delta_\mathrm{vir}(z) \rho_\rmc(z)]}$ is the redshift-dependent halo dynamical time \citep[see][]{Jiang.vdBosch.16}. The maximum time step of 0.06 Gyr is motivated by convergence tests ran during the calibration of our subhalo mass-loss model, which we discuss in Section~\ref{sssec:massloss}.

Both host haloes and subhaloes at accretion are assumed to follow a Navarro-Frenk-White density profile \citep[hereafter \citetalias{Navarro.etal.97};][]{Navarro.etal.97} with a concentration parameter, $c_\mathrm{vir}$,\footnote{We use $c_\mathrm{vir,h}$ and $c_\mathrm{vir,s}$ to refer to host- and subhalo concentrations, respectively.} that depends on mass and redshift (or time) according to the model introduced by \citet{Zhao2009}:
\begin{equation}\label{eqn:conc}
    c_\mathrm{vir} (M_\mathrm{vir}, t) = 4.0 \left[ 1 + \left( \frac{t}{3.75 t_{0.04}} \right)^{8.4} \right]^{1/8}.
\end{equation}
Thus, the concentration of the halo at a proper time, $t$, is determined based on the time at which its main progenitor has accumulated a mass of $0.04 M_\mathrm{vir}(t)$, denoted $t_{0.04}$. Each branch of the merger tree has its own virial mass accretion history, $M_\mathrm{vir}(t)$, that tracks the halo from the time that it attains a mass of $0.04 M_\mathrm{res} = 0.04 \psi_\mathrm{res} M_0$ until the time that it merges into a more massive halo. Note that in order to have well-defined concentrations for all progenitor haloes down to a `leaf mass' of $M_\mathrm{res}$, we track the main progenitor branch of each leaf further back in time down to $0.04M_\mathrm{res}$.

\texttt{SatGen} tracks subhaloes of \textit{all orders}. The main branch, which follows the main progenitor of the $z=z_0$ host halo back in time, is considered to be order-$0$. Subhaloes that are directly accreted onto the main host are order-$1$. These subhaloes themselves can host sub-subhaloes, which are order-$2$, and so on. We use an inclusive mass definition in our merger trees, which means that the summed mass of all order-$k$ subhaloes is included in the mass of their order-$(k-1)$ host. In some of our results (e.g., the SHMFs), we consider subhaloes of all orders; however, due to the inclusive mass definition, we only consider order-$1$ subhaloes for other results (e.g., the substructure mass fraction).

\subsection{Orbit integration}\label{ssec:orbitint}

Upon accretion, the initial orbital configuration (i.e., location on the virial sphere, orientation of the orbital plane, and the initial velocity vector) of each subhalo is drawn at random using the state-of-the-art universal infall model of \citet[][see Appendix~\ref{app:initorbits} for details]{Li2020}. Note that this is a significant and important improvement over the approach taken in the original \texttt{SatGen} paper, where it was assumed that all subhaloes initially have an orbital energy of $E_{\rm orb} = V^2_{\rm vir,h}/2 + \Phi_\rmh(r_{\rm vir})$, where $V_{\rm vir,h}$ and $\Phi_\rmh$ are the instantaneous virial velocity and potential of the host halo, and a specific orbital angular momentum of $L_{\rm orb} = \eta r_{\rm vir} V_{\rm vir}$, where $\eta \in [0,1]$ is drawn from a simple sinusoidal probability distribution, $p(\eta)= \pi \sin(\pi \eta) / 2$.

Subhalo orbits are subsequently integrated according to the evolving potential of the immediate host and a simple prescription for dynamical friction. In particular, subhaloes are treated as point masses with phase space coordinates that are updated at each time step by integrating the following equation of motion:
\begin{equation}
    \ddot{\boldsymbol{r}} = -\nabla \Phi_\rmh + \boldsymbol{a}_\mathrm{DF} .
\end{equation}
The integration is performed using a fourth-order Runge--Kutta method. Here, $\boldsymbol{r}$ is the host-centric position vector of the subhalo and $\boldsymbol{a}_\mathrm{DF}$ is the acceleration due to dynamical friction (DF). The latter is modeled using the standard approach of \citet{Chandrasekhar.43}, which gives the acceleration as 
\begin{equation}\label{DF}
    \boldsymbol{a}_\mathrm{DF} = - 4\pi G^2 m \ln\Lambda \rho(\boldsymbol{r}) F(<v) \frac{\boldsymbol{v}}{v^3}
\end{equation}
\citep[see][]{MBW10}. Here, $\ln \Lambda = \ln(M/m)$ is the Coulomb logarithm, $M$ and $m$ are the instantaneous masses of the host and subhalo, respectively, $\rho(\boldsymbol{r})$ is the host \citetalias{Navarro.etal.97} density profile, $\boldsymbol{v}$ is the relative velocity of the subhalo with respect to the host, and $F(<v)$ is the fraction of local host particles contributing to dynamical friction. The velocity distribution of the background particles is assumed to be Maxwellian and isotropic such that
\begin{equation}
    F(< v_\mathrm{rel}) = \mathrm{erf}(X) - \frac{2 X}{\sqrt{\pi}} \rme^{-X^2}\,.
\end{equation}
Here, $X \equiv v_\mathrm{rel} / (\sqrt{2} \sigma)$, where $\sigma(\boldsymbol{r})$ is the one-dimensional isotropic velocity dispersion of the host, which we compute using the Jeans equation for hydrostatic equilibrium in a spherical system \citep[e.g.,][]{Binney.Tremaine.08}. We use the orbital velocity of the subhalo for $\boldsymbol{v}_\mathrm{rel}$, ignoring the spin of the host halo.

Because of its simplicity and ability to produce results in reasonable agreement with simulations, equation~\eqref{DF} has long been the standard approach for capturing dynamical friction in semi-analytical models. However, it is based on a number of assumptions (i.e., a point particle moving in an isotropic, homogeneous background of field particles) that are clearly not justified when modeling the orbital evolution of dark matter subhaloes. In order to account for these \citep[and other, see][]{MBW10} inherent shortcomings, we multiply $\boldsymbol{a}_\mathrm{DF}$ by a corrective factor, $\beta_{\rm DF}$, of order unity. We treat $\beta_{\rm DF}$ as a free parameter, which allows us to adjust the overall strength of dynamical friction (see Section~\ref{ssec:dfstrength}).

\subsection{Tidal stripping}
\label{ssec:tidstrip}

As a subhalo orbits its host, it is subjected to tidal stripping and tidal shock heating. As discussed in detail in \citet{vdBosch.etal.2018a}, neither of these processes can be rigorously treated analytically. Consequently, all previous semi-analytical models of subhalo evolution have calibrated their treatments using cosmological simulations, thereby inheriting any shortcomings present within such simulations (i.e., artificial disruption). The primary goal of this work is to build a semi-analytical model of DM substructure evolution that is calibrated in a way such that its results are not sensitive to such numerical artifacts. We achieve this by calibrating our model against \textit{DASH}, a large suite of idealized, high-resolution $N$-body simulations that track individual, live $N$-body subhaloes as they orbit a fixed, analytical host halo potential \citep[][]{Ogiya2019}. Both the host halo and the initial $N$-body subhalo in \textit{DASH} are modeled as spherical NFW haloes. \textit{DASH} consists of 2,253 simulations spanning a wide range of relevant parameter space, including initial orbital energy and angular momentum, as well as the concentration parameters of both the host- and subhalo. The library consists of various data products generated from each simulation, including the phase space coordinates of the subhalo centre-of-mass, the subhalo radial density profile, $\rho(l,t)$, and the bound mass fraction, $\fbound \equiv m(t)/m_\mathrm{acc}$, where $m_\mathrm{acc}$ is the initial subhalo virial mass (i.e., the subhalo mass at accretion), each of which are recorded over 301 snapshots of time. Below, we use these results to calibrate a model that describes the evolution of the density profiles (Section~\ref{sssec:eshdp}) and bound masses (Section~\ref{sssec:massloss}) of subhaloes as they orbit their host (note that the former is required for modeling the latter).

\subsubsection{The evolved subhalo density profile (ESHDP)}
\label{sssec:eshdp}

In \citetalias{Green.vdBosch.19}, we used \textit{DASH} to calibrate a model that describes how the internal structure of a subhalo evolves in response to tidal stripping and heating. In particular, motivated by the work of \citet{Hayashi.etal.03} and \citet{Penarrubia.etal.10}, \citetalias{Green.vdBosch.19} present a `transfer function' that describes the density profile of a tidally stripped subhalo as a function of its initial density profile and its instantaneous bound mass fraction, $\fbound$. Consequently, the density profile of a subhalo at any time, $t$, is given by
\begin{equation}\label{eqn:eshdp}
 \rho_\rms(l,t) = H(l|\, \fbound(t), c_\mathrm{vir,s}) \, \rho_\rms(l,t_{\rm acc}),
\end{equation}
where $c_\mathrm{vir,s}$ is the concentration of the subhalo \textit{at accretion} and $t_\mathrm{acc}$ denotes the time of accretion. The \textit{DASH}-calibrated transfer function is given by
\begin{equation}\label{eqn:tf}
    H(l |\, \fbound, c_\mathrm{vir,s}) = \frac{f_{\rm te}}{1 + \left( \frac{l}{l_\rms} \left[\frac{l_\mathrm{vir} - l_{\rm te}}{l_\mathrm{vir}l_{\rm te}}\right]\right)^\delta} .
\end{equation}
Here, $f_\mathrm{te}$, $l_\mathrm{te}$, and $\delta$ are all expressed as fitting functions that depend on both $\fbound$ and $c_\mathrm{vir,s}$ (see equations~[6]--[8] and Table~1 of \citetalias{Green.vdBosch.19}; note that $l_{\rm te} \equiv r_{\rm te}$), whereas $l_\rms$ and $l_\mathrm{vir}$ are the scale radius and virial radius of the \citetalias{Navarro.etal.97} subhalo \textit{at accretion}.\footnote{The dependence on $c_\mathrm{vir,s}$ went unnoticed in \citet{Hayashi.etal.03} and \citet{Penarrubia.etal.10}, both of which only studied subhaloes with a single concentration ($c_\mathrm{vir,s}=10$ and $23.1$, respectively).} The transfer function describes how the outer density profile of the subhalo steepens from $\rmd\ln\rho/\rmd\ln l = -3$ (i.e., the outer slope of the initial \citetalias{Navarro.etal.97} profile) to roughly $-(5-6)$ as the initial subhalo mass is stripped away. In addition, the central density of the subhalo is lowered as $\fbound$ decreases, which is primarily a consequence of re-virialization in response to mass loss.

\subsubsection{Mass-loss rate}
\label{sssec:massloss}

A common approach to modeling the combined impact of tidal stripping and heating \citep[e.g.,][]{Taffoni.etal.03, Zentner.Bullock.03, Oguri.Lee.04, Zentner.etal.05, Pullen.etal.14}, which we adopt as well, is to assume that over each time step, $\Delta t$, some portion, $\Delta m$, of the subhalo mass outside of its instantaneous tidal radius, $l_\rmt$, is stripped away. In particular, we set
\begin{equation}\label{eqn:massloss}
    \Delta m = - \alpha \frac{\Delta t}{t_\mathrm{char}} m(> l_\rmt)\,.
\end{equation}
Here, $\alpha$ is a fudge factor that controls the stripping efficiency,
\begin{equation}\label{eqn:tchar}
    t_\mathrm{char} = \sqrt{\frac{3 \pi}{16 G \bar{\rho}_\rmh(r)}}\,
\end{equation}
is the characteristic orbital time of the subhalo (identical to the dynamical time introduced in Section~\ref{ssec:trees}), with $r$ the instantaneous, host-centric radius of the subhalo and $\bar{\rho}_\rmh(r)$ the mean density of the host halo within $r$, and 
\begin{equation}\label{eqn:rt}
    l_\rmt = r \Bigg[ \frac{m(<l_\rmt)/M(<r)}{2+\frac{\Omega^2(t) r^3}{GM(<r)} - \frac{\rmd \ln M}{\rmd \ln r}\big|_{r}} \Bigg]^{1/3}
\end{equation}
\citep[][]{King.62}, with $\Omega(t) = \vert \br \times \boldsymbol{v} \vert/ r^2$ the instantaneous angular orbital velocity of the subhalo. We have also experimented with other definitions of $t_\mathrm{char}$ and $l_\rmt$ but find that this combination, when used in conjunction with equation~(\ref{eqn:massloss}), is able to reproduce the \textit{DASH} results most accurately.\footnote{The tidal radius is only an approximation of the zero-velocity surface, which itself is neither spherical nor infinitesimally thin, and different authors often adopt different definitions. See  \citet{Read.etal.06b}, \citet{Tollet.etal.17}, and \citet{vdBosch.etal.2018a} for detailed discussions.}

We use the $m(t)/ m_\mathrm{acc}$ trajectories from the \textit{DASH} simulations to calibrate $\alpha$ as follows. Given the data products from a particular \textit{DASH} simulation, we create interpolators for $r(t)$, $\Omega(t)$, and $m(t)/m_\mathrm{acc}$. In order to avoid transient behavior in the simulations that results from the instantaneous introduction of a subhalo into its host potential \citep[see][]{Ogiya2019}, we initialize our model based on the properties of the \textit{DASH} subhalo at the beginning of its second orbit (i.e., after it has returned to apocentre for the first time). Given a choice for $\Delta t$ and $\alpha$, we evolve $m(t)/m_\mathrm{acc}$ using equation~\eqref{eqn:massloss}, where we set $m(> l_\rmt) = m(t) - m(< l_\rmt)$. Here, $m(< l_\rmt)$ is computed using the ESHDP of equation~(\ref{eqn:eshdp}), which depends on the instantaneous value of $m(t)/m_\mathrm{acc}$ and the initial $c_\mathrm{vir,s}$,\footnote{This enclosed mass profile is not analytical. Hence, in \texttt{SatGen}, we provide an interpolator for $m(<l)/m_\mathrm{acc}$ (and $\sigma(l)$, the one-dimensional isotropic velocity dispersion), which is itself a function of $l$, $m(t)/m_\mathrm{acc}$, and $c_\mathrm{vir,s}$. We interpolate over $\log[m(t)/m_\mathrm{acc}]$ and $\log(c_\mathrm{vir,s})$ using cubic B-splines and patch the surfaces together in $\log(l)$-space linearly.} and we demand that $\Delta m \geq 0$ such that the subhalo mass decreases monotonically. For each combination of simulation (indexed by $i$) and $\alpha$ value, we compute a cost function, $C(i | \alpha)$, which is simply the mean squared residual in $\log[m(t_j) / m_\mathrm{acc}]$ between our model and \textit{DASH} averaged over all $n_{\mathrm{apo},i}$ apocentric passages subsequent to the initialization of our model (indexed by $j$). We then determine the total cost for a given $\alpha$ by computing the mean of the $C(i | \alpha)$ taken over all of the \textit{DASH} simulations, which can be written explicitly as
\begin{equation}
    C(\alpha) = \sum_i^{n_\mathrm{sim}} \frac{C(i | \alpha)}{n_\mathrm{sim}} = \sum_i^{n_\mathrm{sim}} \, \sum_j^{n_{\mathrm{apo},i}} \frac{\log^2[m_{\mathrm{model},i}(t_j) / m_{\mathrm{DASH},i}(t_j)]}{n_\mathrm{sim} n_{\mathrm{apo},i}}.
\end{equation}
We emphasize that this cost function weighs each simulation equally, which is motivated by the fact that \textit{DASH} samples the parameter space of orbits and halo concentrations according to a cosmological simulation-inferred joint probability distribution. The cost function depends somewhat on the time step used to integrate the model predictions (see equation~[\ref{eqn:massloss}]), but we find that the results converge with $\Delta t =0.06$ Gyr, which we adopt throughout as the \textit{maximum} time step for integrating the evolution of the subhalo in \texttt{SatGen}.

We find that $C(\alpha)$ is minimized for $\alpha \simeq 0.6$, for which the root-mean-square error in the apocentric mass predictions is 0.097 dex. In order to look for any secondary parametric dependence that the optimal $\alpha$ may have, we determine the best-fit $\alpha$ on a per-simulation basis, which we denote $\alpha_i$. We then look at the correlation between $\alpha_i$ and the concentrations of the host- and subhalo as well as with the orbital parameters. We find that $\alpha_i$ depends strongly on $c_\mathrm{vir,s}/c_\mathrm{vir,h}$. By binning the simulations by $c_\mathrm{vir,s}/c_\mathrm{vir,h}$ and taking the median $\alpha_i$ in each bin, we find a power-law relation that is well fit by
\begin{equation}\label{eqn:alpha}
    \alpha = 0.55 \Big( \frac{c_\mathrm{vir,s}/c_\mathrm{vir,h}}{2}\Big)^{-1/3} .
\end{equation}
This relation captures the fact that subhaloes that are more compact relative to their host are more resilient to stripping. We adopt this parametrization of $\alpha$ in \texttt{SatGen}, emphasizing that, for typical values of $c_\mathrm{vir,s} / c_\mathrm{vir,h}$, the concentration-dependence has a $\lta 30\%$ effect. In determining $\alpha$, we use the instantaneous host $c_\mathrm{vir,h}$ (which evolves as long as the host itself has not yet become a subhalo) whereas the subhalo $c_\mathrm{vir,s}$ is fixed to its value at infall. 

Although it is tempting to compare our best-fit value for $\alpha$ to that of previous semi-analytical models that rely on equation~(\ref{eqn:massloss}), such a comparison is frustrated by the fact that different studies have used different forms for $t_\mathrm{char}$ and/or $l_\rmt$ (see \citealt{vdBosch.etal.2018a, Drakos2020} for detailed discussions). In addition, none of the previous studies have accounted for the detailed evolution of the subhalo density profile (as in, e.g., equation~[\ref{eqn:eshdp}]), rendering such a comparison moot. We do emphasize, though, that by calibrating our model to the idealized \textit{DASH} simulations, rather than to cosmological simulations, such as in \citet{Zentner.etal.05} and \citet{Pullen.etal.14}, our calibration is not adversely impacted by potential issues resulting from artificial disruption.

\begin{figure}
    \centering
    \includegraphics[width=0.47\textwidth]{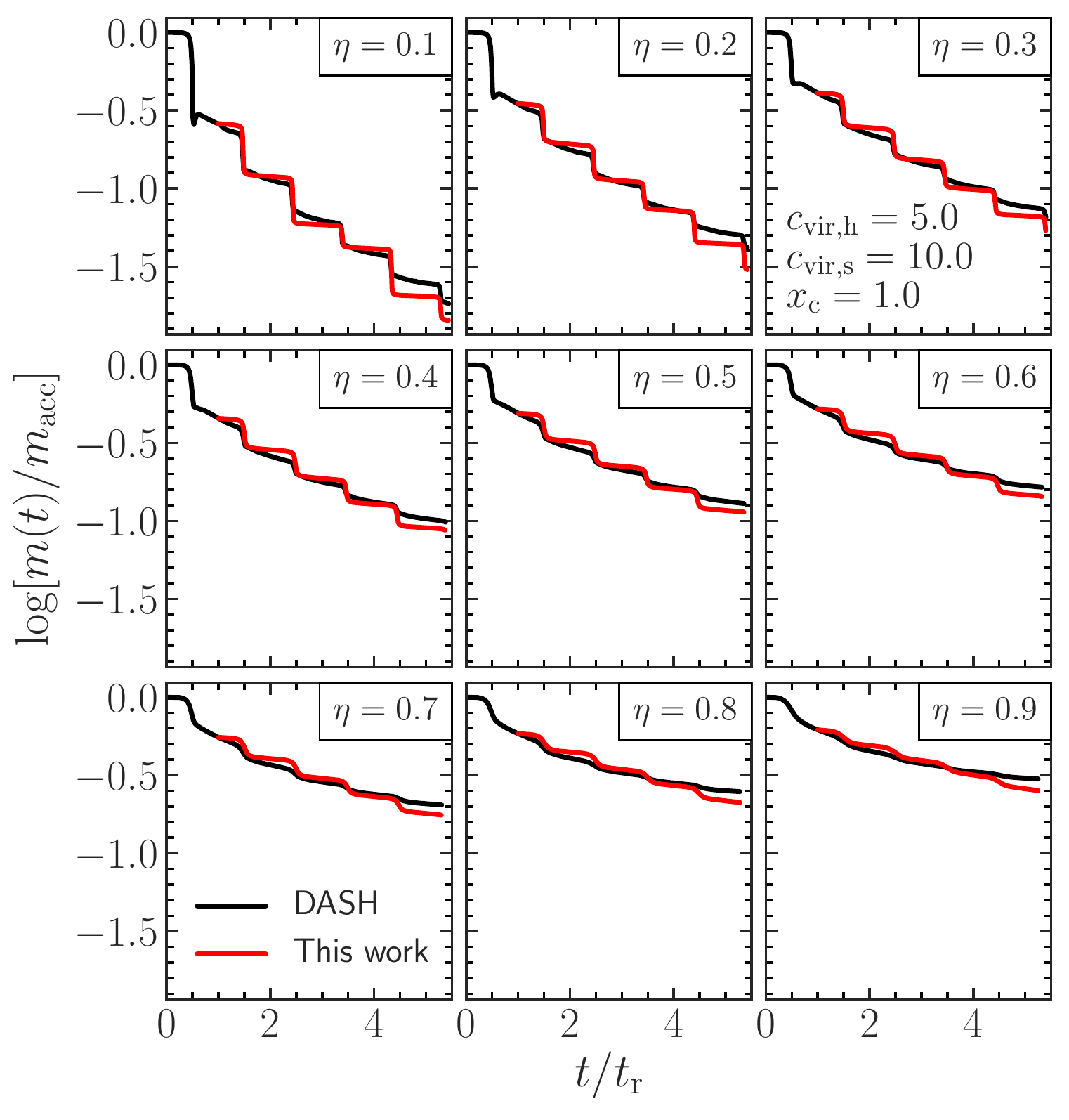}
    \caption{A comparison between our calibrated mass-loss model predictions and the \textit{DASH} $m(t)/m_\mathrm{acc}$ trajectories of several simulations. The times are normalized by the radial orbital period, $t_\rmr$. We fix $c_\mathrm{vir,h}=5$, $c_\mathrm{vir,s}=10$, and $x_\rmc=1$ \citep[these parameters are typical of systems seen in cosmological simulations; see, e.g.,][]{Ogiya2019}, demonstrating that the model performance is strong over a wide range of circularity values, ranging from highly elliptical ($\eta = 0.1$) to nearly circular orbits ($\eta = 0.9$).}
    \label{fig:fbt}
\end{figure}

Fig.~\ref{fig:fbt} compares the $m(t)/m_\mathrm{acc}$ trajectories of several \textit{DASH} simulations (black lines) to predictions based on our mass-loss model (red lines). In each case, $c_\mathrm{vir,h}=5$, $c_\mathrm{vir,s}=10$, and the orbital energy, $E$, is that of a circular orbit at the virial radius of the host (i.e., $x_\rmc \equiv  r_\rmc (E) / r_\mathrm{vir} = 1$, where $r_\rmc (E)$ is the radius of a circular orbit with energy $E$). Different panels correspond to different values of the orbital circularity, $\eta \equiv L/L_\rmc(E)$, as indicated, where $L$ is the orbital angular momentum and $L_\rmc(E)$ is the angular momentum of a circular orbit with the same orbital energy as that of the subhalo. Clearly, our model tracks the \textit{DASH} $m(t)/m_\mathrm{acc}$ curves quite faithfully over ${\sim}5$ radial orbital periods. Importantly, the performance of the model is strong over the full range of $\eta$, spanning from orbits that are close to radial ($\eta = 0.1$) to those that are close to circular ($\eta = 0.9$), a feat that has proven difficult for previous semi-analytical models of subhalo mass evolution \citep[cf.][]{Penarrubia.etal.10, Drakos2020}. Although not shown, we emphasize that the model performs comparably for other configurations as well. In particular, the concentration dependence built into the parametrization of the stripping efficiency (i.e., equation~[\ref{eqn:alpha}]) considerably improves the predictions made for systems with $c_\mathrm{vir,s}/ c_\mathrm{vir,h}$ ratios that deviate significantly from two. 

\begin{figure}
    \centering
    \includegraphics[width=0.47\textwidth]{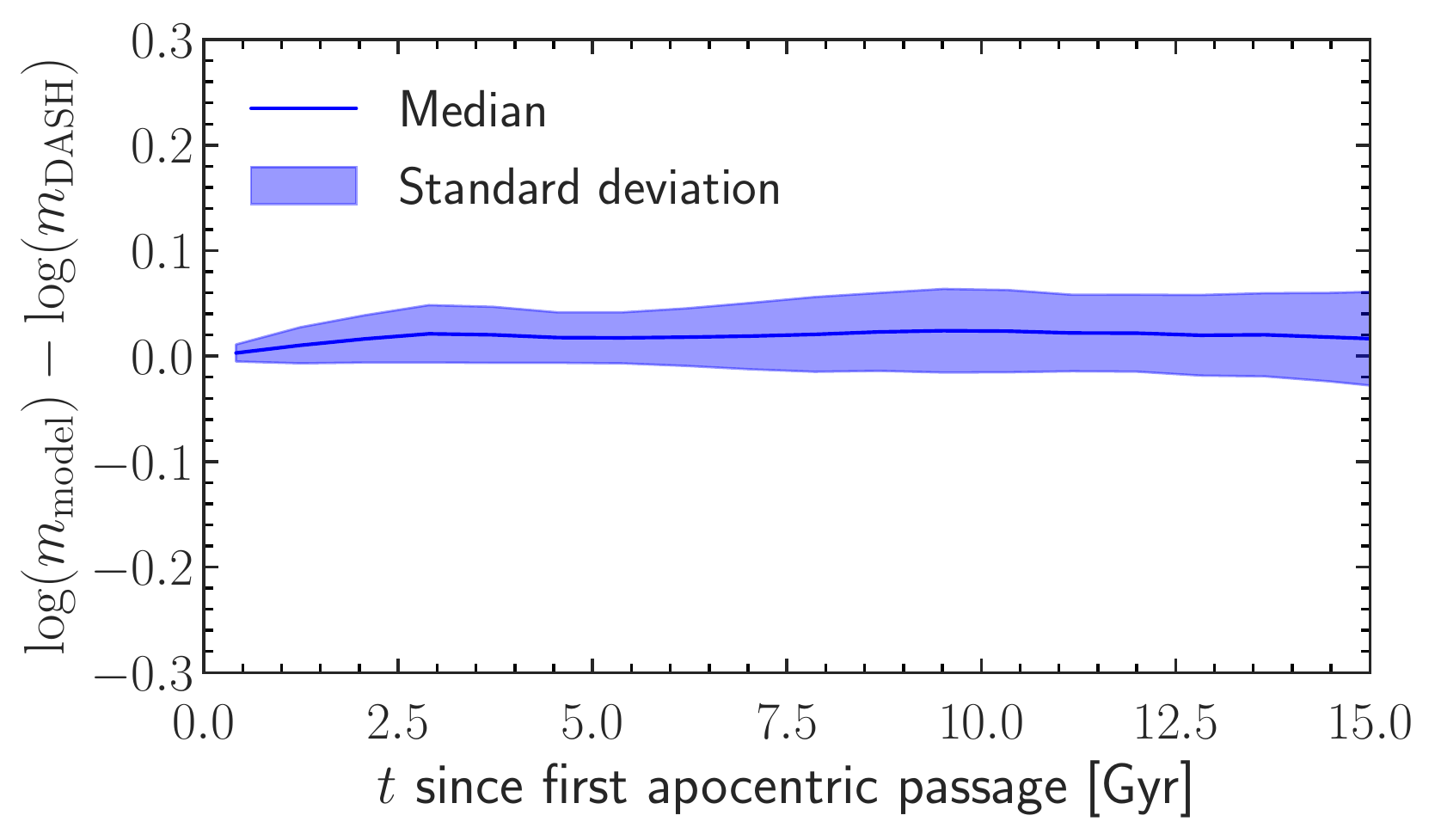}
    \caption{The time evolution of the median and standard deviation of the log-residuals between our mass-loss model predictions and the simulated mass trajectories taken over the ensemble of DASH simulations. The model performs well over the full parameter space, with minimal bias and scatter for longer than a Hubble time.}
    \label{fig:mass_residuals}
\end{figure}

We use the mass-loss model to predict the mass evolution of every simulated DASH subhalo. In Fig.~\ref{fig:mass_residuals}, we plot the time evolution of the median and standard deviation of the log-residuals between our model predictions and the DASH mass trajectories. We find that the mass-loss model performs well over the full parameter space, with minimal bias and scatter for longer than a Hubble time. After 15 Gyr of evolution, the scatter in the log-residuals of our mass-loss model reaches only 0.04 dex; hence, the impact of mass evolution error will be subdominant to the intrinsic halo-to-halo variance in our quantities of interest.

\subsubsection{Stripping of higher-order substructure}
\label{ssec:higherorder}

In addition to the treatment of subhalo mass loss, \texttt{SatGen} also implements a procedure for the splashback release of higher-order subhaloes. Specifically, each time step that an order-$k$ subhalo lies outside of the tidal radius of its order-$(k-1)$ host, it has a probability of $\min[\alpha \Delta t / t_\mathrm{char}(r), 1]$ of being released from its host and becoming an order-$(k-1)$ subhalo. Here, $\alpha$ and $t_\mathrm{char}(r)$ are computed for the order-$(k-1)$ host with respect to its order-$(k-2)$ parent, which is responsible for stripping off the order-$k$ subhalo. In the event of release, the phase space coordinates of the subhalo with respect to its new, order-$(k-2)$ host are the superposition of its original coordinates with respect to its old, order-$(k-1)$ host and those of the old host with respect to the order-$(k-2)$ system. The remaining bound mass of the original, order-$k$ subhalo is instantaneously removed from the mass of its old, order-$(k-1)$ parent in order to enforce mass conservation.

\subsubsection{Resolution limits}
\label{ssec:reslim}

As discussed in Section~\ref{ssec:trees}, \texttt{SatGen} has a merger tree resolution limit, which sets the smallest subhalo mass at accretion to $\psi_\mathrm{res} M_0$. Such a limit is necessary in order to maintain computational feasibility, as the size of the merger tree grows exponentially with decreasing $\psi_\mathrm{res}$. However, once accreted, a subhalo is evolved in \texttt{SatGen} for as long as its mass $m\geq \phi_\mathrm{res} \, m_\mathrm{acc}$. Here, $\phi_\mathrm{res}$ is the imposed resolution limit for the bound mass fraction. Our default is to set $\phi_\mathrm{res} = \psi_\mathrm{res}$, which ensures that the least (most) massive subhaloes are tracked down to $m = \phi_\mathrm{res} \psi_\mathrm{res} M_0$ ($m = \psi_\mathrm{res} M_0$). In what follows, both resolution limits are adjusted depending on the specific topic that is under investigation.

\subsection{Artificial disruption}\label{ssec:artdisr}

Recently, \citet{vdBosch.etal.2018a} and \citet{vdBosch.etal.2018b} carried out a comprehensive analytical and numerical study focused on subhalo disruption. Using simple, physical arguments, the authors demonstrate that the inner remnant of a \citetalias{Navarro.etal.97} subhalo should survive even when tidal shock heating has injected an amount of energy that is many multiples of the binding energy of the subhalo and/or tidal stripping has removed more than 99.9\% of the initial subhalo mass. This claim is confirmed using idealized $N$-body simulations of subhalo evolution (similar to \textit{DASH}), with the authors concluding that the majority of subhalo disruption seen in cosmological simulations is numerical in nature. 

Let us use \textit{Bolshoi} as our example cosmological simulation for considering the rate of artificial disruption. \Citet{vdBosch.17} used the merger trees from \textit{Bolshoi} to separate subhalo evolution into several unique channels. Of these channels, the disruption (D) and withering (W) branches pertain specifically to numerical subhalo disruption. A subhalo in one snapshot that evolves along the D channel has no descendent at \textit{any} subsequent snapshot. On the other hand, a subhalo that evolves along the W channel has a descendent in the subsequent snapshot that falls below the 50 particle resolution limit imposed by the author. By studying these branches, \citet{vdBosch.17} concludes that artificial disruption (D) occurs at a rate of $2.4\%/$Gyr and falling below the mass limit (W) occurs at a rate of ${\sim}10\%/$Gyr. When combined, the total numerical disruption (W + D) rate in \text{Bolshoi} is roughly $13\%/$Gyr, resulting in ${\sim}65\%$ of subhaloes accreted at $z=1$ being numerically disrupted by the present day, in good agreement with independent estimates made by \cite{Han.etal.16} and \cite{Jiang.vdBosch.17}. As long as simulations have a finite number of particles, the W channel will exist. However, its significance diminishes as simulation resolution limits move toward smaller halo masses that are below all scales of interest. The D channel, on the other hand, is more alarming, since it represents subhaloes, often \textit{well} above the mass limit, that simply disappear from the merger tree. The $2.4\%/$Gyr of the D channel translates to roughly 20\% of subhaloes accreted at $z=1$ being (artificially) disrupted by $z=0$.

In order to assess the overall significance of numerical disruption, we aim to model both the impact of the W branch in isolation as well as the impact of both the W and D channels in combination on the \texttt{SatGen} results. As introduced in Section~\ref{sssec:massloss}, the W branch subhaloes in \texttt{SatGen} are simply those with a final mass that has fallen below the merger tree resolution limit, $\psi_\mathrm{res} M_0$. Although even \texttt{SatGen} has an imposed resolution limit on how far down in $m/m_{\rm acc}$ it tracks a subhalo, we can nevertheless make reasonable predictions in the absence of withering by considering all subhaloes with  $m/m_\mathrm{acc} \geq \phi_\mathrm{res} = 10^{-5}$, which we refer to as the ``wither-free'' fiducial model. Whenever withering is considered, the subhalo mass limit is set to $\psi_\mathrm{res} M_0$ instead.

A key goal of this work is to assess the impact of artificial disruption on the subhalo demographics in cosmological simulations. We are able to do so by adding a model of artificial disruption into \texttt{SatGen} and adjusting its strength (if needed) such that the \texttt{SatGen} predictions (which are inherently free of artificial disruption) reproduce the abundance of subhaloes in a simulation such as \textit{Bolshoi}. This feat also requires properly accounting for the mass resolution limit (withering) of the simulation of interest. We implement a version of the artificial disruption mechanism used in \citet{Jiang.vdBosch.16}, which itself is based on the prescription of \citet{Taylor.Babul.04}. A subhalo is marked as artificially disrupted when its mass, $m(t)$, falls below its `disruption mass', given by
\begin{equation}\label{eqn:mdis}
    m_\mathrm{dis} = m_\mathrm{acc}(<f_\mathrm{dis} l_\rms) = m_\mathrm{acc} \frac{f(f_\mathrm{dis})}{f(c_\mathrm{vir,s})} .
\end{equation}
Here, $m_\mathrm{acc}(<l)$ denotes the enclosed  \citetalias{Navarro.etal.97} mass profile of the subhalo at accretion, and $f(x) = \ln(1+x)-x/(1+x)$. The sensitivity of haloes to artificial disruption is set by $f_\mathrm{dis}$, which represents the effective radius that a halo can be stripped down to before being disrupted. Under this prescription, haloes with a larger initial concentration are more resilient to disruption. This approach to modeling (artificial) disruption has been employed in previous semi-analytical models \citep[e.g.,][]{Hayashi.etal.03, Taylor.Babul.04, Zentner.etal.05}, with $f_\mathrm{dis}$ ranging from $0.1$ to $2.0$. 

\begin{figure}
    \centering
    \includegraphics[width=0.47\textwidth]{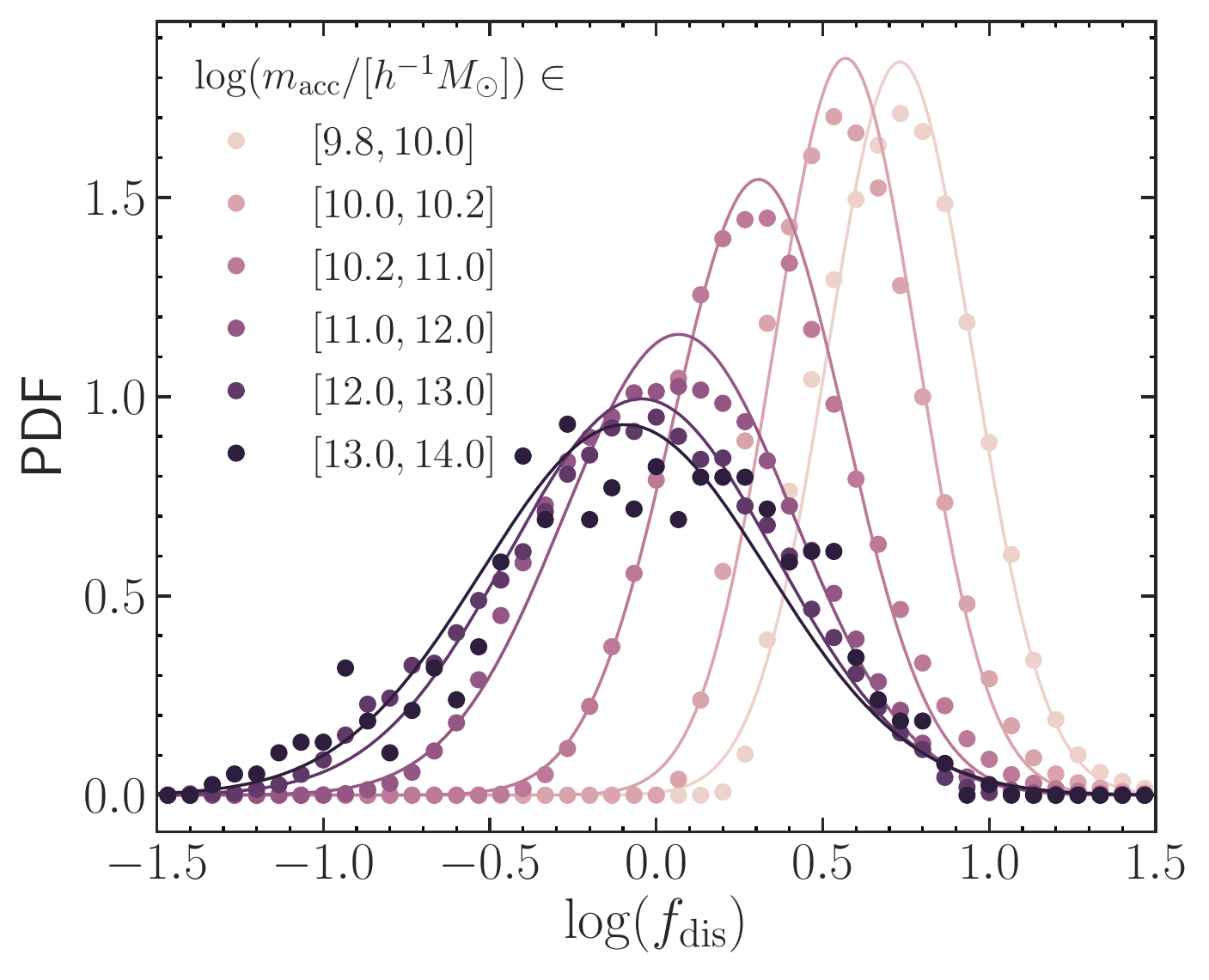}
    \caption{The $\log(f_\mathrm{dis})$ distribution of disrupted \textit{Bolshoi} subhaloes. Here, $f_\mathrm{dis}$ is a proxy for the mass below which a particular subhalo is artificially disrupted in the simulation (see equation~[\ref{eqn:mdis}]). Each color denotes a different $m_\mathrm{acc}$ bin. The points are calculated using all \textit{Bolshoi} subhaloes that disrupt at $z \geq 0.0148$. The solid curves correspond to our model that is fit to the \textit{Bolshoi} disruption data (equation~[\ref{eqn:dismodel}]). The $f_\mathrm{dis}$ are distributed log-normal, with $\mu$ decreasing (and $\sigma$ increasing) as $m_\mathrm{acc}$ is increased (up to a saturation point, above which the distribution remains fixed).}
    \label{fig:fdis_model}
\end{figure}

Rather than select a fixed value for $f_\mathrm{dis}$, \citet{Jiang.vdBosch.16} randomly sampled $f_\mathrm{dis}$ for each subhalo from a universal log-normal distribution. We augment this approach by calibrating a more general model of $f_\mathrm{dis}$ that takes into account a dependence on $m_\mathrm{acc}$ that we identify in the \textit{Bolshoi} subhaloes. Using all halo catalogues from \textit{Bolshoi}\footnote{Available at \href{http://www.slac.stanford.edu/~behroozi/Bolshoi_Catalogs/}{http://www.slac.stanford.edu/~behroozi/Bolshoi\_Catalogs/}} with $z \geq 0.0148$,\footnote{We omit using the several snapshots closer to $z=0$ in order to avoid contaminating the D branch with instances of snapshot-limited failed phantom-patching \citep[see discussion in][]{vdBosch.17}.} we extract $m_\mathrm{acc}$, $c_\mathrm{vir,s}$, and $m_\mathrm{dis}$ from all D channel subhaloes from which the $f_\mathrm{dis}$ of each corresponding subhalo is calculated. We find that the distribution of $f_\mathrm{dis}$ has minimal dependence on redshift and host halo mass, but has a strong dependence on $m_\mathrm{acc}$. As shown in Fig.~\ref{fig:fdis_model}, when binned by $m_\mathrm{acc}$, the $f_\mathrm{dis}$ distribution is roughly log-normal with a log-mean, $\mu$, and log-variance, $\sigma^2$, that increases and decreases, respectively, with decreasing $m_\mathrm{acc}$. This indicates that subhaloes that are more massive at accretion are less likely to undergo artificial disruption. However, note that this trend in $m_\mathrm{acc}$-space appears to saturate at the massive end. Motivated by these findings, we model $f_\mathrm{dis}(m_\mathrm{acc})$ as a log-normal with
\begin{equation}\label{eqn:dismodel}
  \begin{split}
    & \mu = A + B \, \left[1+\big(\log(m_\mathrm{acc})+C\big)^{-2}\right]^{-1/2},\;\;\mathrm{and} \\
    & \sigma =  D + E \mu + F \mu^2\,. 
  \end{split}
\end{equation}
Using maximum likelihood estimation, we obtain the best-fit parameters of $(A,B,C,D,E,F) = (3.08, -3.26, -8.89, 0.38, -0.51, 0.40)$. The corresponding best-fit model is indicated as solid lines in Fig.~\ref{fig:fdis_model} and captures all of the salient details of the data.

When modeling artificial disruption in \texttt{SatGen}, we randomly draw a value of $f_\mathrm{dis}$ from the log-normal distribution described by equation~\eqref{eqn:dismodel} for each subhalo at accretion. Subsequently, the subhalo is marked as artificially disrupted once its mass drops below its assigned $m_\mathrm{dis}$, which is computed using equation~\eqref{eqn:mdis}. By applying this artificial disruption mechanism, \texttt{SatGen} is able to faithfully reproduce the statistics of the \textit{Bolshoi} D branch subhaloes. We caution that this particular treatment of artificial disruption is only applicable to \textit{Bolshoi}. Readers interested in modeling artificial disruption in another simulation must first characterize the corresponding $f_\mathrm{dis}$ statistics of the particular simulation.

\subsection{Dynamical friction strength}
\label{ssec:dfstrength}

In order to calibrate the overall efficiency of dynamical friction, which we quantify through the correction factor, $\beta_\mathrm{DF}$, we seek a measurement made from  cosmological simulations that is both sensitive to dynamical friction and insensitive to any underlying artificial disruption. In \citet{vdBosch.etal.16}, the authors study the segregation of subhaloes in \textit{Bolshoi}. They measure the mean host-centric radius of subhaloes, $\langle r/ r_\mathrm{vir} \rangle$, as a function of their redshift of accretion, $z_\mathrm{acc}$. 

\begin{figure}
    \centering
    \includegraphics[width=0.47\textwidth]{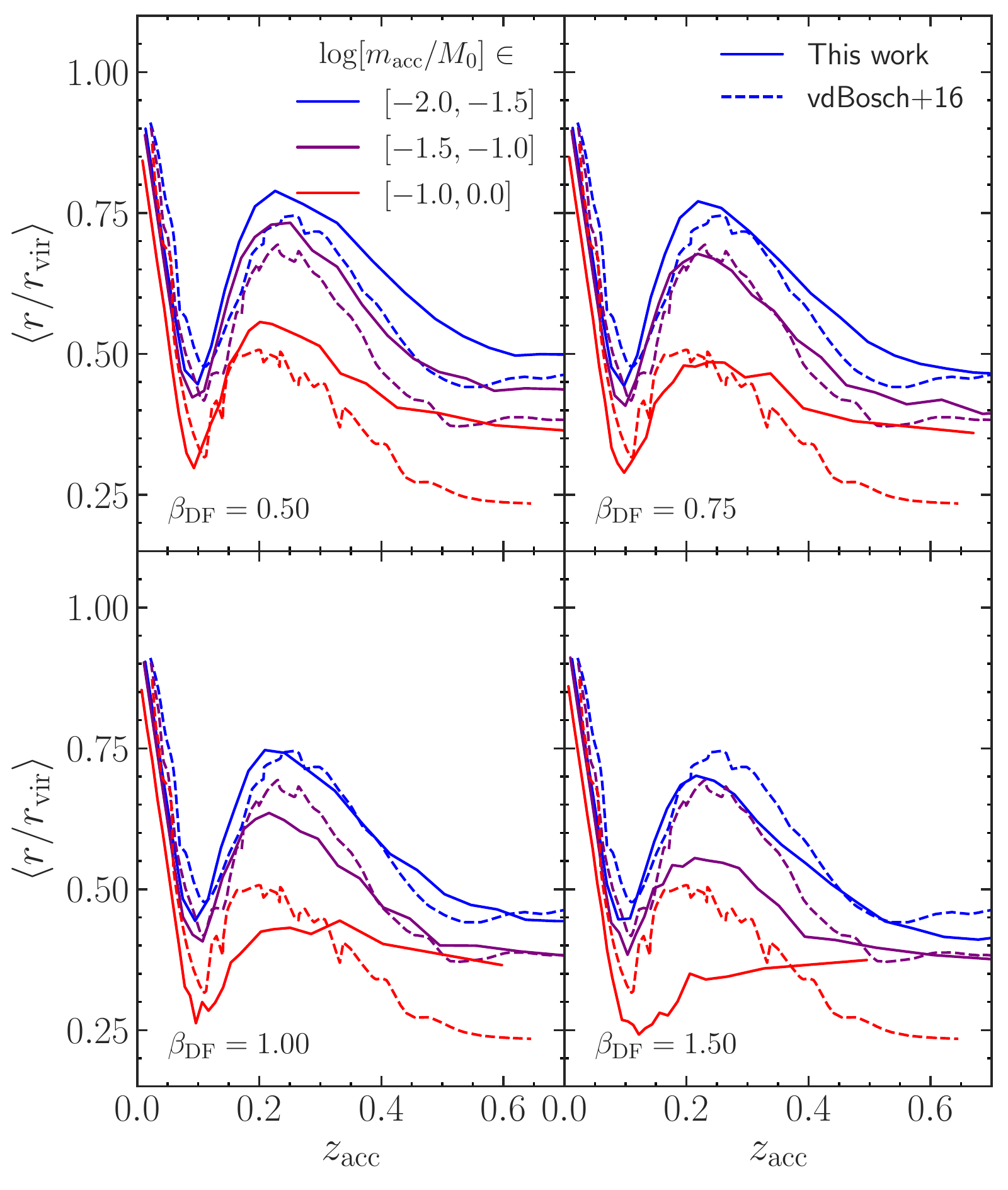}
    \caption{A comparison between the $\langle r/ r_\mathrm{vir} \rangle - z_\mathrm{acc}$ relation of \textit{Bolshoi} subhaloes binned by infall mass relative to $z=0$ host mass, $m_\mathrm{acc}/M_0$ \citep[dashed curves, reproduced from Fig.~10 in][]{vdBosch.etal.16}, and analogous predictions by \texttt{SatGen} (solid curves). Each panel corresponds to a different value of $\beta_\mathrm{DF}$ (indicated in the bottom-left of each panel), which controls the strength of dynamical friction (see Section~\ref{ssec:orbitint}). We adopt $\beta_\mathrm{DF}=0.75$ as the fiducial value used in \texttt{SatGen}, since this yields the best agreement with respect to the peak values of $\langle r/ r_\mathrm{vir} \rangle$ at $z_{\rm acc} \sim 0.25$, which corresponds to the first apocentric passage since infall.}
    \label{fig:rrvir_vs_zacc_binned}
\end{figure}

Plotting $\langle r/ r_\mathrm{vir} \rangle$ (averaged over thousands of subhaloes) as a function of $z_{\rm acc}$ \citep[see Fig.~7 in][]{vdBosch.etal.16} reveals the characteristics of an orbit (for $z_{\rm acc} \lta 0.5$). Subhaloes accreted at $z_{\rm acc} \sim 0.1$ have just reached pericentre for the first time, while those that are at their first apocentric passage since accretion typically were accreted around $z_{\rm acc} \sim 0.25$. Note that phase mixing, which is primarily driven by variance in the orbital periods of subhaloes at infall,\footnote{The efficiency of phase mixing is further enhanced by dynamical friction, which impacts the subhalo orbit differently depending on $m_\mathrm{acc}$, and variance in the host mass accretion history, which itself affects the evolution of the subhalo orbit between infall and the present day.} results in a lack of orbital coherence for subhaloes accreted before $z_\mathrm{acc} \sim 0.5$; this is made apparent by the lack of clear apo- or pericentric passages in $\langle r/ r_\mathrm{vir} \rangle$ at high $z_\mathrm{acc}$. Interestingly,  the $\langle r/ r_\mathrm{vir} \rangle(z_{\rm acc})$ curves show a clear dependence on $m_\mathrm{acc}/M_0$. In particular, subhaloes with larger $m_\mathrm{acc}/M_0$ reach a smaller apocentric $\langle r/ r_\mathrm{vir} \rangle$ at $z_{\rm acc} \sim 0.25$ than their less massive counterparts \citep[see Fig.~10 in][which is reproduced as the dashed lines in Fig.~\ref{fig:rrvir_vs_zacc_binned}]{vdBosch.etal.16}. This is a manifestation of dynamical friction, which allows us to calibrate $\beta_{\rm DF}$ as follows.\footnote{Since artificial disruption is rare for subhaloes that were only accreted recently, this feature is not significantly impacted by artificial disruption.}

We construct a set of ${\sim}$45,000 merger trees (with $\psi_\mathrm{res}=10^{-3}$) with host masses consistent with the ${\sim}$9,000 host halo sample used in \citet{vdBosch.etal.16} --- we augment our sample by generating five trees per unique host mass. We evolve the subhaloes with \texttt{SatGen}, repeating the procedure for several values of $\beta_\mathrm{DF}$ covering the range $[0, 1.5]$. We apply the same selection function as used in \citet{vdBosch.etal.16}: we only consider subhaloes with $m(z=0)/M_0 \geq 10^{-3}$, $m_\mathrm{acc}/M_0 \geq 10^{-2}$, and $m(z=0)/m_\mathrm{acc} \geq 10^{-1}$. We first bin the subhaloes by $m_\mathrm{acc}/M_0$ and then compute $\langle r/ r_\mathrm{vir} \rangle$ in $z_\mathrm{acc}$ bins, which are chosen such that the number of subhaloes in each bin is the same.

Fig.~\ref{fig:rrvir_vs_zacc_binned} shows the resulting $\langle r/ r_\mathrm{vir} \rangle - z_\mathrm{acc}$ relation for four values of $\beta_\mathrm{DF}$ as indicated. Clearly, when $\beta_\mathrm{DF}=0.75$, \texttt{SatGen} is able to very closely reproduce the simulation results. For $\beta_\mathrm{DF}=0.5$ ($1.0$), \texttt{SatGen} yields apocentric $\langle r/ r_\mathrm{vir} \rangle$ that are too large (small) relative to \textit{Bolshoi}, with the disagreement being more significant for the subhaloes with larger $m_\mathrm{acc}/M_0$ that are more strongly influenced by dynamical friction. These findings are independent of whether or not we incorporate artificial disruption using the method described in Section~\ref{ssec:artdisr}, which is consistent with the notion that the $\langle r/ r_\mathrm{vir} \rangle - z_\mathrm{acc}$ relation should be relatively insensitive to artificial disruption (at least for $z_{\rm acc} \lta 0.5$). Hence, in what follows, we adopt $\beta_\mathrm{DF}=0.75$ as our fiducial dynamical friction strength. In Section~\ref{ssec:fsub}, we quantify the impact of $\beta_\mathrm{DF}$ on our substructure mass fraction predictions (by comparing to the `natural' case of $\beta_\mathrm{DF}=1$), demonstrating that our results are insensitive to its exact value.

\section{Results}
\label{sec:results}

Given a host halo mass, $M_0$, target redshift, $z_0$, and requested number of individual trees, $N_\mathrm{tree}$, \texttt{SatGen} produces $N_\mathrm{tree}$ subhalo catalogs at each redshift time step until $z_0$. These catalogs trace the mass and phase space coordinates of each subhalo over its evolution. In this section, we present the results of these \texttt{SatGen} subhalo catalogs and make comparisons to \textit{Bolshoi}. We begin by studying SHMFs (and subhalo maximum circular velocity functions), comparing \texttt{SatGen} results with and without the artificial disruption mechanism and discuss the significant impact of splashback subhaloes (Section~\ref{ssec:shmf}). In Section~\ref{ssec:rad}, we proceed to incorporate position data by calculating the radial profile and the (projected) enclosed substructure mass fraction, $F_\mathrm{sub}(< R)$. In Section~\ref{ssec:fsub}, we quantify how $f_\mathrm{sub}(< r_\mathrm{vir})$ varies with both $M_0$ and resolution limit, $\psi_\mathrm{res}$. We also quantify the impact of model parameters (i.e., stripping efficiency and DF strength) on $f_\mathrm{sub}$ predictions. Lastly, in Section~\ref{ssec:disrate}, we estimate the total rate of numerical disruption that occurs via the W and D channels modeled by \texttt{SatGen}, which we compare to the numerical disruption rate of \textit{Bolshoi} haloes \citep[as measured by][]{vdBosch.17}.

\subsection{Subhalo mass/velocity functions}\label{ssec:shmf}

\begin{figure*}
    \centering
    \includegraphics[width=\textwidth]{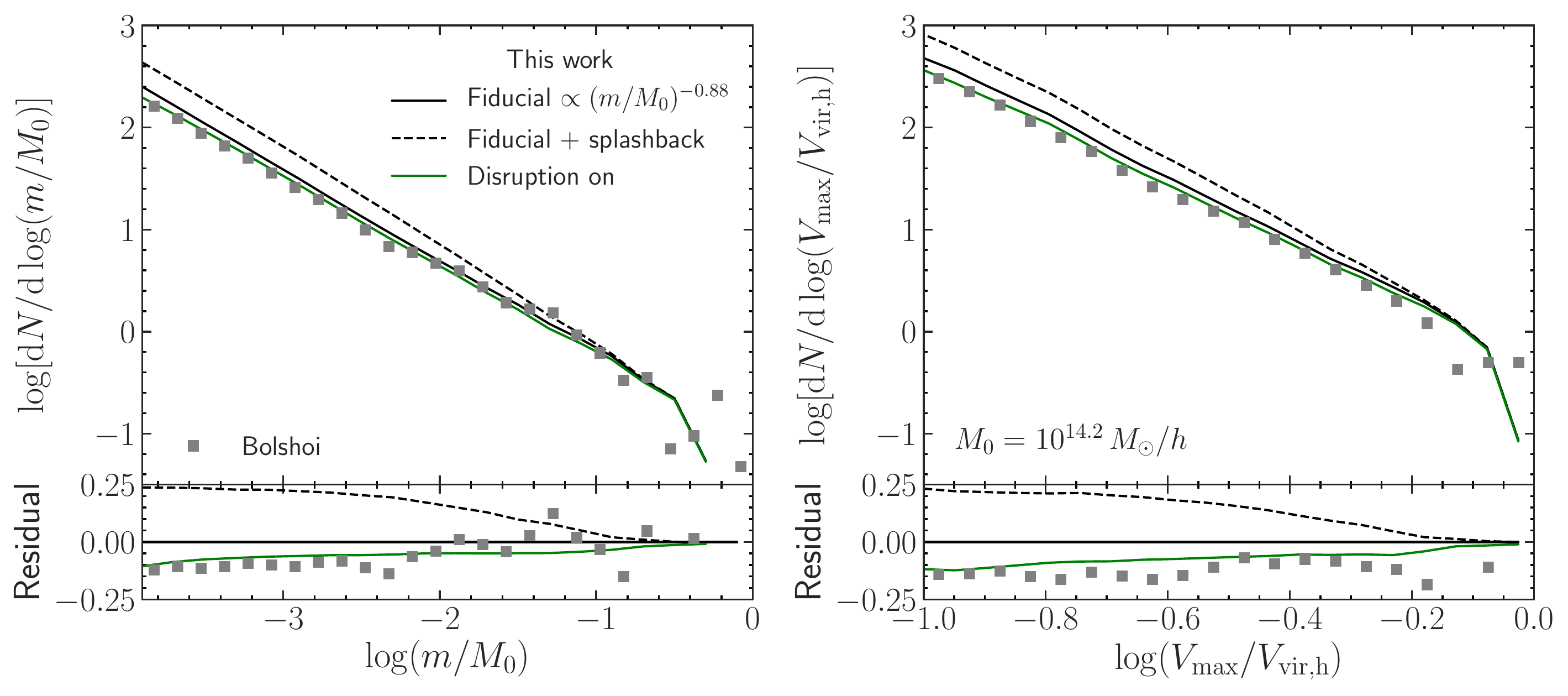}
    \caption{The subhalo mass function (SHMF; \textit{left}) and subhalo maximum circular velocity function (SHVF; \textit{right}) predictions for a host halo with $M_0 = 10^{14.2}\, \Msunh$ at $z=0$ (and virial velocity at $z=0$ denoted by $V_\mathrm{vir,h}$). The \texttt{SatGen} results are averages taken over 10,000 merger trees generated with $\psi_\mathrm{res}=10^{-4}$. These results are compared to the same quantities computed from 282 \textit{Bolshoi} host haloes with $\log(M_0/[\Msunh]) \in [14.0, 14.5]$ (with a mean of 14.2), which are shown as gray squares. The fiducial \texttt{SatGen} predictions (black lines) are used as the baseline for comparison in the residual plots. In the ``fiducial + splashback'' case (dashed black lines), we include subhaloes in the SHMF that are in the merger tree but instantaneously lie outside of the host $r_\mathrm{vir}$ at $z=0$. Lastly, the ``disruption on'' case (green lines) demonstrates the impact of our artificial disruption mechanism (Section~\ref{ssec:artdisr}), which is calibrated to reproduce the statistical properties of \textit{Bolshoi} subhalo disruption (D channel). At the low-$m/M_0$ end, artificial disruption suppresses the \texttt{SatGen} SHMF by $\sim 0.05 - 0.1$ dex, which brings our predictions into good agreement with \textit{Bolshoi}. For $m/M_0 \lesssim 10^{-2.5}$, nearly half of the subhaloes lie outside $r_\mathrm{vir}$ \citep[consistent with][]{Bakels2021}.}
    \label{fig:shmf}
\end{figure*}

We turn our attention to the \texttt{SatGen} predictions of the subhalo mass function for a $10^{14.2}\, \Msunh$ host. In a cosmological simulation, the SHMF, $\rmd N / \rmd \log(m/M_0)$, is calculated using subhaloes \textit{of all orders} enclosed within the virial radius of the host. Note that since we use an inclusive mass definition and consider all orders of substructure, the total substructure mass is \textit{not} the mass-weighted integral of the SHMF. The left-hand panel of Fig.~\ref{fig:shmf} shows the mean SHMF computed from 10,000 trees (with $\psi_\mathrm{res}=10^{-4}$). For comparison, the filled symbols indicate the mean SHMF of the 282 \textit{Bolshoi} host haloes with $\log(M_0/[\Msunh]) \in [14.0, 14.5]$ (with a mean of 14.2). On the high-$m/M_0$ end, the \textit{Bolshoi} SHMF is somewhat noisy due to limited halo statistics. However, a comparison at the low-$m/M_0$ end illustrates that \texttt{SatGen} predicts a $\sim 0.1$ dex enhancement in the SHMF relative to \textit{Bolshoi}.

If the primary cause of the disagreement between the \texttt{SatGen} and \textit{Bolshoi} SHMFs is artificial disruption, then the application of our artificial disruption mechanism (Section~\ref{ssec:artdisr}) should result in better agreement between the model and simulation results. Indeed, Fig.~\ref{fig:shmf} shows that ``turning on'' D channel disruption suppresses the \texttt{SatGen} SHMF by $\sim 0.05 - 0.1$ dex at low $m/M_0$, bringing it into closer agreement with \textit{Bolshoi}. Restricting to $-3.9 \leq \log(m/M_0) \leq -0.8$, we fit a power-law to the SHMF of the form $\rmd N / \rmd \log(m/M_0) \propto A (m/M_0)^B$. For the fiducial \texttt{SatGen} results, we find $A=-1.066$ and $B=-0.885$, whereas the disruption mechanism slightly suppresses both the normalization and the magnitude of the slope, resulting in $A=-1.085$ and $B=-0.868$.\footnote{These SHMF slopes are consistent with previous work, which typically find $-1.0 \lesssim B \lesssim -0.8$ \citep[e.g.,][]{Boylan-Kolchin.etal.10, Gao.etal.12, vdBosch.Jiang.16}.} The reduced slope is a consequence of the $m_\mathrm{acc}$-dependence of our artificial disruption model. Note that the \textit{Bolshoi} SHMF is too noisy to compute a reliable estimate of the slope over the same mass range, but it agrees well visually with the \texttt{SatGen} ``disruption on'' results.

Thus far, these results suggest that artificial disruption has, at most, a $\sim 20\%$ impact on the SHMF of well-resolved host haloes, with the difference being strongest at low $m/M_0$. We discuss more quantitatively the impact of disruption and its dependence on halo mass relative to the simulation resolution limit in Section~\ref{ssec:fsub}, which focuses on the substructure mass fraction. The modest decrease in the SHMF normalization due to disruption predicted by \texttt{SatGen} is considerably smaller than the factor-of-two suppression suggested by the \citetalias{Green.vdBosch.19}-interpretation of the \citet{Jiang.vdBosch.16} model. This is because their orbit-averaged model did not take into account the impact of splashback haloes, which are subhaloes that have previously fallen within the host $r_\mathrm{vir}$ (thus becoming included in the halo merger tree) but instantaneously lie outside of $r_\mathrm{vir}$ at $z=0$ (and therefore are typically not included in simulation-based measurements of the SHMF). \citet{Benson2017} briefly discusses this limitation of standard EPS-based approaches to substructure modeling, concluding that a full dynamical model (such as \texttt{SatGen}) is necessary in order to properly account for splashback haloes. In Fig.~\ref{fig:shmf}, we illustrate that when splashback haloes are included in the SHMF, the subhalo abundance is enhanced by $\sim 0.2 - 0.25$ dex on the low-$m/M_0$ end relative to the fiducial model. When the ``fiducial + splashback'' curve is compared directly to \textit{Bolshoi}, we find the same $\sim 0.3$ dex (factor of two) difference as \citetalias{Green.vdBosch.19}. This highlights the importance of properly accounting for splashback haloes by integrating subhalo orbits. Consistent with these predictions, \citet{Bakels2021} recently reported that roughly half of all subhaloes lie outside of $1.2r_\mathrm{200c}$ (approximately $r_\mathrm{vir}$) in a sample of galaxy- to group-mass host haloes studied in a cosmological simulation \citep[consistent with previous work by, e.g.,][]{Gill.etal.04, Ludlow2009}.

In addition to mass, another property of subhaloes that is often used \citep[especially in subhalo abundance matching, e.g.,][]{Trujillo-Gomez.etal.11, Reddick.etal.13, Hearin.etal.13, Zentner.etal.14} is its maximum circular velocity, $V_{\rm max}$. Hence, we also present results for the subhalo maximum circular velocity function (SHVF), $\rmd N/\rmd\ln(V_\mathrm{max}/V_\mathrm{vir,h})$, where $V_\mathrm{vir,h}$ denotes the virial velocity of the host halo at $z=0$. Since the enclosed mass profile corresponding to the \citetalias{Green.vdBosch.19} ESHDP is not analytical, we compute $V_\mathrm{max}$ by multiplying the subhalo's maximum circular velocity at accretion, 
\begin{equation}
    V_\mathrm{max,acc} = \sqrt{\frac{G m_\mathrm{acc}}{l_\mathrm{vir}} \times \frac{0.216 \, c_\mathrm{vir,s}}{f(c_\mathrm{vir,s})}}
\end{equation}
\citep[][]{Bullock.etal.01}, by the `tidal track' \citep[][]{Penarrubia2008} expression for $V_\mathrm{max} / V_\mathrm{max,acc}$, given by equation~(11) in \citetalias{Green.vdBosch.19}. This tidal track itself is a function of both $m/m_\mathrm{acc}$ and $c_\mathrm{vir,s}$, as given by equations (12) and (13) in \citetalias{Green.vdBosch.19}.

Using the same 10,000 trees as those used to compute the SHMF, we obtain the SHVF shown in the right-hand panel of Fig.~\ref{fig:shmf}. Just as for the SHMF, the filled symbols indicate the corresponding result computed from the 282 \textit{Bolshoi} host haloes with $\log(M_0/[\Msunh]) \in [14.0, 14.5]$. As is evident, the abundance of subhaloes with $\log(V_\mathrm{max}/V_\mathrm{vir,h}) \lta -0.4$ predicted by \texttt{SatGen} is about 0.15 dex higher than that of \textit{Bolshoi}. However, when including artificial disruption, the \texttt{SatGen} predictions once again agree closely with the simulation results.

\subsection{Radial profiles}
\label{ssec:rad}

Having looked at the subhalo mass and velocity functions, we proceed to incorporate additional spatial information by considering several other quantities of interest. First, we measure the subhalo radial distribution, $\rmd \tilde{N} / \rmd x^3 |_\mathrm{sub}$, as the number of subhaloes per unit shell volume as a function of $x = r / r_\mathrm{vir}$, which we normalize to unity at $r_\mathrm{vir}$. We assess the radial bias of the subhaloes by comparing $\rmd \tilde{N} / \rmd x^3 |_\mathrm{sub}$ to the \citetalias{Navarro.etal.97} profile of the host halo, $\rmd \tilde{N} / \rmd x^3 |_\mathrm{NFW}$, which we also write as a function of $x$ and normalize to unity at $r_\mathrm{vir}$. The `bias function' is simply the ratio between $\rmd \tilde{N} / \rmd x^3 |_\mathrm{sub}$ and $\rmd \tilde{N} / \rmd x^3 |_\mathrm{NFW}$, which tends to unity when the subhalo distribution is unbiased with respect to the density profile of the host. We incorporate subhaloes of all orders when computing $\rmd \tilde{N} / \rmd x^3 |_\mathrm{sub}$. The second quantity of interest is the fraction of mass enclosed within a given {\it projected} host-centric radius that is bound in subhaloes. We define this quantity as
\begin{equation}\label{eqn:Fsub}
    F_\mathrm{sub}(< X) = \frac{1}{M(< X)} \, \sum_{X_i < X} m_i \,,
\end{equation}
where $X = R / R_\mathrm{vir}$, $R$ is the projected radius, the sum runs over all \textit{first-order} subhaloes (due to the inclusive mass definition) with projected radii within $R$, and $M(<R)$ is the projected mass profile of the \citetalias{Navarro.etal.97} host halo \citep[see][]{Golse2002}. Finally, in Section~\ref{ssec:fsub}, we focus on $f_\mathrm{sub}(< r_\mathrm{vir})$, which is computed in the same way as $F_\mathrm{sub}$ except that three-dimensional radii are used instead.

\subsubsection{Number density and radial bias profiles}

\begin{figure*}
    \centering
    \includegraphics[width=\textwidth]{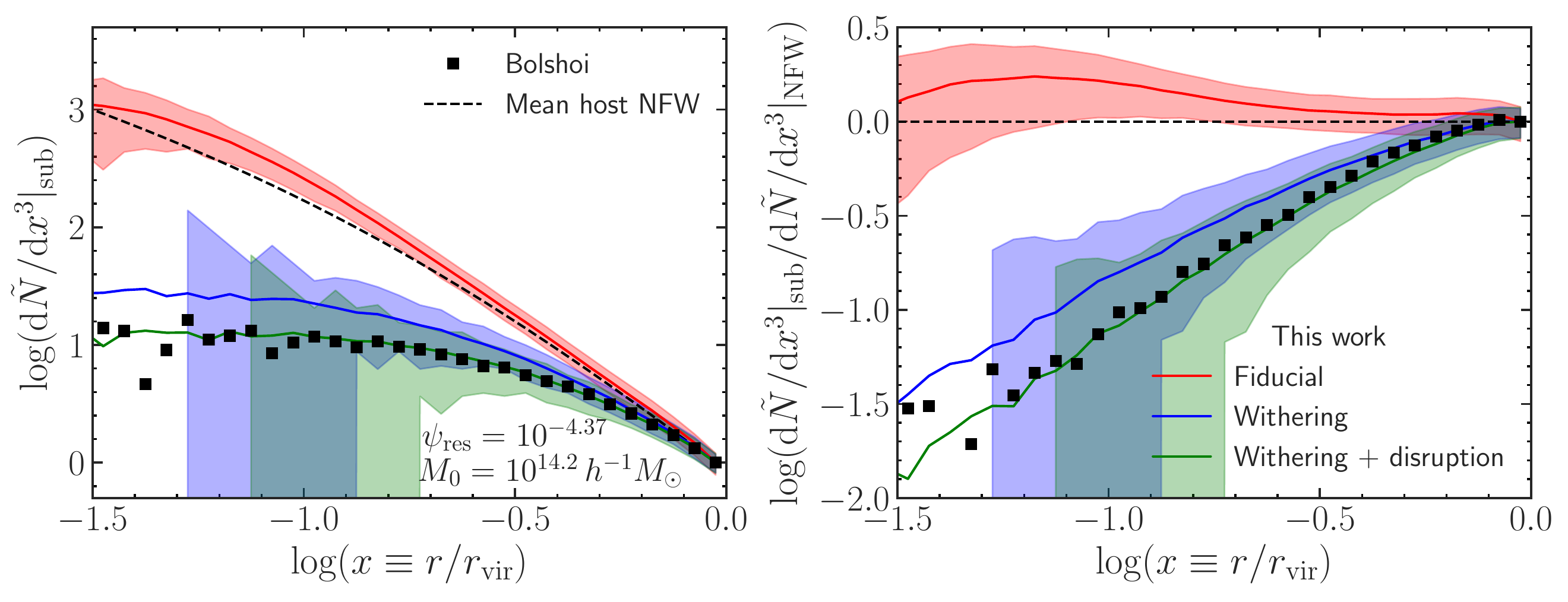}
    \caption{The subhalo radial profile (including subhaloes of all orders), $\rmd \tilde{N} / \rmd x^3|_\mathrm{sub}$ (\textit{left}), normalized to unity at $r_\mathrm{vir}$ and the bias function (\textit{right}), which quantifies how the radial profile differs from the host density profile. The \texttt{SatGen} results are computed from 2,000 merger trees of systems with $M_0 = 10^{14.2}\, \Msunh$ at $z=0$ and a merger tree resolution limit of $\psi_\mathrm{res}=10^{-4.37}$, consistent with the \textit{Bolshoi} resolution limit for hosts of the same mass. The lines represent the sample means and the shaded regions denote the 16--84 percentiles taken over the sample, which quantify the halo-to-halo variance. The fiducial result (red) includes all subhaloes with $m/m_\mathrm{acc} \geq 10^{-5}$ (approximating the lack of a resolution limit), whereas the ``withering'' result (blue) mimics the \textit{Bolshoi} mass limit by only including subhaloes down to $m = \psi_\mathrm{res}M_0$. Lastly, ``withering + disruption'' (green) additionally includes the statistical treatment of artificial disruption (Section~\ref{ssec:artdisr}). The same quantities are computed from the 282 \textit{Bolshoi} host haloes with $\log(M_0/[\Msunh]) \in [14.0, 14.5]$ (black squares). When artificial disruption and withering are taken into account, \texttt{SatGen} is able to exquisitely reproduce the \textit{Bolshoi} bias function. In the absence of such numerical limitations, \texttt{SatGen} predicts a nearly unbiased radial profile \citep[in agreement with][]{Han.etal.16}.}
    \label{fig:spatial_dist}
\end{figure*}

We begin by studying $\rmd \tilde{N} / \rmd x^3|_\mathrm{sub}$ and the corresponding bias function in Fig.~\ref{fig:spatial_dist}. Since we aim to make direct comparisons to \textit{Bolshoi}, we set $\psi_\mathrm{res} = m_\mathrm{res,B} / M_0$, where $m_\mathrm{res,B}=10^{9.83}\, \Msunh$ corresponds to the 50-particle halo limit that we impose on the \textit{Bolshoi} results. For $M_0 = 10^{14.2}\, \Msunh$, this corresponds to $\log(\psi_\mathrm{res}) = -4.37$. We compare the mean results obtained from 2,000 \texttt{SatGen} trees with the mean of the 282 \textit{Bolshoi} host haloes with $\log(M_0/[\Msunh]) \in [14.0, 14.5]$. The shaded regions denote the 16--84 percentiles of the halo-to-halo variance. The \texttt{SatGen} results are shown for three cases. The ``fiducial'' result considers all subhaloes with $m/m_\mathrm{acc} \geq \phi_\mathrm{res}=10^{-5}$, whereas the ``withering'' result is limited to subhaloes with $m \geq \psi_\mathrm{res} M_0$. Lastly, the ``withering + disruption'' result includes the impact of the artificial disruption mechanism and thus can be compared directly to \textit{Bolshoi}.

The radial profile of \textit{Bolshoi} subhaloes becomes increasingly biased towards the central region of the host, something that has been pointed out in numerous previous studies \citep[e.g.,][]{Diemand.Moore.Stadel.04, Springel.etal.08, Han.etal.16}. The \textit{Bolshoi} radial profile and bias function are reproduced exquisitely by \texttt{SatGen}, but only when the impact of both withering and artificial disruption are included. Modeling the simulation mass limit alone is sufficient to reproduce the \textit{Bolshoi} mean curves within the halo-to-halo variance of \texttt{SatGen}; however, when artificial disruption is also taken into account, the mean curves are brought into near perfect agreement. Artificial disruption further suppresses the mean $\rmd \tilde{N} / \rmd x^3|_\mathrm{sub}$ by roughly a factor of two in the central region of the host. When all subhaloes can instead evolve down to $m / m_\mathrm{acc} = 10^{-5}$, regardless of $m_\mathrm{acc}$, we find that the radial bias is completely eliminated. In fact, we obtain a slight overabundance of subhaloes towards the centre. This is due to dynamical friction, as we obtain a fully unbiased radial profile when we set $\beta_\mathrm{DF}=0$ (i.e., no dynamical friction). We note that \citet{Han.etal.16} report a similar finding in the \textit{Aquarius} simulations \citep{Springel.etal.08}. By following the most-bound particle at accretion of all subhaloes (regardless of whether or not the subhalo survives to the present day), they find a dynamical friction-driven overabundance of subhalo remnants in the halo centre that decreases towards a fully unbiased profile as $m_\mathrm{acc}$ decreases. Taken together with \texttt{SatGen}, these results demonstrate that the chief cause of the dearth of subhaloes in the central regions of haloes is the limiting mass resolution of the simulation. It is neither physical nor primarily a manifestation of artificial disruption; the latter only makes a relatively modest impact.

\subsubsection{Projected enclosed substructure fraction}

\begin{figure}
    \centering
    \includegraphics[width=0.47\textwidth]{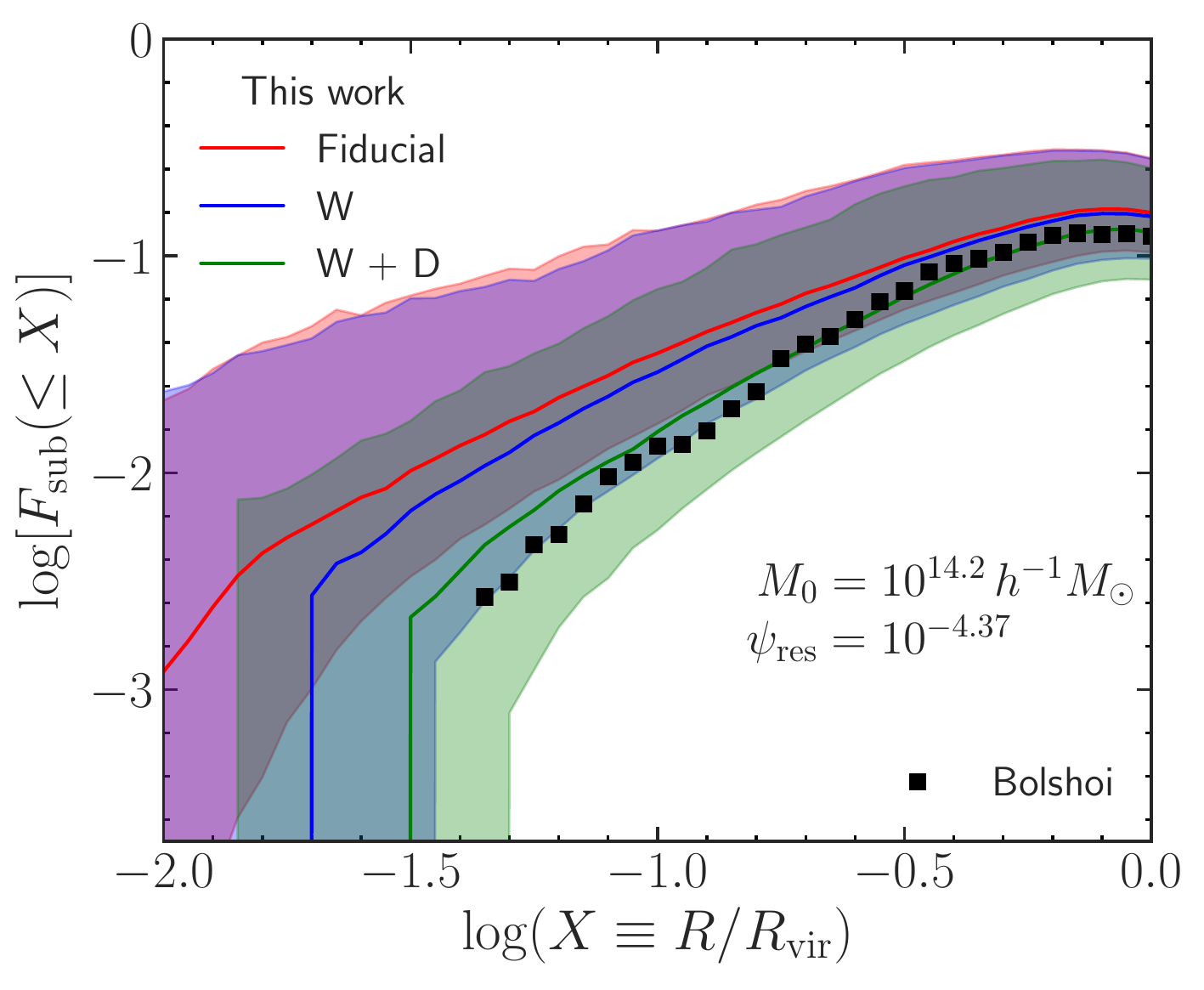}
    \caption{The fraction of mass enclosed within a given projected host-centric radius that is bound in \textit{first-order} subhaloes, $F_\mathrm{sub}(< X)$, as defined in equation~\eqref{eqn:Fsub}. The same \texttt{SatGen} predictions, \textit{Bolshoi} results, and plotting conventions are used as in Fig.~\ref{fig:spatial_dist}, with the exception being that the curves/points instead correspond to the sample medians. When both withering and artificial disruption are emulated, \texttt{SatGen} closely reproduces the \textit{Bolshoi} $F_\mathrm{sub}(< X)$ profile. The substructure mass fraction is only weakly enhanced by the additional resolution in $m/m_\mathrm{acc}$-space afforded by the fiducial model, but it is reasonably sensitive to $\psi_\mathrm{res}$ (see Section~\ref{ssec:fsub}).}
    \label{fig:Fsub}
\end{figure}

Fig.~\ref{fig:Fsub} compares the $F_\mathrm{sub}(< X)$ predictions of \texttt{SatGen} to the results of \textit{Bolshoi}.\footnote{Since the simulation halo catalogs are constructed such that subhaloes must be instantaneously located within the virial radius of their host, we also only consider \texttt{SatGen} subhaloes within the three-dimensional virial extent of the host halo when computing $F_\mathrm{sub}(< X)$.} We use the same \texttt{SatGen} data, simulation data, and plotting conventions as in Fig.~\ref{fig:spatial_dist}, with the only difference being that the curves/points correspond to sample \textit{medians}. Since the (projected) enclosed substructure mass fraction, $F_\mathrm{sub}(< X)$, is simply a mass-weighted radial profile, and since \texttt{SatGen} reproduces both the SHMF and radial profile of \textit{Bolshoi} subhaloes, it should come as little surprise that the model also succeeds at predicting $F_\mathrm{sub}(< X)$. Once again, when we include the effects of both withering and artificial disruption, the model predictions are in nearly perfect agreement with \textit{Bolshoi}. Without accounting for artificial disruption, the median \textit{Bolshoi} $F_\mathrm{sub}(< X)$ curve barely lies within the halo-to-halo variance of the withering-only prediction (for small $X$). Similar to $\rmd \tilde{N} / \rmd x^3|_\mathrm{sub}$, at $X \approx 0.1$, artificial disruption suppresses the median $F_\mathrm{sub}$ by roughly a factor of two. The difference between the fiducial and withering-only model prediction is quite small, which lies in stark contrast to the number density profile. The reason for this is that the enhanced resolution of the fiducial model predominantly results in an increased abundance of highly stripped low-$m_\mathrm{acc}$ subhaloes, which contribute little to the total substructure mass but make up a substantial portion of the number density. As we discuss in Section~\ref{ssec:fsub}, the substructure mass fraction is primarily sensitive to the merger tree resolution ($\psi_\mathrm{res}$).

\subsubsection{Dependence of $\rmd \tilde{N} / \rmd x^3|_\mathrm{sub}$ on subhalo properties}

\begin{figure*}
    \centering
    \includegraphics[width=\textwidth]{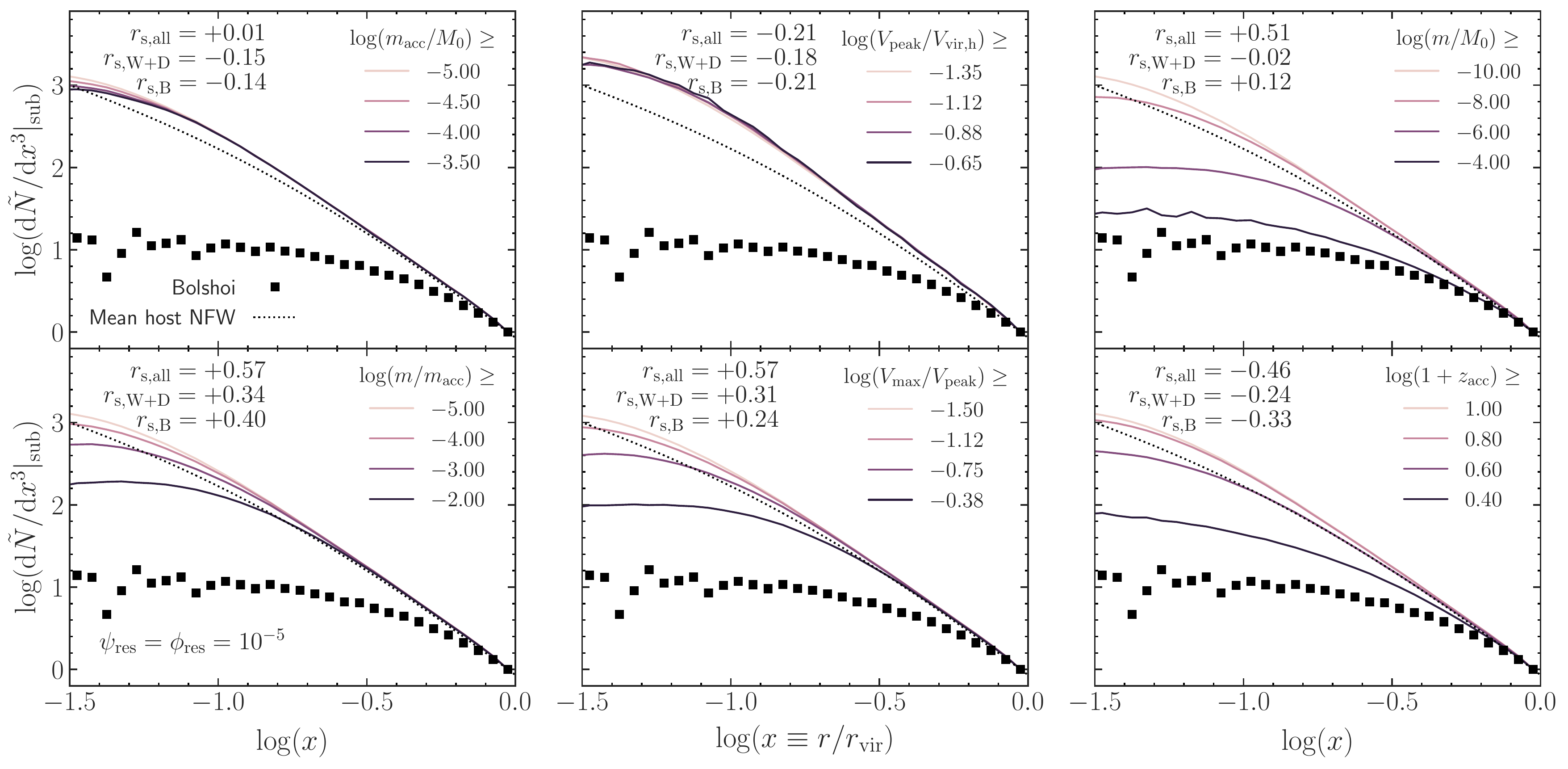}
    \caption{The subhalo radial profiles for 2,000 \texttt{SatGen} trees with $M_0 = 10^{14.2}\, \Msunh$ at $z=0$ and $\psi_\mathrm{res} = \phi_\mathrm{res} = 10^{-5}$. In each panel, the subhaloes are segmented by a different property and the mean $\rmd \tilde{N} / \rmd x^3|_\mathrm{sub}$ is computed for each lower bound. In order to assess the amount of bias, we plot the mean density profile of the host in each panel (dotted line). We overplot the mean \textit{Bolshoi} radial profile of all subhaloes in each panel (black squares). We compute the Spearman coefficient between $r/r_\mathrm{vir}$ and each property for all subhaloes ($r_\mathrm{s,all}$) and the subhaloes that would survive \textit{Bolshoi} withering and artificial disruption ($r_\mathrm{s,W+D}$). There is little dependence on $m_\mathrm{acc}/M_0$ and $V_\mathrm{peak}/V_\mathrm{vir,h}$. As evidenced by the $m/m_\mathrm{acc}$ and $V_\mathrm{max}/V_\mathrm{peak}$ panels, highly-stripped subhaloes follow the host density profile with little bias whereas minimally-stripped systems are less commonly found in the halo centre. Similarly, massive, recently accreted subhaloes are biased towards the outer halo whereas the inclusion of older, less massive subhaloes leads to a more unbiased profile. Withering and artificial disruption tend to weaken (or reverse) the Spearman correlation between each property and $r/r_\mathrm{vir}$, bringing our $r_\mathrm{s,W+D}$ into good agreement with \textit{Bolshoi} \citep[$r_\mathrm{s,B}$, as computed in][]{vdBosch.etal.16}.}
    \label{fig:abundance_dep}
\end{figure*}

We have demonstrated that by properly modeling the effects of withering and artificial disruption on the subhalo population, \texttt{SatGen} can successfully reproduce the radial distribution of simulated subhaloes. We now take a closer look at the $\rmd \tilde{N} / \rmd x^3|_\mathrm{sub}$ predictions of our fiducial model in the absence of these numerical depletion channels. Here, we analyze the results of 2,000 \texttt{SatGen} trees with $M_0 = 10^{14.2}\, \Msunh$ at $z=0$ and $\psi_\mathrm{res} = \phi_\mathrm{res} = 10^{-5}$. Thus, the lowest-$m_\mathrm{acc}$ subhaloes are tracked all the way down to $10^{4.2}\, \Msunh$. In Fig.~\ref{fig:abundance_dep}, we plot the mean $\rmd \tilde{N} / \rmd x^3|_\mathrm{sub}$ computed using subhaloes from these trees while varying the lower limit of several properties: (i) $m_\mathrm{acc} / M_0$, (ii) $m/m_\mathrm{acc}$, (iii) $V_\mathrm{peak} / V_\mathrm{vir,h}$, where $V_\mathrm{peak}$ is the peak $V_\mathrm{max}$ attained by the subhalo over its life (in \texttt{SatGen}, this is equivalent to the $V_\mathrm{max}$ at accretion, $V_\mathrm{max,acc}$), (iv) $V_\mathrm{max} / V_\mathrm{peak}$, (v) $m/M_0$, and (vi) $\log(1 + z_\mathrm{acc})$. For comparison, we also plot the mean \textit{Bolshoi} $\rmd \tilde{N} / \rmd x^3|_\mathrm{sub}$ computed using all subhaloes (i.e., the same as in Fig.~\ref{fig:spatial_dist}) in each panel. Lastly, in order to facilitate a comparison with the segregation study of \citet{vdBosch.etal.16}, we also compute the Spearman rank correlation coefficient, $r_\rms$, between $r/r_\mathrm{vir}$ and each of the six properties computed with all subhaloes (denoted $r_\mathrm{s,all}$) and with subhaloes that would survive \textit{Bolshoi} withering ($m > m_\mathrm{res,B}$) and artificial disruption (denoted $r_\mathrm{s,W+D}$, which can be directly compared to the \textit{Bolshoi} results, $r_\mathrm{s,B}$).

The normalized radial profile is nearly independent of $m_\mathrm{acc} / M_0$ and $V_\mathrm{peak}/V_\mathrm{vir,h}$. When both withering and artificial disruption are taken into account, we find that $r_\rms$ for each of these properties is consistent with the corresponding \textit{Bolshoi} result reported by \citet{vdBosch.etal.16}. The $m/m_\mathrm{acc}$, $m/M_0$, $V_\mathrm{max} / V_\mathrm{peak}$, and $\log(1 + z_\mathrm{acc})$ panels all tell a similar story: older, less massive, and highly-stripped subhaloes follow the host potential with minimal bias. However, recently accreted, massive, and minimally-stripped systems are biased towards the halo outskirts. These trends are weakened by withering and artificial disruption, bringing the $r_\mathrm{s,W+D}$ for each into good agreement with \citet{vdBosch.etal.16}.

Taken together, \texttt{SatGen} predicts that the full subhalo population should exhibit little bias with respect to the host. The dearth of subhaloes in the halo centre, which is found universally in dark matter-only simulations \citep[e.g.,][]{Ghigna.etal.98, Springel.etal.01, Diemand.Moore.Stadel.04, Springel.etal.08, Han.etal.16}, is a result of inadequate resolution that causes the non-physical elimination of old, highly-stripped subhalo remnants that should be abundant in the host core.

\subsection{Substructure mass fractions}\label{ssec:fsub}

We denote the fraction of matter bound into subhaloes within the virial radius of the host as $f_\mathrm{sub}(< r_\mathrm{vir})$. In this section, we study how the \texttt{SatGen} predictions of $f_\mathrm{sub}(< r_\mathrm{vir})$ vary with resolution limit, set by $\psi_\mathrm{res}$, and how they are affected by artificial disruption. In what follows, we write $f_\mathrm{sub} (\psi_\mathrm{res})$ to represent the value of $f_\mathrm{sub}(< r_\mathrm{vir})$ computed from first-order subhaloes with $m > \psi_\mathrm{res} M_0$. Written explicitly,
\begin{equation}\label{eqn:fsub}
    f_\mathrm{sub}(\psi_\mathrm{res}) = \frac{1}{M_0} \sum_{\substack{r < r_\mathrm{vir} \\ m > \psi_\mathrm{res}M_0}} m_i\,,
\end{equation}
where the summation runs over first-order subhaloes only. We conclude the section by demonstrating that $f_\mathrm{sub}$ is insensitive to small changes in the stripping efficiency parameter, $\alpha$, and the dynamical friction strength, $\beta_\mathrm{DF}$.

\subsubsection{Comparison of $f_\mathrm{sub}(\psi_\mathrm{res})$ to \textit{Bolshoi}}

\begin{figure}
    \centering
    \includegraphics[width=0.47\textwidth]{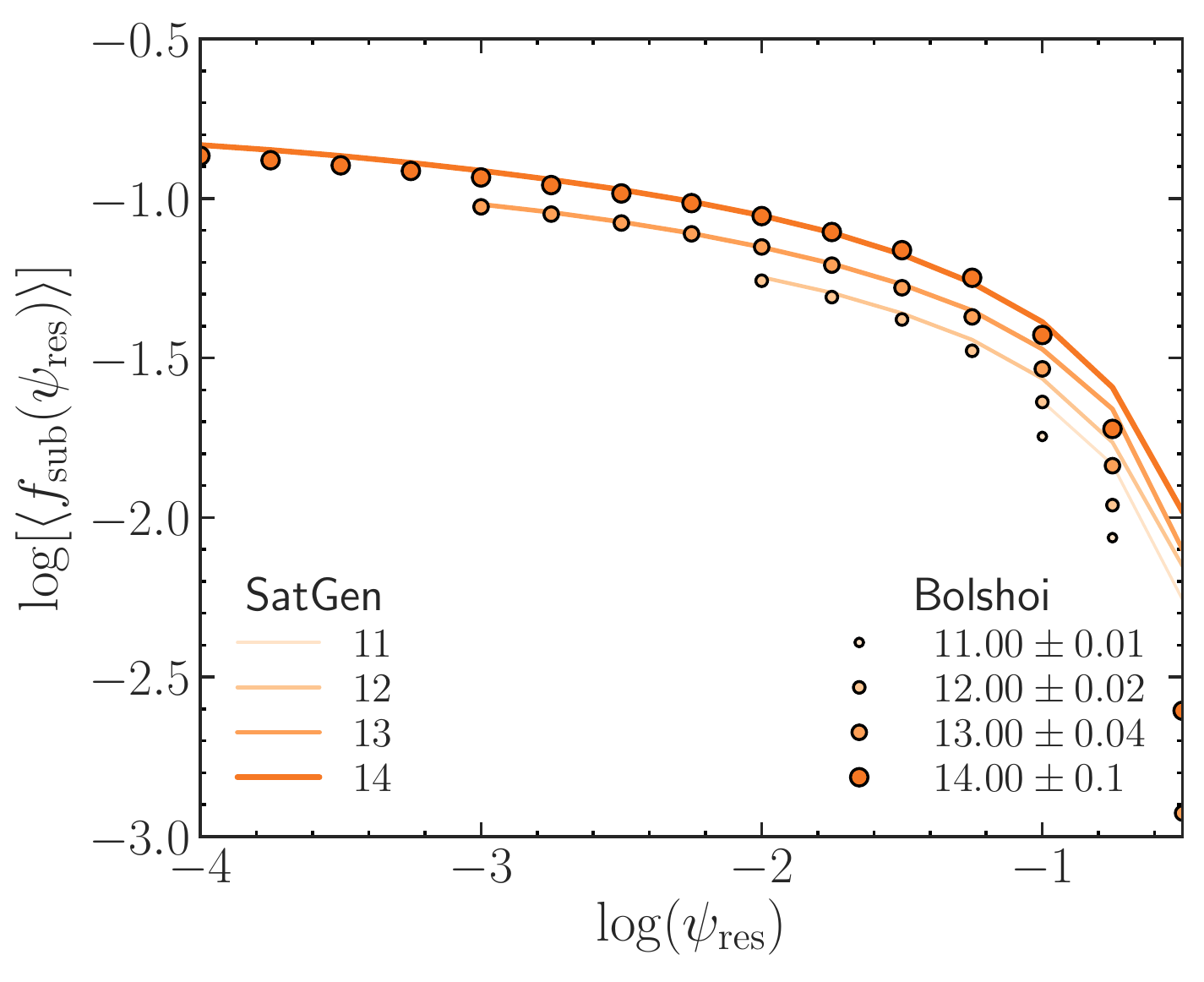}
    \caption{The average fraction of mass bound in subhaloes with $m > \psi_\mathrm{res}M_0$ within $r_\mathrm{vir}$ of host haloes of a given $M_0$, as defined in equation~\eqref{eqn:fsub}. The \texttt{SatGen} predictions are generated using 10,000 trees for each halo mass and are suppressed via the artificial disruption model (Section~\ref{ssec:artdisr}). The masses in the legend are reported in $\log(M_0 / [\Msunh])$. We plot the curves down to the $\psi_\mathrm{res}$ that corresponds to the \textit{Bolshoi} 50-particle mass limit for each $M_0$. The model predictions agree well with the simulation results over a range of $M_0$.}
    \label{fig:fsub_vs_res_wd}
\end{figure}

We begin by demonstrating the $\psi_\mathrm{res}$-dependence of $f_\mathrm{sub}$. Here, we include artificial disruption in the \texttt{SatGen} predictions in order to facilitate comparisons with \textit{Bolshoi}. In Fig.~\ref{fig:fsub_vs_res_wd}, we plot $f_\mathrm{sub}(\psi_\mathrm{res})$ for several different halo masses. The \texttt{SatGen} predictions are obtained using 10,000 trees (with $\psi_\mathrm{res}=10^{-4}$) of haloes with $M_0 = 10^{11 - 14}\, \Msunh$ at $z=0$. The \textit{Bolshoi} results are computed from the 8815, 4713, 1138, and 244 host haloes with $M_0 = 10^{11 \pm 0.01}$, $10^{12 \pm 0.02}$, $10^{13 \pm 0.04}$, and $10^{14 \pm 0.1}\, \Msunh$ at $z=0$. For each halo mass, we only show results down to the $\psi_\mathrm{res}$ that corresponds to the 50-particle \textit{Bolshoi} mass limit. As is evident, when combined with the \textit{Bolshoi}-calibrated artificial disruption mechanism, \texttt{SatGen} is able to accurately reproduce the subhalo statistics (and their resolution dependence) of simulated haloes over several orders of magnitude in mass. There is some tension for $\psi_\mathrm{res} \gta 0.1$, indicating that the SHMFs of \texttt{SatGen} and \textit{Bolshoi} disagree at the massive end. However, this likely reflects uncertainties with the (sub)halo finder used to analyze the simulation results rather than a shortcoming of \texttt{SatGen} (see \citealt{vdBosch.Jiang.16} for a detailed discussion). 

\subsubsection{Mass-dependence and halo-to-halo variance of $f_\mathrm{sub}$}

\begin{figure*}
    \centering
    \includegraphics[width=\textwidth]{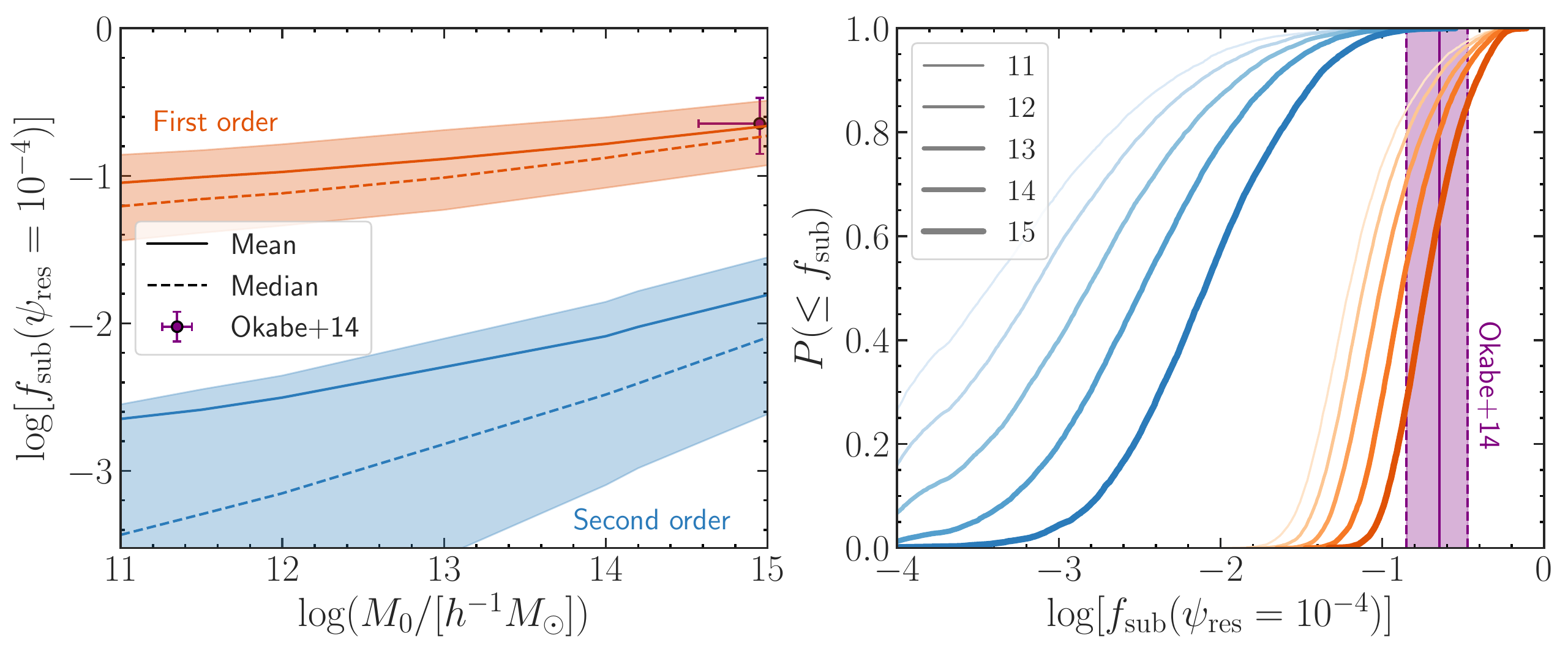}
    \caption{The total (first-order) and second-order $f_\mathrm{sub}(\psi_\mathrm{res} = 10^{-4})$ predictions of \texttt{SatGen} in the absence of artificial disruption. The mean, median, and 16--84 percentile halo-to-halo variance of $f_\mathrm{sub}$ (\textit{left}) as well as the corresponding cumulative distribution function (\textit{right}) are computed using 10,000 trees of haloes with $M_0 = 10^{11 - 15}\, \Msunh$ at $z=0$. The masses in the legend are reported in $\log(M_0 / [\Msunh])$. Due to the inclusive mass definition, the first-order $f_\mathrm{sub}$ includes the mass of subhaloes of all-orders, whereas the second-order $f_\mathrm{sub}$ includes the mass of subhaloes of order-2 and higher. For comparison, we plot the gravitational lensing estimate of $f_\mathrm{sub}(\psi_\mathrm{res} = 10^{-3})$ for the Coma cluster measured by \citet{Okabe14}, finding excellent agreement with our model predictions.}
    \label{fig:fsub_v_mass_fid}
\end{figure*}

Fig.~\ref{fig:fsub_v_mass_fid} plots $f_\mathrm{sub} (\psi_\mathrm{res} = 10^{-4})$ as a function of host halo mass. These results have been obtained using 10,000 trees each (with $\psi_\mathrm{res}=10^{-4}$) for haloes with $\log(M_0/[\Msunh]) \in [11, 15]$ at $z=0$. Note that we have not included our treatment of artificial disruption here and the results are thus intended to reflect estimates of the true subhalo mass fractions in the absence of numerical artifacts. The left-hand panel shows the mean, median and 16--84 percentiles for both first- and second-order subhaloes, as indicated, whereas the right-hand panel plots the corresponding cumulative distribution functions.

Overall, the trends shown are consistent with the orbit-averaged model used by \citet{Jiang.vdBosch.17}: $f_\mathrm{sub}$ increases with $M_0$ and the halo-to-halo variance decreases slightly with $M_0$. As discussed in detail in \citet{Jiang.vdBosch.17}, this halo-to-halo variance is predominately driven by variance in the halo mass accretion histories \citep[see also e.g.,][]{Giocoli.etal.10, Green2020}. The second-order $f_\mathrm{sub}$ also increases with $M_0$, has much larger log-scatter than the total $f_\mathrm{sub}$, and its mean is smaller by a factor of $\approx 15 - 30$. This difference between first- and second-order $f_\mathrm{sub}$ is considerably larger than predicted by \citet{Jiang.vdBosch.17}, which is primarily due to the fact that \texttt{SatGen} allows higher-order subhaloes to be stripped from their parent subhalo (see Section~\ref{ssec:higherorder}). 

For comparison, we also plot the result of \citet{Okabe14}, who used weak gravitational lensing to infer $f_\mathrm{sub} = 0.226\substack{+0.111 \\ -0.085}$ for the Coma cluster,\footnote{This subhalo mass fraction is measured with $\psi_\mathrm{res} = 10^{-3}$, rather than $10^{-4}$. As shown in Fig.~\ref{fig:fsub_vs_res_wd}, the mean $f_\mathrm{sub}(\psi_\mathrm{res} = 10^{-3})$ is $\lta 0.1$ dex smaller than $f_\mathrm{sub}(\psi_\mathrm{res} = 10^{-4})$ for high-mass host haloes, which is negligible compared to both the halo-to-halo variance and the measurement error.} which is assumed to have a mass of $M_0 = 8.92\substack{+20.05 \\ -5.17} \times 10^{14} \Msunh$ \citep[][]{Okabe2010}. Our \texttt{SatGen} predictions are in excellent agreement with this measurement, demonstrating consistency between observations and the $\Lambda$CDM paradigm.

\subsubsection{Impact of disruption on $f_\mathrm{sub}$}

Overall, the results of previous subsections illustrate that artificial disruption impacts subhalo statistics less significantly than the factor of two suggested by \citetalias{Green.vdBosch.19}. We now formalize this by comparing \texttt{SatGen} predictions of $f_\mathrm{sub}(\psi_\mathrm{res} = m_\mathrm{res,B} / M_0)$ with and without the impact of artificial disruption included (but with the same degree of withering in both cases since $\psi_\mathrm{res}$ is fixed). For this test, we use 10,000 trees with $\psi_\mathrm{res}=10^{-4}$ and $M_0 = 10^{12}$ and $10^{13}\, \Msunh$ as well as 2,000 trees with $\psi_\mathrm{res}=10^{-5}$ and $M_0 = 10^{14.2}\, \Msunh$ in order to estimate $f_\mathrm{sub}(\psi_\mathrm{res} = m_\mathrm{res,B} / M_0)$ with and without disruption.\footnote{Note that we need additional resolution for the high-mass case in order to resolve the merger trees down to $m_\mathrm{res,B}$.} We find that artificial disruption results in a relative suppression of $f_\mathrm{sub}(\psi_\mathrm{res} = m_\mathrm{res,B} / M_0)$ by $8\%$, $10\%$, and $12\%$ for $M_0 = 10^{12}$, $10^{13}$, and $10^{14.2}\, \Msunh$, respectively. Indeed, this level of suppression is significantly less than a factor of two (i.e., 50\%). As already discussed in Section~\ref{ssec:shmf}, the primary reason that the \citetalias{Green.vdBosch.19} estimate of the artificial disruption impact is much larger is that the orbit-averaged model on which their estimate is based does not account for splashback haloes (i.e., the fact that at any moment in time about half of all haloes ever accreted by the host are located outside of the virial radius).

\subsubsection{Insensitivity of $f_\mathrm{sub}$ to the model parameter choices}

The substructure mass fraction is a useful summary statistic for illustrating how sensitive \texttt{SatGen} is to our model parameters (the stripping efficiency, $\alpha$, and the dynamical friction strength, $\beta_\mathrm{DF}$). For this test, we once again focus on $M_0 = 10^{14.2}\, \Msunh$ haloes. We use 10,000 trees with $\psi_\mathrm{res}=10^{-4}$ and evolve the subhaloes using each of the following cases: (i) our fiducial parameters ($\beta_\mathrm{DF} = 0.75$ and $\alpha=\alpha(c_\mathrm{vir,s}/c_\mathrm{vir,h})$ described by equation~[\ref{eqn:alpha}]), (ii) fiducial $\beta_\mathrm{DF}=0.75$ and fixed $\alpha=0.6$, and (iii) fiducial $\alpha=\alpha(c_\mathrm{vir,s}/c_\mathrm{vir,h})$ and the `natural' $\beta_\mathrm{DF}=1.0$ (i.e., Chandrasekhar dynamical friction without a correction factor). As our benchmark, we consider the fractional change in the mean $f_\mathrm{sub}(\psi_\mathrm{res} = 10^{-4})$ relative to the fiducial case. Setting $\alpha=0.6$ results in a $2\%$ relative increase in $f_\mathrm{sub}$ relative to fiducial. Increasing $\beta_\mathrm{DF}$ from 0.75 to 1.0 results in a $\sim 4\%$ relative decrease in $f_\mathrm{sub}$. The level of impact on other statistics (i.e., SHMF, radial profiles) is comparable. Hence, we conclude that our model predictions are reliable at the level of a few percent and that the uncertainties are small in comparison to the halo-to-halo variance. The sensitivity to these parameters is also significantly smaller than the impact of artificial disruption on the results of cosmological simulations, making \texttt{SatGen} a more reliable alternative for studying the substructure of dark matter haloes.

\subsection{Total W + D disruption rate}
\label{ssec:disrate}

The artificial disruption mechanism of Section~\ref{ssec:artdisr} is constructed such that the D channel population of \texttt{SatGen} subhaloes has an $f_\mathrm{dis}$ distribution consistent with that of \textit{Bolshoi}. However, this alone is not sufficient to guarantee that the W + D numerical disruption \textit{rate} of \texttt{SatGen} subhaloes is in agreement with the $13\%/$Gyr that \citet{vdBosch.17} measured from the W + D channel \textit{Bolshoi} subhaloes. In order to make a fair comparison between the \texttt{SatGen} W + D disruption rate and \textit{Bolshoi}, we run the following test. Starting with the same sample of \textit{Bolshoi} host halo masses (at $z\sim 0$) as used in \citet{vdBosch.17}, we randomly sub-sample 40,000 masses from the total of ${\sim}$160,000. Rather than use a fixed $\psi_\mathrm{res}$, we instead set $\psi_\mathrm{res} = m_\mathrm{res,B} / M_0$, where $m_\mathrm{res,B} = 10^{9.83} \Msunh$ is the 50-particle \textit{Bolshoi} resolution mass. Following the procedure of \citet{vdBosch.17}, we determine the W + D disruption rate by measuring the fraction of the subhaloes present at $z=0.0174$ (i.e., 240 Myr ago) that have been disrupted (either via the W or D channel) by $z=0$. In particular, our $z=0.0174$ sample consists of all subhaloes that have merged with the host, have a mass above both $m_\mathrm{res,B}$ and the assigned $m_\mathrm{dis}$ (i.e., it has neither disrupted nor withered by $z=0.0174$), and have an instantaneous orbital radius within $r_\mathrm{vir}$ of the host centre.

The subset of this sample with a $z=0$ mass below either $m_\mathrm{res,B}$ or its assigned $m_\mathrm{dis}$ are counted as having numerically disrupted between $z=0.0174$ and $z=0$. We convert this disruption fraction into a rate by dividing it by the 240 Myr time interval considered. Using this approach, we determine that the combination of withering and our artificial disruption mechanism yields a W + D numerical disruption rate of ${\sim}16.7\%/$Gyr, which is only slightly larger than the $13\%/$Gyr that \citet{vdBosch.17} measured in \textit{Bolshoi}. Hence, we conclude that our implementation of artificial disruption in \texttt{SatGen} accurately reproduces this numerical artifact in the \textit{Bolshoi} simulation. However, we caution that it may not adequately describe artificial disruption in other simulations, each of which is likely to have subtly different disruption statistics. The real strength of \texttt{SatGen} is not its ability to reproduce the results of cosmological simulations but rather to make reliable predictions that are free from the numerical limitations that hamper such simulations.

\section{Summary and Discussion}\label{sec:summary}

This work represents the culmination of several previous studies aimed at quantifying the impact of artificial disruption on state-of-the-art dark matter-only cosmological simulations. Studying the evolution of \textit{Bolshoi} subhaloes, \citet{vdBosch.17} found that the combined effect of the finite mass resolution (i.e., withering) and artificial disruption results in rapid depletion of the subhalo population. In the follow-up studies of \citet{vdBosch.etal.2018a} and \citet{vdBosch.etal.2018b}, the authors used a combination of analytical arguments and idealized numerical experiments to demonstrate that complete physical disruption of $\Lambda$CDM subhalo remnants is exceedingly rare, concluding that the majority of disruption seen in cosmological simulations must be numerical in nature. Following this, \citet{Ogiya2019} released the \textit{DASH} library of high-resolution idealized simulations of halo mergers. This data release marked the beginning of a research program focused on developing a new semi-analytical model of subhalo evolution that is calibrated independently of cosmological simulations, enabling its predictions to be free of the effects of artificial disruption. Thus, \citetalias{Green.vdBosch.19} used \textit{DASH} to construct an accurate model of the evolved subhalo density profile, which is a simple function of the initial profile and the fraction of mass lost since infall \citep[similar to the approaches of, e.g.,][]{Hayashi.etal.03, Penarrubia.etal.10, Drakos.etal.17, Errani2020b}. Additionally, using the orbit-averaged subhalo evolution model and artificial disruption mechanism of \citet{Jiang.vdBosch.16}, \citetalias{Green.vdBosch.19} inferred that artificial disruption could potentially be responsible for suppressing the SHMF normalization by as much as a factor of two. Recently, \citet{Jiang2021} released the \texttt{SatGen} library, a new semi-analytical modeling framework for studying subhalo and satellite galaxy evolution in a full dynamical context (i.e., the orbits of individual subhaloes are integrated instead of using an orbit-averaged approach). 

In the present paper, we used \texttt{SatGen} as a scaffolding to develop a comprehensive model of substructure evolution that is not adversely impacted by the limitations of artificial disruption and simulation resolution limits. To this end, we made several modifications and improvements to \texttt{SatGen}, which we summarize below:
\begin{itemize}
    \item The initial orbits of infalling subhaloes are sampled using the state-of-the-art model of \citet{Li2020} (see Appendix~\ref{app:initorbits}). This model marks an improvement over previous approaches \citep[e.g.,][]{Zentner.etal.05, Wetzel.11, Jiang.etal.15} because it is expressed as a general function of the host halo peak height and host-to-subhalo mass ratio. Furthermore, the free parameters of the model were fit using a large simulation suite.
    \item The evolved subhalo density profiles (ESHDPs) are characterized using the model of \citetalias{Green.vdBosch.19} (Section~\ref{sssec:eshdp}). At infall, subhaloes are assumed to have \citetalias{Navarro.etal.97} profiles. However, as mass is stripped and $m/m_\mathrm{acc}$ decreases, the profile becomes tidally truncated in a manner consistent with the evolution of \textit{DASH} subhaloes.
    \item In line with the original \texttt{SatGen} implementation, the instantaneous subhalo mass-loss rate (Section~\ref{sssec:massloss}) is written according to equation~\eqref{eqn:massloss}, which depends on the \citet{King.62} tidal radius (computed using the ESHDPs), the local dynamical time, and the ``stripping efficiency'', $\alpha$. We re-calibrated $\alpha$ (equation~[\ref{eqn:alpha}]) so that the mass-loss model accurately reproduces the $m(t)/m_\mathrm{acc}$ trajectories of \textit{DASH} subhaloes.
    \item The strength of the (Chandrasekhar) dynamical friction is controlled by a correction factor, $\beta_\mathrm{DF}$, which we calibrate such that \texttt{SatGen} reproduces the $m_\mathrm{acc}/M_0$-dependence of the $\langle r / r_\mathrm{vir} \rangle - z_\mathrm{acc}$ relation of \textit{Bolshoi} subhaloes (Section~\ref{ssec:dfstrength}). We have demonstrated that the resulting best-fit value ($\beta_\mathrm{DF}=0.75$) is not adversely affected by artificial disruption in the \textit{Bolshoi} simulation.
    \item In order to assess the impact of artificial disruption on simulations, we developed a model that reproduces the statistical properties of disruption in \textit{Bolshoi} that can be optionally applied to \texttt{SatGen} results. We found that the $f_\mathrm{dis}$ distribution of disrupted (D channel) \textit{Bolshoi} subhaloes is well-described by a family of log-normal distributions, the parameters of which are functions of $m_\mathrm{acc}$ (Section~\ref{ssec:artdisr}).
    \item \texttt{SatGen} is ideally suited to assess the impact of the resolution limit of numerical simulations by only including subhaloes with a final mass that lies above the merger tree resolution (i.e., $m > \psi_\mathrm{res}M_0$). In addition, by instead allowing each subhalo to evolve down to arbitrary $\phi_\mathrm{res} = m/m_\mathrm{acc}$ (here, we have used values as low as $\phi_\mathrm{res} = 10^{-5}$), \texttt{SatGen} can model the subhalo population with an effectively ``arbitrary resolution''.
\end{itemize}

We used this updated model to predict subhalo mass and maximum circular velocity functions, number density profiles, radial bias profiles, and substructure mass fractions. We considered the effect of both the simulation mass limit and artificial disruption on each quantity and studied the dependence of $f_\mathrm{sub}$ on host halo mass. We summarize our most notable findings below:
\begin{itemize}
    \item When the effects of both withering and artificial disruption are included, \texttt{SatGen} yields subhalo demographics in excellent agreement with \textit{Bolshoi}.
    \item Artificial disruption only results in a $\sim 8 - 12\%$ suppression of $f_\mathrm{sub}(<r_\mathrm{vir})$ and a $\sim 20\%$ suppression of the SHMF. While still significant, this greatly ameliorates previous concerns that the overall abundance of dark matter subhaloes could be artificially suppressed by a factor of two. However, the impact of artificial disruption is more pronounced at smaller host-centric radii, where it halves both $F_\mathrm{sub}(<X)$ and $\rmd \tilde{N} / \rmd x^3|_\mathrm{sub}$ within $\sim 0.1 r_{\rm vir}$. 
    \item By comparing the SHMF computed by including only subhaloes within $r_\mathrm{vir}$ (i.e., consistent with simulation approaches) to the SHMF computed by including \textit{all} surviving subhaloes ever accreted by the host, we infer that splashback haloes make up roughly half of the total subhalo population. This is in good agreement with results from several simulation studies \citep[e.g.,][]{Gill.etal.04, Ludlow2009, Bakels2021}. Hence, it is essential that semi-analytical models of subhalo and satellite galaxy evolution properly account for the splashback population. This is naturally achieved with full dynamical models, such as \texttt{SatGen}, which integrate the orbits of individual subhaloes. At the same time, it indicates a serious limitation of orbit-averaged approaches, such as those used in \citet{vdBosch.etal.05} and \citet{Jiang.vdBosch.16}. 
    \item We have demonstrated that the radial bias in the subhalo number density (i.e., the dearth of subhaloes in the halo centre relative to the host density profile), a feature that is consistently present in dark matter-only simulations \citep[e.g.,][]{Ghigna.etal.98, Springel.etal.01, Diemand.Moore.Stadel.04, Springel.etal.08, Han.etal.16}, is predominantly an artifact of the simulation mass resolution (at least in the absence of baryonic processes) and not of artificial disruption. The latter only slightly enhances the bias and is subdominant to the impact of the mass resolution. By allowing subhaloes to evolve down to arbitrarily low $m/m_\mathrm{acc}$ (as opposed to having a fixed absolute mass limit), the radial bias is completely eliminated. In fact, dynamical friction causes a slight enhancement of the subhalo number density relative to the host profile near the halo centre, which marks a complete reversal of the trend seen in simulations. 
\end{itemize}

Although the model presented here is able to accurately reproduce the subhalo statistics of a cosmological simulation when its numerical limitations are properly taken into account, the true strength of the updated version of \texttt{SatGen} presented here lies in the fact that it can be used to predict subhalo demographics with an arbitrarily high resolution and in the absence of artificial disruption. We have therefore made the updated code publicly available in the hope that it will enable/accommodate a wide variety of future research programs. For example, \texttt{SatGen} could prove a powerful tool to investigate claimed discrepancies between simulations and observations regarding the abundance and central concentration of dark matter substructure \citep[e.g.,][]{Carlsten2020, Meneghetti2020} and/or the dark matter deficiency of associated satellite galaxies \citep[e.g.,][]{Ogiya2018, Jackson2021}.

\section*{Acknowledgements}

The authors thank Uddipan Banik, Nicole Drakos, Dhruba Dutta Chowdhury, Zhaozhou Li, and Go Ogiya for helpful conversations throughout the development of this work. SBG is supported by the US National Science Foundation Graduate Research Fellowship under Grant No. DGE-1752134. FCvdB is supported by the National Aeronautics and Space Administration through Grant No. 17-ATP17-0028 issued as part of the Astrophysics Theory Program. FJ is supported by the Troesh Fellowship from the California Institute of Technology.

\section*{Data availability}

The \textit{DASH} simulation data is available online.\footnote{\href{https://cosmo.oca.eu/dash/}{https://cosmo.oca.eu/dash/}} The updated \texttt{SatGen} library is available in the \texttt{sheridan} branch of the \texttt{SatGen} GitHub repository.\footnote{\href{https://github.com/shergreen/SatGen/tree/sheridan}{https://github.com/shergreen/SatGen/tree/sheridan}}


\bibliographystyle{mnras}
\bibliography{references_vdb}

\begin{thebibliography}{}
\makeatletter
\relax
\def\mn@urlcharsother{\let\do\@makeother \do\$\do\&\do\#\do\^\do\_\do\%\do\~}
\def\mn@doi{\begingroup\mn@urlcharsother \@ifnextchar [ {\mn@doi@}
  {\mn@doi@[]}}
\def\mn@doi@[#1]#2{\def\@tempa{#1}\ifx\@tempa\@empty \href
  {http://dx.doi.org/#2} {doi:#2}\else \href {http://dx.doi.org/#2} {#1}\fi
  \endgroup}
\def\mn@eprint#1#2{\mn@eprint@#1:#2::\@nil}
\def\mn@eprint@arXiv#1{\href {http://arxiv.org/abs/#1} {{\tt arXiv:#1}}}
\def\mn@eprint@dblp#1{\href {http://dblp.uni-trier.de/rec/bibtex/#1.xml}
  {dblp:#1}}
\def\mn@eprint@#1:#2:#3:#4\@nil{\def\@tempa {#1}\def\@tempb {#2}\def\@tempc
  {#3}\ifx \@tempc \@empty \let \@tempc \@tempb \let \@tempb \@tempa \fi \ifx
  \@tempb \@empty \def\@tempb {arXiv}\fi \@ifundefined
  {mn@eprint@\@tempb}{\@tempb:\@tempc}{\expandafter \expandafter \csname
  mn@eprint@\@tempb\endcsname \expandafter{\@tempc}}}

\bibitem[\protect\citeauthoryear{{Aung}, {Nagai}, {Rozo}  \&
  {Garc{\'\i}a}}{{Aung} et~al.}{2021}]{Aung2021}
{Aung} H.,  {Nagai} D.,  {Rozo} E.,   {Garc{\'\i}a} R.,  2021, \mn@doi [\mnras]
  {10.1093/mnras/staa3994}, \href
  {https://ui.adsabs.harvard.edu/abs/2021MNRAS.502.1041A} {502, 1041}

\bibitem[\protect\citeauthoryear{{Bakels}, {Ludlow}  \& {Power}}{{Bakels}
  et~al.}{2021}]{Bakels2021}
{Bakels} L.,  {Ludlow} A.~D.,   {Power} C.,  2021, \mn@doi [\mnras]
  {10.1093/mnras/staa3979}, \href
  {https://ui.adsabs.harvard.edu/abs/2021MNRAS.501.5948B} {501, 5948}

\bibitem[\protect\citeauthoryear{{Behroozi}, {Wechsler}  \&
  {Conroy}}{{Behroozi} et~al.}{2013}]{Behroozi.etal.13c}
{Behroozi} P.~S.,  {Wechsler} R.~H.,   {Conroy} C.,  2013, \mn@doi [\apj]
  {10.1088/0004-637X/770/1/57}, \href
  {http://adsabs.harvard.edu/abs/2013ApJ...770...57B} {770, 57}

\bibitem[\protect\citeauthoryear{{Benson}}{{Benson}}{2017}]{Benson2017}
{Benson} A.~J.,  2017, \mn@doi [\mnras] {10.1093/mnras/stx343}, \href
  {https://ui.adsabs.harvard.edu/abs/2017MNRAS.467.3454B} {467, 3454}

\bibitem[\protect\citeauthoryear{{Benson}}{{Benson}}{2020}]{Benson2020}
{Benson} A.~J.,  2020, \mn@doi [\mnras] {10.1093/mnras/staa341}, \href
  {https://ui.adsabs.harvard.edu/abs/2020MNRAS.493.1268B} {493, 1268}

\bibitem[\protect\citeauthoryear{{Benson}, {Frenk}, {Baugh}, {Cole}  \&
  {Lacey}}{{Benson} et~al.}{2001}]{Benson.etal.01}
{Benson} A.~J.,  {Frenk} C.~S.,  {Baugh} C.~M.,  {Cole} S.,   {Lacey} C.~G.,
  2001, \mn@doi [\mnras] {10.1046/j.1365-8711.2001.04824.x}, \href
  {http://adsabs.harvard.edu/abs/2001MNRAS.327.1041B} {327, 1041}

\bibitem[\protect\citeauthoryear{{Berlind} et~al.,}{{Berlind}
  et~al.}{2003}]{Berlind.etal.03}
{Berlind} A.~A.,  et~al., 2003, \mn@doi [\apj] {10.1086/376517}, \href
  {http://adsabs.harvard.edu/abs/2003ApJ...593....1B} {593, 1}

\bibitem[\protect\citeauthoryear{{Binney} \& {Tremaine}}{{Binney} \&
  {Tremaine}}{2008}]{Binney.Tremaine.08}
{Binney} J.,  {Tremaine} S.,  2008, {Galactic Dynamics: Second Edition}.
Princeton University Press

\bibitem[\protect\citeauthoryear{{Bonaca} et~al.,}{{Bonaca}
  et~al.}{2020}]{Bonaca2020}
{Bonaca} A.,  et~al., 2020, \mn@doi [\apjl] {10.3847/2041-8213/ab800c}, \href
  {https://ui.adsabs.harvard.edu/abs/2020ApJ...892L..37B} {892, L37}

\bibitem[\protect\citeauthoryear{{Bose} et~al.,}{{Bose}
  et~al.}{2017}]{Bose.etal.17}
{Bose} S.,  et~al., 2017, \mn@doi [\mnras] {10.1093/mnras/stw2686}, \href
  {http://adsabs.harvard.edu/abs/2017MNRAS.464.4520B} {464, 4520}

\bibitem[\protect\citeauthoryear{{Boylan-Kolchin}, {Springel}, {White}  \&
  {Jenkins}}{{Boylan-Kolchin} et~al.}{2010}]{Boylan-Kolchin.etal.10}
{Boylan-Kolchin} M.,  {Springel} V.,  {White} S.~D.~M.,   {Jenkins} A.,  2010,
  \mn@doi [\mnras] {10.1111/j.1365-2966.2010.16774.x}, \href
  {http://adsabs.harvard.edu/abs/2010MNRAS.406..896B} {406, 896}

\bibitem[\protect\citeauthoryear{{Bryan} \& {Norman}}{{Bryan} \&
  {Norman}}{1998}]{Bryan.Norman.98}
{Bryan} G.~L.,  {Norman} M.~L.,  1998, \mn@doi [\apj] {10.1086/305262}, \href
  {http://adsabs.harvard.edu/abs/1998ApJ...495...80B} {495, 80}

\bibitem[\protect\citeauthoryear{{Bullock}, {Kolatt}, {Sigad}, {Somerville},
  {Kravtsov}, {Klypin}, {Primack}  \& {Dekel}}{{Bullock}
  et~al.}{2001}]{Bullock.etal.01}
{Bullock} J.~S.,  {Kolatt} T.~S.,  {Sigad} Y.,  {Somerville} R.~S.,  {Kravtsov}
  A.~V.,  {Klypin} A.~A.,  {Primack} J.~R.,   {Dekel} A.,  2001, \mn@doi
  [\mnras] {10.1046/j.1365-8711.2001.04068.x}, \href
  {https://ui.adsabs.harvard.edu/abs/2001MNRAS.321..559B} {321, 559}

\bibitem[\protect\citeauthoryear{Burkert}{Burkert}{2000}]{Burkert2000}
Burkert A.,  2000, \mn@doi [\apj] {10.1086/312674}, 534, L143

\bibitem[\protect\citeauthoryear{{Campbell}, {van den Bosch}, {Padmanabhan},
  {Mao}, {Zentner}, {Lange}, {Jiang}  \& {Villarreal}}{{Campbell}
  et~al.}{2018}]{Campbell.etal.18}
{Campbell} D.,  {van den Bosch} F.~C.,  {Padmanabhan} N.,  {Mao} Y.-Y.,
  {Zentner} A.~R.,  {Lange} J.~U.,  {Jiang} F.,   {Villarreal} A.,  2018,
  \mn@doi [\mnras] {10.1093/mnras/sty495}, \href
  {http://adsabs.harvard.edu/abs/2018MNRAS.477..359C} {477, 359}

\bibitem[\protect\citeauthoryear{{Carlberg}}{{Carlberg}}{2012}]{Carlberg2012}
{Carlberg} R.~G.,  2012, \mn@doi [\apj] {10.1088/0004-637X/748/1/20}, \href
  {http://cdsads.u-strasbg.fr/abs/2012ApJ...748...20C} {748, 20}

\bibitem[\protect\citeauthoryear{{Carlsten}, {Greene}, {Peter}, {Greco}  \&
  {Beaton}}{{Carlsten} et~al.}{2020}]{Carlsten2020}
{Carlsten} S.~G.,  {Greene} J.~E.,  {Peter} A. H.~G.,  {Greco} J.~P.,
  {Beaton} R.~L.,  2020, \mn@doi [\apj] {10.3847/1538-4357/abb60b}, \href
  {https://ui.adsabs.harvard.edu/abs/2020ApJ...902..124C} {902, 124}

\bibitem[\protect\citeauthoryear{{Chandrasekhar}}{{Chandrasekhar}}{1943}]{Chandrasekhar.43}
{Chandrasekhar} S.,  1943, \mn@doi [\apj] {10.1086/144517}, \href
  {http://adsabs.harvard.edu/abs/1943ApJ....97..255C} {97, 255}

\bibitem[\protect\citeauthoryear{{Chaves-Montero}, {Angulo}, {Schaye},
  {Schaller}, {Crain}, {Furlong}  \& {Theuns}}{{Chaves-Montero}
  et~al.}{2016}]{Chaves2016}
{Chaves-Montero} J.,  {Angulo} R.~E.,  {Schaye} J.,  {Schaller} M.,  {Crain}
  R.~A.,  {Furlong} M.,   {Theuns} T.,  2016, \mn@doi [\mnras]
  {10.1093/mnras/stw1225}, \href
  {https://ui.adsabs.harvard.edu/abs/2016MNRAS.460.3100C} {460, 3100}

\bibitem[\protect\citeauthoryear{{Cole}, {Lacey}, {Baugh}  \& {Frenk}}{{Cole}
  et~al.}{2000}]{Cole2000}
{Cole} S.,  {Lacey} C.~G.,  {Baugh} C.~M.,   {Frenk} C.~S.,  2000, \mn@doi
  [\mnras] {10.1046/j.1365-8711.2000.03879.x}, \href
  {https://ui.adsabs.harvard.edu/abs/2000MNRAS.319..168C} {319, 168}

\bibitem[\protect\citeauthoryear{{Col{\'{\i}}n}, {Avila-Reese},
  {Gonz{\'a}lez-Samaniego}  \& {Vel{\'a}zquez}}{{Col{\'{\i}}n}
  et~al.}{2015}]{Colin.etal.15}
{Col{\'{\i}}n} P.,  {Avila-Reese} V.,  {Gonz{\'a}lez-Samaniego} A.,
  {Vel{\'a}zquez} H.,  2015, \mn@doi [\apj] {10.1088/0004-637X/803/1/28}, \href
  {http://adsabs.harvard.edu/abs/2015ApJ...803...28C} {803, 28}

\bibitem[\protect\citeauthoryear{{Conroy}, {Wechsler}  \& {Kravtsov}}{{Conroy}
  et~al.}{2006}]{Conroy.etal.06}
{Conroy} C.,  {Wechsler} R.~H.,   {Kravtsov} A.~V.,  2006, \mn@doi [\apj]
  {10.1086/503602}, \href {http://adsabs.harvard.edu/abs/2006ApJ...647..201C}
  {647, 201}

\bibitem[\protect\citeauthoryear{{Dalal} \& {Kochanek}}{{Dalal} \&
  {Kochanek}}{2002}]{Dalal.Kochanek.02}
{Dalal} N.,  {Kochanek} C.~S.,  2002, \mn@doi [\apj] {10.1086/340303}, \href
  {http://adsabs.harvard.edu/abs/2002ApJ...572...25D} {572, 25}

\bibitem[\protect\citeauthoryear{{Delos}}{{Delos}}{2019}]{Delos2019}
{Delos} M.~S.,  2019, \mn@doi [\prd] {10.1103/PhysRevD.100.063505}, \href
  {https://ui.adsabs.harvard.edu/abs/2019PhRvD.100f3505D} {100, 063505}

\bibitem[\protect\citeauthoryear{{Diemand}, {Moore}  \& {Stadel}}{{Diemand}
  et~al.}{2004}]{Diemand.Moore.Stadel.04}
{Diemand} J.,  {Moore} B.,   {Stadel} J.,  2004, \mn@doi [\mnras]
  {10.1111/j.1365-2966.2004.07940.x}, \href
  {http://adsabs.harvard.edu/abs/2004MNRAS.352..535D} {352, 535}

\bibitem[\protect\citeauthoryear{{Diemand}, {Kuhlen}  \& {Madau}}{{Diemand}
  et~al.}{2007}]{Diemand.etal.07a}
{Diemand} J.,  {Kuhlen} M.,   {Madau} P.,  2007, \mn@doi [\apj]
  {10.1086/520573}, \href {http://adsabs.harvard.edu/abs/2007ApJ...667..859D}
  {667, 859}

\bibitem[\protect\citeauthoryear{{Diemer}}{{Diemer}}{2020a}]{Diemer2020b}
{Diemer} B.,  2020a, arXiv e-prints, \href
  {https://ui.adsabs.harvard.edu/abs/2020arXiv200710992D} {p. arXiv:2007.10992}

\bibitem[\protect\citeauthoryear{{Diemer}}{{Diemer}}{2020b}]{Diemer2020a}
{Diemer} B.,  2020b, \mn@doi [\apjs] {10.3847/1538-4365/abbf51}, \href
  {https://ui.adsabs.harvard.edu/abs/2020ApJS..251...17D} {251, 17}

\bibitem[\protect\citeauthoryear{{Drakos}, {Taylor}  \& {Benson}}{{Drakos}
  et~al.}{2017}]{Drakos.etal.17}
{Drakos} N.~E.,  {Taylor} J.~E.,   {Benson} A.~J.,  2017, \mn@doi [\mnras]
  {10.1093/mnras/stx652}, \href
  {http://adsabs.harvard.edu/abs/2017MNRAS.468.2345D} {468, 2345}

\bibitem[\protect\citeauthoryear{{Drakos}, {Taylor}  \& {Benson}}{{Drakos}
  et~al.}{2020}]{Drakos2020}
{Drakos} N.~E.,  {Taylor} J.~E.,   {Benson} A.~J.,  2020, \mn@doi [\mnras]
  {10.1093/mnras/staa760}, \href
  {https://ui.adsabs.harvard.edu/abs/2020MNRAS.494..378D} {494, 378}

\bibitem[\protect\citeauthoryear{{Erkal}, {Belokurov}, {Bovy}  \&
  {Sanders}}{{Erkal} et~al.}{2016}]{Erkal.etal.16}
{Erkal} D.,  {Belokurov} V.,  {Bovy} J.,   {Sanders} J.~L.,  2016, \mn@doi
  [\mnras] {10.1093/mnras/stw1957}, \href
  {http://adsabs.harvard.edu/abs/2016MNRAS.463..102E} {463, 102}

\bibitem[\protect\citeauthoryear{{Errani} \& {Navarro}}{{Errani} \&
  {Navarro}}{2020}]{Errani2020b}
{Errani} R.,  {Navarro} J.~F.,  2020, arXiv e-prints, \href
  {https://ui.adsabs.harvard.edu/abs/2020arXiv201107077E} {p. arXiv:2011.07077}

\bibitem[\protect\citeauthoryear{{Facchinetti}, {Lavalle}  \&
  {Stref}}{{Facchinetti} et~al.}{2020}]{Facchinetti2020}
{Facchinetti} G.,  {Lavalle} J.,   {Stref} M.,  2020, arXiv e-prints, \href
  {https://ui.adsabs.harvard.edu/abs/2020arXiv200710392F} {p. arXiv:2007.10392}

\bibitem[\protect\citeauthoryear{{Fong} \& {Han}}{{Fong} \&
  {Han}}{2020}]{Fong2020}
{Fong} M.,  {Han} J.,  2020, arXiv e-prints, \href
  {https://ui.adsabs.harvard.edu/abs/2020arXiv200803477F} {p. arXiv:2008.03477}

\bibitem[\protect\citeauthoryear{{Gan}, {Kang}, {van den Bosch}  \&
  {Hou}}{{Gan} et~al.}{2010}]{Gan2010}
{Gan} J.,  {Kang} X.,  {van den Bosch} F.~C.,   {Hou} J.,  2010, \mn@doi
  [\mnras] {10.1111/j.1365-2966.2010.17266.x}, \href
  {http://cdsads.u-strasbg.fr/abs/2010MNRAS.408.2201G} {408, 2201}

\bibitem[\protect\citeauthoryear{{Gao}, {White}, {Jenkins}, {Stoehr}  \&
  {Springel}}{{Gao} et~al.}{2004}]{Gao.etal.04}
{Gao} L.,  {White} S.~D.~M.,  {Jenkins} A.,  {Stoehr} F.,   {Springel} V.,
  2004, \mn@doi [\mnras] {10.1111/j.1365-2966.2004.08360.x}, \href
  {http://adsabs.harvard.edu/abs/2004MNRAS.355..819G} {355, 819}

\bibitem[\protect\citeauthoryear{{Gao}, {Navarro}, {Frenk}, {Jenkins},
  {Springel}  \& {White}}{{Gao} et~al.}{2012}]{Gao.etal.12}
{Gao} L.,  {Navarro} J.~F.,  {Frenk} C.~S.,  {Jenkins} A.,  {Springel} V.,
  {White} S.~D.~M.,  2012, \mn@doi [\mnras] {10.1111/j.1365-2966.2012.21564.x},
  \href {https://ui.adsabs.harvard.edu/abs/2012MNRAS.425.2169G} {425, 2169}

\bibitem[\protect\citeauthoryear{{Garrison-Kimmel} et~al.,}{{Garrison-Kimmel}
  et~al.}{2017}]{Garrison-Kimmel.etal.17}
{Garrison-Kimmel} S.,  et~al., 2017, \mn@doi [\mnras] {10.1093/mnras/stx1710},
  \href {https://ui.adsabs.harvard.edu/abs/2017MNRAS.471.1709G} {471, 1709}

\bibitem[\protect\citeauthoryear{{Ghigna}, {Moore}, {Governato}, {Lake},
  {Quinn}  \& {Stadel}}{{Ghigna} et~al.}{1998}]{Ghigna.etal.98}
{Ghigna} S.,  {Moore} B.,  {Governato} F.,  {Lake} G.,  {Quinn} T.,   {Stadel}
  J.,  1998, \mn@doi [\mnras] {10.1046/j.1365-8711.1998.01918.x}, \href
  {http://adsabs.harvard.edu/abs/1998MNRAS.300..146G} {300, 146}

\bibitem[\protect\citeauthoryear{{Gill}, {Knebe}  \& {Gibson}}{{Gill}
  et~al.}{2004}]{Gill.etal.04}
{Gill} S.~P.~D.,  {Knebe} A.,   {Gibson} B.~K.,  2004, \mn@doi [\mnras]
  {10.1111/j.1365-2966.2004.07786.x}, \href
  {http://adsabs.harvard.edu/abs/2004MNRAS.351..399G} {351, 399}

\bibitem[\protect\citeauthoryear{{Gilman}, {Birrer}  \& {Treu}}{{Gilman}
  et~al.}{2020}]{Gilman2020}
{Gilman} D.,  {Birrer} S.,   {Treu} T.,  2020, \mn@doi [\aap]
  {10.1051/0004-6361/202038829}, \href
  {https://ui.adsabs.harvard.edu/abs/2020A&A...642A.194G} {642, A194}

\bibitem[\protect\citeauthoryear{{Giocoli}, {Tormen}, {Sheth}  \& {van den
  Bosch}}{{Giocoli} et~al.}{2010}]{Giocoli.etal.10}
{Giocoli} C.,  {Tormen} G.,  {Sheth} R.~K.,   {van den Bosch} F.~C.,  2010,
  \mn@doi [\mnras] {10.1111/j.1365-2966.2010.16311.x}, \href
  {http://adsabs.harvard.edu/abs/2010MNRAS.404..502G} {404, 502}

\bibitem[\protect\citeauthoryear{{Golse} \& {Kneib}}{{Golse} \&
  {Kneib}}{2002}]{Golse2002}
{Golse} G.,  {Kneib} J.~P.,  2002, \mn@doi [\aap] {10.1051/0004-6361:20020639},
  \href {https://ui.adsabs.harvard.edu/abs/2002A&A...390..821G} {390, 821}

\bibitem[\protect\citeauthoryear{{Green} \& {van den Bosch}}{{Green} \& {van
  den Bosch}}{2019}]{Green.vdBosch.19}
{Green} S.~B.,  {van den Bosch} F.~C.,  2019, \mn@doi [\mnras]
  {10.1093/mnras/stz2767}, \href
  {https://ui.adsabs.harvard.edu/abs/2019MNRAS.490.2091G} {490, 2091}

\bibitem[\protect\citeauthoryear{{Green}, {Aung}, {Nagai}  \& {van den
  Bosch}}{{Green} et~al.}{2020}]{Green2020}
{Green} S.~B.,  {Aung} H.,  {Nagai} D.,   {van den Bosch} F.~C.,  2020, \mn@doi
  [\mnras] {10.1093/mnras/staa1712}, \href
  {https://ui.adsabs.harvard.edu/abs/2020MNRAS.496.2743G} {496, 2743}

\bibitem[\protect\citeauthoryear{{Griffen}, {Ji}, {Dooley}, {G{\'o}mez},
  {Vogelsberger}, {O'Shea}  \& {Frebel}}{{Griffen}
  et~al.}{2016}]{Griffen.etal.16}
{Griffen} B.~F.,  {Ji} A.~P.,  {Dooley} G.~A.,  {G{\'o}mez} F.~A.,
  {Vogelsberger} M.,  {O'Shea} B.~W.,   {Frebel} A.,  2016, \mn@doi [\apj]
  {10.3847/0004-637X/818/1/10}, \href
  {http://adsabs.harvard.edu/abs/2016ApJ...818...10G} {818, 10}

\bibitem[\protect\citeauthoryear{{Guo}, {White}, {Li}  \&
  {Boylan-Kolchin}}{{Guo} et~al.}{2010}]{Guo.etal.10}
{Guo} Q.,  {White} S.,  {Li} C.,   {Boylan-Kolchin} M.,  2010, \mn@doi [\mnras]
  {10.1111/j.1365-2966.2010.16341.x}, \href
  {http://adsabs.harvard.edu/abs/2010MNRAS.404.1111G} {404, 1111}

\bibitem[\protect\citeauthoryear{{Han}, {Cole}, {Frenk}  \& {Jing}}{{Han}
  et~al.}{2016}]{Han.etal.16}
{Han} J.,  {Cole} S.,  {Frenk} C.~S.,   {Jing} Y.,  2016, \mn@doi [\mnras]
  {10.1093/mnras/stv2900}, \href
  {http://adsabs.harvard.edu/abs/2016MNRAS.457.1208H} {457, 1208}

\bibitem[\protect\citeauthoryear{Hayashi, {Navarro}, {Taylor}, {Stadel}  \&
  {Quinn}}{Hayashi et~al.}{2003}]{Hayashi.etal.03}
Hayashi E.,  {Navarro} J.~F.,  {Taylor} J.~E.,  {Stadel} J.,   {Quinn} T.,
  2003, \mn@doi [\apj] {10.1086/345788}, \href
  {http://adsabs.harvard.edu/abs/2003ApJ...584..541H} {584, 541}

\bibitem[\protect\citeauthoryear{{Hayashi}, {Ichikawa}, {Matsumoto}, {Ibe},
  {Ishigaki}  \& {Sugai}}{{Hayashi} et~al.}{2016}]{Hayashi2016}
{Hayashi} K.,  {Ichikawa} K.,  {Matsumoto} S.,  {Ibe} M.,  {Ishigaki} M.~N.,
  {Sugai} H.,  2016, \mn@doi [\mnras] {10.1093/mnras/stw1457}, \href
  {http://cdsads.u-strasbg.fr/abs/2016MNRAS.461.2914H} {461, 2914}

\bibitem[\protect\citeauthoryear{{Hearin}, {Zentner}, {Berlind}  \&
  {Newman}}{{Hearin} et~al.}{2013}]{Hearin.etal.13}
{Hearin} A.~P.,  {Zentner} A.~R.,  {Berlind} A.~A.,   {Newman} J.~A.,  2013,
  \mn@doi [\mnras] {10.1093/mnras/stt755}, \href
  {http://adsabs.harvard.edu/abs/2013MNRAS.433..659H} {433, 659}

\bibitem[\protect\citeauthoryear{{Hezaveh} et~al.,}{{Hezaveh}
  et~al.}{2016}]{Hezaveh2016}
{Hezaveh} Y.~D.,  et~al., 2016, \mn@doi [\apj] {10.3847/0004-637X/823/1/37},
  \href {http://cdsads.u-strasbg.fr/abs/2016ApJ...823...37H} {823, 37}

\bibitem[\protect\citeauthoryear{{Hiroshima}, {Ando}  \&
  {Ishiyama}}{{Hiroshima} et~al.}{2018}]{Hiroshima2018}
{Hiroshima} N.,  {Ando} S.,   {Ishiyama} T.,  2018, \mn@doi [\prd]
  {10.1103/PhysRevD.97.123002}, \href
  {http://cdsads.u-strasbg.fr/abs/2018PhRvD..97l3002H} {97, 123002}

\bibitem[\protect\citeauthoryear{{Jackson} et~al.,}{{Jackson}
  et~al.}{2021}]{Jackson2021}
{Jackson} R.~A.,  et~al., 2021, \mn@doi [\mnras] {10.1093/mnras/stab093}, \href
  {https://ui.adsabs.harvard.edu/abs/2021MNRAS.502.1785J} {502, 1785}

\bibitem[\protect\citeauthoryear{{Jiang} \& {van den Bosch}}{{Jiang} \& {van
  den Bosch}}{2014}]{Jiang.vdBosch.14}
{Jiang} F.,  {van den Bosch} F.~C.,  2014, \mn@doi [\mnras]
  {10.1093/mnras/stu280}, \href
  {https://ui.adsabs.harvard.edu/abs/2014MNRAS.440..193J} {440, 193}

\bibitem[\protect\citeauthoryear{{Jiang} \& {van den Bosch}}{{Jiang} \& {van
  den Bosch}}{2016}]{Jiang.vdBosch.16}
{Jiang} F.,  {van den Bosch} F.~C.,  2016, \mn@doi [\mnras]
  {10.1093/mnras/stw439}, \href
  {http://adsabs.harvard.edu/abs/2016MNRAS.458.2848J} {458, 2848}

\bibitem[\protect\citeauthoryear{{Jiang} \& {van den Bosch}}{{Jiang} \& {van
  den Bosch}}{2017}]{Jiang.vdBosch.17}
{Jiang} F.,  {van den Bosch} F.~C.,  2017, \mn@doi [\mnras]
  {10.1093/mnras/stx1979}, \href
  {https://ui.adsabs.harvard.edu/abs/2017MNRAS.472..657J} {472, 657}

\bibitem[\protect\citeauthoryear{{Jiang}, {Cole}, {Sawala}  \& {Frenk}}{{Jiang}
  et~al.}{2015}]{Jiang.etal.15}
{Jiang} L.,  {Cole} S.,  {Sawala} T.,   {Frenk} C.~S.,  2015, \mn@doi [\mnras]
  {10.1093/mnras/stv053}, \href
  {http://adsabs.harvard.edu/abs/2015MNRAS.448.1674J} {448, 1674}

\bibitem[\protect\citeauthoryear{{Jiang}, {Dekel}, {Freundlich}, {van den
  Bosch}, {Green}, {Hopkins}, {Benson}  \& {Du}}{{Jiang}
  et~al.}{2021}]{Jiang2021}
{Jiang} F.,  {Dekel} A.,  {Freundlich} J.,  {van den Bosch} F.~C.,  {Green}
  S.~B.,  {Hopkins} P.~F.,  {Benson} A.,   {Du} X.,  2021, \mn@doi [\mnras]
  {10.1093/mnras/staa4034}, \href
  {https://ui.adsabs.harvard.edu/abs/2021MNRAS.502..621J} {502, 621}

\bibitem[\protect\citeauthoryear{{Kampakoglou} \& {Benson}}{{Kampakoglou} \&
  {Benson}}{2007}]{Kampakoglou.Benson.07}
{Kampakoglou} M.,  {Benson} A.~J.,  2007, \mn@doi [\mnras]
  {10.1111/j.1365-2966.2006.11223.x}, \href
  {http://adsabs.harvard.edu/abs/2007MNRAS.374..775K} {374, 775}

\bibitem[\protect\citeauthoryear{{Keeton} \& {Moustakas}}{{Keeton} \&
  {Moustakas}}{2009}]{Keeton.Moustakas.09}
{Keeton} C.~R.,  {Moustakas} L.~A.,  2009, \mn@doi [\apj]
  {10.1088/0004-637X/699/2/1720}, \href
  {http://adsabs.harvard.edu/abs/2009ApJ...699.1720K} {699, 1720}

\bibitem[\protect\citeauthoryear{{King}}{{King}}{1962}]{King.62}
{King} I.,  1962, \mn@doi [\aj] {10.1086/108756}, \href
  {http://adsabs.harvard.edu/abs/1962AJ.....67..471K} {67, 471}

\bibitem[\protect\citeauthoryear{{Klypin}, {Trujillo-Gomez}  \&
  {Primack}}{{Klypin} et~al.}{2011}]{Klypin.etal.11}
{Klypin} A.~A.,  {Trujillo-Gomez} S.,   {Primack} J.,  2011, \mn@doi [\apj]
  {10.1088/0004-637X/740/2/102}, \href
  {http://adsabs.harvard.edu/abs/2011ApJ...740..102K} {740, 102}

\bibitem[\protect\citeauthoryear{{Klypin}, {Prada}, {Yepes}, {He{\ss}}  \&
  {Gottl{\"o}ber}}{{Klypin} et~al.}{2015}]{Klypin.etal.15}
{Klypin} A.,  {Prada} F.,  {Yepes} G.,  {He{\ss}} S.,   {Gottl{\"o}ber} S.,
  2015, \mn@doi [\mnras] {10.1093/mnras/stu2685}, \href
  {http://adsabs.harvard.edu/abs/2015MNRAS.447.3693K} {447, 3693}

\bibitem[\protect\citeauthoryear{{Knebe}, {Arnold}, {Power}  \&
  {Gibson}}{{Knebe} et~al.}{2008}]{Knebe.etal.08}
{Knebe} A.,  {Arnold} B.,  {Power} C.,   {Gibson} B.~K.,  2008, \mn@doi
  [\mnras] {10.1111/j.1365-2966.2008.13102.x}, \href
  {http://adsabs.harvard.edu/abs/2008MNRAS.386.1029K} {386, 1029}

\bibitem[\protect\citeauthoryear{{Knebe} et~al.,}{{Knebe}
  et~al.}{2013}]{Knebe.etal.13}
{Knebe} A.,  et~al., 2013, \mn@doi [\mnras] {10.1093/mnras/stt1403}, \href
  {http://adsabs.harvard.edu/abs/2013MNRAS.435.1618K} {435, 1618}

\bibitem[\protect\citeauthoryear{{Kravtsov}, {Berlind}, {Wechsler}, {Klypin},
  {Gottl{\"o}ber}, {Allgood}  \& {Primack}}{{Kravtsov}
  et~al.}{2004}]{Kravtsov.etal.04}
{Kravtsov} A.~V.,  {Berlind} A.~A.,  {Wechsler} R.~H.,  {Klypin} A.~A.,
  {Gottl{\"o}ber} S.,  {Allgood} B.,   {Primack} J.~R.,  2004, \mn@doi [\apj]
  {10.1086/420959}, \href {http://adsabs.harvard.edu/abs/2004ApJ...609...35K}
  {609, 35}

\bibitem[\protect\citeauthoryear{{Lange}, {van den Bosch}, {Zentner}, {Wang}
  \& {Villarreal}}{{Lange} et~al.}{2019}]{Lange2019}
{Lange} J.~U.,  {van den Bosch} F.~C.,  {Zentner} A.~R.,  {Wang} K.,
  {Villarreal} A.~S.,  2019, \mn@doi [\mnras] {10.1093/mnras/stz1466}, \href
  {https://ui.adsabs.harvard.edu/abs/2019MNRAS.487.3112L} {487, 3112}

\bibitem[\protect\citeauthoryear{{Li}, {Zhao}, {Jing}, {Han}  \& {Dong}}{{Li}
  et~al.}{2020}]{Li2020}
{Li} Z.-Z.,  {Zhao} D.-H.,  {Jing} Y.~P.,  {Han} J.,   {Dong} F.-Y.,  2020,
  \mn@doi [\apj] {10.3847/1538-4357/abc481}, \href
  {https://ui.adsabs.harvard.edu/abs/2020ApJ...905..177L} {905, 177}

\bibitem[\protect\citeauthoryear{{Lovell}, {Frenk}, {Eke}, {Jenkins}, {Gao}  \&
  {Theuns}}{{Lovell} et~al.}{2014}]{Lovell.etal.14}
{Lovell} M.~R.,  {Frenk} C.~S.,  {Eke} V.~R.,  {Jenkins} A.,  {Gao} L.,
  {Theuns} T.,  2014, \mn@doi [\mnras] {10.1093/mnras/stt2431}, \href
  {http://adsabs.harvard.edu/abs/2014MNRAS.439..300L} {439, 300}

\bibitem[\protect\citeauthoryear{{Ludlow}, {Navarro}, {Springel}, {Jenkins},
  {Frenk}  \& {Helmi}}{{Ludlow} et~al.}{2009}]{Ludlow2009}
{Ludlow} A.~D.,  {Navarro} J.~F.,  {Springel} V.,  {Jenkins} A.,  {Frenk}
  C.~S.,   {Helmi} A.,  2009, \mn@doi [\apj] {10.1088/0004-637X/692/1/931},
  \href {https://ui.adsabs.harvard.edu/abs/2009ApJ...692..931L} {692, 931}

\bibitem[\protect\citeauthoryear{{Ludlow}, {Schaye}  \& {Bower}}{{Ludlow}
  et~al.}{2019}]{Ludlow2019}
{Ludlow} A.~D.,  {Schaye} J.,   {Bower} R.,  2019, \mn@doi [\mnras]
  {10.1093/mnras/stz1821}, \href
  {https://ui.adsabs.harvard.edu/abs/2019MNRAS.488.3663L} {488, 3663}

\bibitem[\protect\citeauthoryear{{Mansfield} \& {Avestruz}}{{Mansfield} \&
  {Avestruz}}{2020}]{Mansfield2020}
{Mansfield} P.,  {Avestruz} C.,  2020, \mn@doi [\mnras]
  {10.1093/mnras/staa3388}, \href
  {https://ui.adsabs.harvard.edu/abs/2020MNRAS.500.3309M} {500, 3309}

\bibitem[\protect\citeauthoryear{{Meneghetti} et~al.,}{{Meneghetti}
  et~al.}{2020}]{Meneghetti2020}
{Meneghetti} M.,  et~al., 2020, \mn@doi [Science] {10.1126/science.aax5164},
  \href {https://ui.adsabs.harvard.edu/abs/2020Sci...369.1347M} {369, 1347}

\bibitem[\protect\citeauthoryear{{Mo}, {van den Bosch}  \& {White}}{{Mo}
  et~al.}{2010}]{MBW10}
{Mo} H.,  {van den Bosch} F.~C.,   {White} S.,  2010, {Galaxy Formation and
  Evolution}.
Cambridge University Press

\bibitem[\protect\citeauthoryear{{Nadler}, {Mao}, {Green}  \&
  {Wechsler}}{{Nadler} et~al.}{2019}]{Nadler2019}
{Nadler} E.~O.,  {Mao} Y.-Y.,  {Green} G.~M.,   {Wechsler} R.~H.,  2019,
  \mn@doi [\apj] {10.3847/1538-4357/ab040e}, \href
  {https://ui.adsabs.harvard.edu/abs/2019ApJ...873...34N} {873, 34}

\bibitem[\protect\citeauthoryear{{Nadler} et~al.,}{{Nadler}
  et~al.}{2020a}]{Nadler2020b}
{Nadler} E.~O.,  et~al., 2020a, arXiv e-prints, \href
  {https://ui.adsabs.harvard.edu/abs/2020arXiv200800022N} {p. arXiv:2008.00022}

\bibitem[\protect\citeauthoryear{{Nadler} et~al.,}{{Nadler}
  et~al.}{2020b}]{Nadler2020}
{Nadler} E.~O.,  et~al., 2020b, \mn@doi [\apj] {10.3847/1538-4357/ab846a},
  \href {https://ui.adsabs.harvard.edu/abs/2020ApJ...893...48N} {893, 48}

\bibitem[\protect\citeauthoryear{{Navarro}, {Frenk}  \& {White}}{{Navarro}
  et~al.}{1997}]{Navarro.etal.97}
{Navarro} J.~F.,  {Frenk} C.~S.,   {White} S.~D.~M.,  1997, \mn@doi [\apj]
  {10.1086/304888}, \href {http://adsabs.harvard.edu/abs/1997ApJ...490..493N}
  {490, 493}

\bibitem[\protect\citeauthoryear{{Necib} et~al.,}{{Necib}
  et~al.}{2020}]{Necib2020}
{Necib} L.,  et~al., 2020, \mn@doi [Nature Astronomy]
  {10.1038/s41550-020-1131-2}, \href
  {https://ui.adsabs.harvard.edu/abs/2020NatAs...4.1078N} {4, 1078}

\bibitem[\protect\citeauthoryear{{Newton}, {Cautun}, {Jenkins}, {Frenk}  \&
  {Helly}}{{Newton} et~al.}{2018}]{Newton.etal.18}
{Newton} O.,  {Cautun} M.,  {Jenkins} A.,  {Frenk} C.~S.,   {Helly} J.~C.,
  2018, \mn@doi [\mnras] {10.1093/mnras/sty1085}, \href
  {http://adsabs.harvard.edu/abs/2018MNRAS.479.2853N} {479, 2853}

\bibitem[\protect\citeauthoryear{{Ngan} \& {Carlberg}}{{Ngan} \&
  {Carlberg}}{2014}]{Ngan2014}
{Ngan} W.~H.~W.,  {Carlberg} R.~G.,  2014, \mn@doi [\apj]
  {10.1088/0004-637X/788/2/181}, \href
  {http://cdsads.u-strasbg.fr/abs/2014ApJ...788..181N} {788, 181}

\bibitem[\protect\citeauthoryear{{Ogiya}}{{Ogiya}}{2018}]{Ogiya2018}
{Ogiya} G.,  2018, \mn@doi [\mnras] {10.1093/mnrasl/sly138}, \href
  {https://ui.adsabs.harvard.edu/abs/2018MNRAS.480L.106O} {480, L106}

\bibitem[\protect\citeauthoryear{{Ogiya}, {van den Bosch}, {Hahn}, {Green},
  {Miller}  \& {Burkert}}{{Ogiya} et~al.}{2019}]{Ogiya2019}
{Ogiya} G.,  {van den Bosch} F.~C.,  {Hahn} O.,  {Green} S.~B.,  {Miller}
  T.~B.,   {Burkert} A.,  2019, \mn@doi [\mnras] {10.1093/mnras/stz375}, \href
  {https://ui.adsabs.harvard.edu/\#abs/2019MNRAS.485..189O} {485, 189}

\bibitem[\protect\citeauthoryear{{Oguri} \& {Lee}}{{Oguri} \&
  {Lee}}{2004}]{Oguri.Lee.04}
{Oguri} M.,  {Lee} J.,  2004, \mn@doi [\mnras]
  {10.1111/j.1365-2966.2004.08304.x}, \href
  {http://adsabs.harvard.edu/abs/2004MNRAS.355..120O} {355, 120}

\bibitem[\protect\citeauthoryear{{Okabe}, {Okura}  \& {Futamase}}{{Okabe}
  et~al.}{2010}]{Okabe2010}
{Okabe} N.,  {Okura} Y.,   {Futamase} T.,  2010, \mn@doi [\apj]
  {10.1088/0004-637X/713/1/291}, \href
  {https://ui.adsabs.harvard.edu/abs/2010ApJ...713..291O} {713, 291}

\bibitem[\protect\citeauthoryear{{Okabe}, {Futamase}, {Kajisawa}  \&
  {Kuroshima}}{{Okabe} et~al.}{2014}]{Okabe14}
{Okabe} N.,  {Futamase} T.,  {Kajisawa} M.,   {Kuroshima} R.,  2014, \mn@doi
  [\apj] {10.1088/0004-637X/784/2/90}, \href
  {https://ui.adsabs.harvard.edu/abs/2014ApJ...784...90O} {784, 90}

\bibitem[\protect\citeauthoryear{{Onions} et~al.,}{{Onions}
  et~al.}{2012}]{Onions.etal.12}
{Onions} J.,  et~al., 2012, \mn@doi [\mnras]
  {10.1111/j.1365-2966.2012.20947.x}, \href
  {http://adsabs.harvard.edu/abs/2012MNRAS.423.1200O} {423, 1200}

\bibitem[\protect\citeauthoryear{{Parkinson}, {Cole}  \& {Helly}}{{Parkinson}
  et~al.}{2008}]{Parkinson.etal.08}
{Parkinson} H.,  {Cole} S.,   {Helly} J.,  2008, \mn@doi [\mnras]
  {10.1111/j.1365-2966.2007.12517.x}, \href
  {http://adsabs.harvard.edu/abs/2008MNRAS.383..557P} {383, 557}

\bibitem[\protect\citeauthoryear{{Pe{\~n}arrubia} \& {Benson}}{{Pe{\~n}arrubia}
  \& {Benson}}{2005}]{Penarrubia.Benson.05}
{Pe{\~n}arrubia} J.,  {Benson} A.~J.,  2005, \mn@doi [\mnras]
  {10.1111/j.1365-2966.2005.09633.x}, \href
  {http://adsabs.harvard.edu/abs/2005MNRAS.364..977P} {364, 977}

\bibitem[\protect\citeauthoryear{{Pe{\~n}arrubia}, {Navarro}  \&
  {McConnachie}}{{Pe{\~n}arrubia} et~al.}{2008}]{Penarrubia2008}
{Pe{\~n}arrubia} J.,  {Navarro} J.~F.,   {McConnachie} A.~W.,  2008, \mn@doi
  [\apj] {10.1086/523686}, \href
  {https://ui.adsabs.harvard.edu/\#abs/2008ApJ...673..226P} {673, 226}

\bibitem[\protect\citeauthoryear{Pe{\~n}arrubia, {Benson}, {Walker}, {Gilmore},
  {McConnachie}  \& {Mayer}}{Pe{\~n}arrubia et~al.}{2010}]{Penarrubia.etal.10}
Pe{\~n}arrubia J.,  {Benson} A.~J.,  {Walker} M.~G.,  {Gilmore} G.,
  {McConnachie} A.~W.,   {Mayer} L.,  2010, \mn@doi [\mnras]
  {10.1111/j.1365-2966.2010.16762.x}, \href
  {http://adsabs.harvard.edu/abs/2010MNRAS.406.1290P} {406, 1290}

\bibitem[\protect\citeauthoryear{{Pieri}, {Bertone}  \& {Branchini}}{{Pieri}
  et~al.}{2008}]{Pieri.etal.08}
{Pieri} L.,  {Bertone} G.,   {Branchini} E.,  2008, \mn@doi [\mnras]
  {10.1111/j.1365-2966.2007.12828.x}, \href
  {http://adsabs.harvard.edu/abs/2008MNRAS.384.1627P} {384, 1627}

\bibitem[\protect\citeauthoryear{{Pullen}, {Benson}  \& {Moustakas}}{{Pullen}
  et~al.}{2014}]{Pullen.etal.14}
{Pullen} A.~R.,  {Benson} A.~J.,   {Moustakas} L.~A.,  2014, \mn@doi [\apj]
  {10.1088/0004-637X/792/1/24}, \href
  {http://adsabs.harvard.edu/abs/2014ApJ...792...24P} {792, 24}

\bibitem[\protect\citeauthoryear{{Read}, {Wilkinson}, {Evans}, {Gilmore}  \&
  {Kleyna}}{{Read} et~al.}{2006}]{Read.etal.06b}
{Read} J.~I.,  {Wilkinson} M.~I.,  {Evans} N.~W.,  {Gilmore} G.,   {Kleyna}
  J.~T.,  2006, \mn@doi [\mnras] {10.1111/j.1365-2966.2005.09959.x}, \href
  {http://adsabs.harvard.edu/abs/2006MNRAS.367..387R} {367, 387}

\bibitem[\protect\citeauthoryear{{Reddick}, {Wechsler}, {Tinker}  \&
  {Behroozi}}{{Reddick} et~al.}{2013}]{Reddick.etal.13}
{Reddick} R.~M.,  {Wechsler} R.~H.,  {Tinker} J.~L.,   {Behroozi} P.~S.,  2013,
  \mn@doi [\apj] {10.1088/0004-637X/771/1/30}, \href
  {http://adsabs.harvard.edu/abs/2013ApJ...771...30R} {771, 30}

\bibitem[\protect\citeauthoryear{{Rico}}{{Rico}}{2020}]{Rico2020}
{Rico} J.,  2020, \mn@doi [Galaxies] {10.3390/galaxies8010025}, \href
  {https://ui.adsabs.harvard.edu/abs/2020Galax...8...25R} {8, 25}

\bibitem[\protect\citeauthoryear{{Rocha}, {Peter}, {Bullock}, {Kaplinghat},
  {Garrison-Kimmel}, {O{\~n}orbe}  \& {Moustakas}}{{Rocha}
  et~al.}{2013}]{Rocha.etal.13}
{Rocha} M.,  {Peter} A.~H.~G.,  {Bullock} J.~S.,  {Kaplinghat} M.,
  {Garrison-Kimmel} S.,  {O{\~n}orbe} J.,   {Moustakas} L.~A.,  2013, \mn@doi
  [\mnras] {10.1093/mnras/sts514}, \href
  {http://adsabs.harvard.edu/abs/2013MNRAS.430...81R} {430, 81}

\bibitem[\protect\citeauthoryear{{Shu} et~al.,}{{Shu} et~al.}{2015}]{Shu2015}
{Shu} Y.,  et~al., 2015, \mn@doi [\apj] {10.1088/0004-637X/803/2/71}, \href
  {http://cdsads.u-strasbg.fr/abs/2015ApJ...803...71S} {803, 71}

\bibitem[\protect\citeauthoryear{{Somalwar}, {Chang}, {Mishra-Sharma}  \&
  {Lisanti}}{{Somalwar} et~al.}{2021}]{Somalwar2021}
{Somalwar} J.~J.,  {Chang} L.~J.,  {Mishra-Sharma} S.,   {Lisanti} M.,  2021,
  \mn@doi [\apj] {10.3847/1538-4357/abc87d}, \href
  {https://ui.adsabs.harvard.edu/abs/2021ApJ...906...57S} {906, 57}

\bibitem[\protect\citeauthoryear{{Springel}, {White}, {Tormen}  \&
  {Kauffmann}}{{Springel} et~al.}{2001}]{Springel.etal.01}
{Springel} V.,  {White} S. D.~M.,  {Tormen} G.,   {Kauffmann} G.,  2001,
  \mn@doi [\mnras] {10.1046/j.1365-8711.2001.04912.x}, \href
  {https://ui.adsabs.harvard.edu/abs/2001MNRAS.328..726S} {328, 726}

\bibitem[\protect\citeauthoryear{{Springel} et~al.,}{{Springel}
  et~al.}{2008}]{Springel.etal.08}
{Springel} V.,  et~al., 2008, \mn@doi [\mnras]
  {10.1111/j.1365-2966.2008.14066.x}, \href
  {http://adsabs.harvard.edu/abs/2008MNRAS.391.1685S} {391, 1685}

\bibitem[\protect\citeauthoryear{{Strigari}, {Koushiappas}, {Bullock}  \&
  {Kaplinghat}}{{Strigari} et~al.}{2007}]{Strigari.etal.07}
{Strigari} L.~E.,  {Koushiappas} S.~M.,  {Bullock} J.~S.,   {Kaplinghat} M.,
  2007, \mn@doi [\prd] {10.1103/PhysRevD.75.083526}, \href
  {http://adsabs.harvard.edu/abs/2007PhRvD..75h3526S} {75, 083526}

\bibitem[\protect\citeauthoryear{{Taffoni}, {Mayer}, {Colpi}  \&
  {Governato}}{{Taffoni} et~al.}{2003}]{Taffoni.etal.03}
{Taffoni} G.,  {Mayer} L.,  {Colpi} M.,   {Governato} F.,  2003, \mn@doi
  [\mnras] {10.1046/j.1365-8711.2003.06395.x}, \href
  {http://adsabs.harvard.edu/abs/2003MNRAS.341..434T} {341, 434}

\bibitem[\protect\citeauthoryear{{Taylor} \& {Babul}}{{Taylor} \&
  {Babul}}{2001}]{Taylor.Babul.01}
{Taylor} J.~E.,  {Babul} A.,  2001, \mn@doi [\apj] {10.1086/322276}, \href
  {http://adsabs.harvard.edu/abs/2001ApJ...559..716T} {559, 716}

\bibitem[\protect\citeauthoryear{{Taylor} \& {Babul}}{{Taylor} \&
  {Babul}}{2004}]{Taylor.Babul.04}
{Taylor} J.~E.,  {Babul} A.,  2004, \mn@doi [\mnras]
  {10.1111/j.1365-2966.2004.07395.x}, \href
  {http://adsabs.harvard.edu/abs/2004MNRAS.348..811T} {348, 811}

\bibitem[\protect\citeauthoryear{{Tollet}, {Cattaneo}, {Mamon}, {Moutard}  \&
  {van den Bosch}}{{Tollet} et~al.}{2017}]{Tollet.etal.17}
{Tollet} {\'E}.,  {Cattaneo} A.,  {Mamon} G.~A.,  {Moutard} T.,   {van den
  Bosch} F.~C.,  2017, \mn@doi [\mnras] {10.1093/mnras/stx1840}, \href
  {https://ui.adsabs.harvard.edu/abs/2017MNRAS.471.4170T} {471, 4170}

\bibitem[\protect\citeauthoryear{{Tormen}, {Bouchet}  \& {White}}{{Tormen}
  et~al.}{1997}]{Tormen.etal.97}
{Tormen} G.,  {Bouchet} F.~R.,   {White} S.~D.~M.,  1997, \mn@doi [\mnras]
  {10.1093/mnras/286.4.865}, \href
  {http://adsabs.harvard.edu/abs/1997MNRAS.286..865T} {286, 865}

\bibitem[\protect\citeauthoryear{{Trujillo-Gomez}, {Klypin}, {Primack}  \&
  {Romanowsky}}{{Trujillo-Gomez} et~al.}{2011}]{Trujillo-Gomez.etal.11}
{Trujillo-Gomez} S.,  {Klypin} A.,  {Primack} J.,   {Romanowsky} A.~J.,  2011,
  \mn@doi [\apj] {10.1088/0004-637X/742/1/16}, \href
  {http://adsabs.harvard.edu/abs/2011ApJ...742...16T} {742, 16}

\bibitem[\protect\citeauthoryear{{Vale} \& {Ostriker}}{{Vale} \&
  {Ostriker}}{2006}]{Vale.Ostriker.06}
{Vale} A.,  {Ostriker} J.~P.,  2006, \mn@doi [\mnras]
  {10.1111/j.1365-2966.2006.10605.x}, \href
  {http://adsabs.harvard.edu/abs/2006MNRAS.371.1173V} {371, 1173}

\bibitem[\protect\citeauthoryear{{Vattis}, {Toomey}  \& {Koushiappas}}{{Vattis}
  et~al.}{2020}]{Vattis2020}
{Vattis} K.,  {Toomey} M.~W.,   {Koushiappas} S.~M.,  2020, arXiv e-prints,
  \href {https://ui.adsabs.harvard.edu/abs/2020arXiv200811577V} {p.
  arXiv:2008.11577}

\bibitem[\protect\citeauthoryear{{Vegetti}, {Koopmans}, {Auger}, {Treu}  \&
  {Bolton}}{{Vegetti} et~al.}{2014}]{Vegetti.etal.14}
{Vegetti} S.,  {Koopmans} L.~V.~E.,  {Auger} M.~W.,  {Treu} T.,   {Bolton}
  A.~S.,  2014, \mn@doi [\mnras] {10.1093/mnras/stu943}, \href
  {http://adsabs.harvard.edu/abs/2014MNRAS.442.2017V} {442, 2017}

\bibitem[\protect\citeauthoryear{{Vogelsberger}, {Zavala}  \&
  {Loeb}}{{Vogelsberger} et~al.}{2012}]{Vogelsberger.etal.12}
{Vogelsberger} M.,  {Zavala} J.,   {Loeb} A.,  2012, \mn@doi [\mnras]
  {10.1111/j.1365-2966.2012.21182.x}, \href
  {http://adsabs.harvard.edu/abs/2012MNRAS.423.3740V} {423, 3740}

\bibitem[\protect\citeauthoryear{{Wetzel}}{{Wetzel}}{2011}]{Wetzel.11}
{Wetzel} A.~R.,  2011, \mn@doi [\mnras] {10.1111/j.1365-2966.2010.17877.x},
  \href {http://adsabs.harvard.edu/abs/2011MNRAS.412...49W} {412, 49}

\bibitem[\protect\citeauthoryear{{Yang}, {Du}, {Benson}, {Pullen}  \&
  {Peter}}{{Yang} et~al.}{2020}]{Yang2020}
{Yang} S.,  {Du} X.,  {Benson} A.~J.,  {Pullen} A.~R.,   {Peter} A. H.~G.,
  2020, \mn@doi [\mnras] {10.1093/mnras/staa2496}, \href
  {https://ui.adsabs.harvard.edu/abs/2020MNRAS.498.3902Y} {498, 3902}

\bibitem[\protect\citeauthoryear{{Zentner} \& {Bullock}}{{Zentner} \&
  {Bullock}}{2003}]{Zentner.Bullock.03}
{Zentner} A.~R.,  {Bullock} J.~S.,  2003, \mn@doi [\apj] {10.1086/378797},
  \href {http://adsabs.harvard.edu/abs/2003ApJ...598...49Z} {598, 49}

\bibitem[\protect\citeauthoryear{{Zentner}, {Berlind}, {Bullock}, {Kravtsov}
  \& {Wechsler}}{{Zentner} et~al.}{2005}]{Zentner.etal.05}
{Zentner} A.~R.,  {Berlind} A.~A.,  {Bullock} J.~S.,  {Kravtsov} A.~V.,
  {Wechsler} R.~H.,  2005, \mn@doi [\apj] {10.1086/428898}, \href
  {http://adsabs.harvard.edu/abs/2005ApJ...624..505Z} {624, 505}

\bibitem[\protect\citeauthoryear{{Zentner}, {Hearin}  \& {van den
  Bosch}}{{Zentner} et~al.}{2014}]{Zentner.etal.14}
{Zentner} A.~R.,  {Hearin} A.~P.,   {van den Bosch} F.~C.,  2014, \mn@doi
  [\mnras] {10.1093/mnras/stu1383}, \href
  {http://adsabs.harvard.edu/abs/2014MNRAS.443.3044Z} {443, 3044}

\bibitem[\protect\citeauthoryear{{Zhao}, {Jing}, {Mo}  \& {B{\"o}rner}}{{Zhao}
  et~al.}{2009}]{Zhao2009}
{Zhao} D.~H.,  {Jing} Y.~P.,  {Mo} H.~J.,   {B{\"o}rner} G.,  2009, \mn@doi
  [\apj] {10.1088/0004-637X/707/1/354}, \href
  {https://ui.adsabs.harvard.edu/abs/2009ApJ...707..354Z} {707, 354}

\bibitem[\protect\citeauthoryear{van~den Bosch}{van~den
  Bosch}{2017}]{vdBosch.17}
van~den Bosch F.~C.,  2017, \mn@doi [\mnras] {10.1093/mnras/stx520}, \href
  {http://adsabs.harvard.edu/abs/2017MNRAS.468..885V} {468, 885}

\bibitem[\protect\citeauthoryear{{van den Bosch} \& {Jiang}}{{van den Bosch} \&
  {Jiang}}{2016}]{vdBosch.Jiang.16}
{van den Bosch} F.~C.,  {Jiang} F.,  2016, \mn@doi [\mnras]
  {10.1093/mnras/stw440}, \href
  {http://adsabs.harvard.edu/abs/2016MNRAS.458.2870V} {458, 2870}

\bibitem[\protect\citeauthoryear{van~den Bosch \& {Ogiya}}{van~den Bosch \&
  {Ogiya}}{2018}]{vdBosch.etal.2018b}
van~den Bosch F.~C.,  {Ogiya} G.,  2018, \mn@doi [\mnras]
  {10.1093/mnras/sty084}, \href
  {https://ui.adsabs.harvard.edu/\#abs/2018MNRAS.475.4066V} {475, 4066}

\bibitem[\protect\citeauthoryear{{van den Bosch}, {Yang}, {Mo}  \&
  {Norberg}}{{van den Bosch} et~al.}{2005a}]{vdbosch.etal.05b}
{van den Bosch} F.~C.,  {Yang} X.,  {Mo} H.~J.,   {Norberg} P.,  2005a, \mn@doi
  [\mnras] {10.1111/j.1365-2966.2004.08407.x}, \href
  {https://ui.adsabs.harvard.edu/abs/2005MNRAS.356.1233V} {356, 1233}

\bibitem[\protect\citeauthoryear{{van den Bosch}, {Tormen}  \& {Giocoli}}{{van
  den Bosch} et~al.}{2005b}]{vdBosch.etal.05}
{van den Bosch} F.~C.,  {Tormen} G.,   {Giocoli} C.,  2005b, \mn@doi [\mnras]
  {10.1111/j.1365-2966.2005.08964.x}, \href
  {http://adsabs.harvard.edu/abs/2005MNRAS.359.1029V} {359, 1029}

\bibitem[\protect\citeauthoryear{{van den Bosch}, {Jiang}, {Hearin},
  {Campbell}, {Watson}  \& {Padmanabhan}}{{van den Bosch}
  et~al.}{2014}]{vdBosch.etal.14}
{van den Bosch} F.~C.,  {Jiang} F.,  {Hearin} A.,  {Campbell} D.,  {Watson} D.,
    {Padmanabhan} N.,  2014, \mn@doi [\mnras] {10.1093/mnras/stu1872}, \href
  {https://ui.adsabs.harvard.edu/abs/2014MNRAS.445.1713V} {445, 1713}

\bibitem[\protect\citeauthoryear{van~den Bosch, {Jiang}, {Campbell}  \&
  {Behroozi}}{van~den Bosch et~al.}{2016}]{vdBosch.etal.16}
van~den Bosch F.~C.,  {Jiang} F.,  {Campbell} D.,   {Behroozi} P.,  2016,
  \mn@doi [\mnras] {10.1093/mnras/stv2338}, \href
  {http://adsabs.harvard.edu/abs/2016MNRAS.455..158V} {455, 158}

\bibitem[\protect\citeauthoryear{{van den Bosch}, {Ogiya}, {Hahn}  \&
  {Burkert}}{{van den Bosch} et~al.}{2018}]{vdBosch.etal.2018a}
{van den Bosch} F.~C.,  {Ogiya} G.,  {Hahn} O.,   {Burkert} A.,  2018, \mn@doi
  [\mnras] {10.1093/mnras/stx2956}, \href
  {https://ui.adsabs.harvard.edu/\#abs/2018MNRAS.474.3043V} {474, 3043}

\bibitem[\protect\citeauthoryear{{van den Bosch}, {Lange}  \& {Zentner}}{{van
  den Bosch} et~al.}{2019}]{vdBosch2019}
{van den Bosch} F.~C.,  {Lange} J.~U.,   {Zentner} A.~R.,  2019, \mn@doi
  [\mnras] {10.1093/mnras/stz2017}, \href
  {https://ui.adsabs.harvard.edu/abs/2019MNRAS.488.4984V} {488, 4984}

\makeatother
\end{thebibliography}


\appendix

\section{Initial orbits}\label{app:initorbits}

Here, we describe our approach for initializing subhalo orbits. We specify the initial phase space coordinates of the infalling subhalo as
\begin{equation}
    \{r, \theta, \phi, v_r, v_\theta, v_\phi\} = \{r_\mathrm{vir}, \theta, \phi, -v\cos\gamma, v\sin\gamma\cos\delta, v\sin\gamma\sin\delta \} .
\end{equation}
We assume that subhalo infall occurs isotropically, and therefore select an initial azimuthal angle, $\phi$, uniformly from $[0, 2\pi)$ and an initial polar angle, $\theta$, by sampling $\cos\theta$ uniformly from $[0,1)$. In order to determine the initial velocity vector, the degrees of freedom of which are the speed, $v$, the angle between the velocity vector and the (negative of the) position vector, $\gamma$, and an additional angle that sets the orientation of the orbital plane, $\delta$, we use the universal model of \citet{Li2020}, which has been calibrated using a large suite of cosmological simulations.

For all \textit{first infall} events (i.e., for a given subhalo, only considering the first time a subhalo enters into the host virial radius) aggregated across all of the simulations and over a wide range of redshift snapshots, \citet{Li2020} find that $u \equiv v/V_\mathrm{vir,h}$ (here, $V_\mathrm{vir,h}$ denotes the instantaneous virial velocity of the host) is well-described by a universal log-normal distribution that is peaked near unity and is independent of subhalo mass and redshift, $z$:
\begin{equation}\label{eqn:pu}
   p(u) \,  \rmd u = \frac{1}{\sqrt{2\pi}\sigma_1} \exp\Bigg[ -\frac{\ln^2(u/\mu_1)}{2\sigma_1^2}\Bigg] \frac{\rmd u}{u} .
\end{equation}
Here, $\mu_1 = 1.2$ and $\sigma_1 = 0.2$. They also find that mergers with larger $M_\mathrm{vir}$ (instantaneous virial mass of the host) and/or $\xi \equiv m_\mathrm{acc}/M_\mathrm{vir}$ result in more radial subhalo orbits, which is mainly attributed to gravitational focusing. By rewriting the host mass, $M_\mathrm{vir}$, in terms of its corresponding density peak height, $\nu \equiv \delta_\rmc (z) / \sigma(M_\mathrm{vir})$, where $\delta_\rmc(z)$ is the critical overdensity of collapse and $\sigma^2(M)$ is the mass variance, the authors find that the distribution of infall angles is redshift-independent and only depends on $u$, $\nu$, and $\xi$. Specifically, $\cos^2 \gamma$ follows an exponential distribution,
\begin{equation}\label{eqn:pgamma}
    p(\cos^2\gamma \, | \, u, \nu, \xi) \, \rmd \cos^2\gamma = \frac{\zeta}{e^\zeta -1 } \exp\big(\zeta \cos^2\gamma\big) \, \rmd \cos^2\gamma,
\end{equation}
where
\begin{equation}\label{eqn:pars}
\begin{split}
    &\zeta = a_0 \exp\Bigg[ - \frac{\ln^2(u/\mu_2)}{2\sigma_1^2}\Bigg] + A(u+1)+B, \\
    &A = a_1 \nu + a_2 \zeta^c + a_3 \nu \zeta^c,  \, \mathrm{and} \\
    &B= b_0 + b_1 \zeta^c\,,
\end{split}
\end{equation}
and the best-fit parameters are $(a_0, a_1, a_2, a_3, b_0, b_1, \mu_2, c) = (0.89, 0.3, -3.33, 0.56, -1.44, 9.60, 1.04, 0.43)$.

We use equations~\eqref{eqn:pu}--\eqref{eqn:pars} to sample the initial $v$ and $\gamma$ for each subhalo at infall. In order to set the orientation of the orbital plane, we assume isotropy and therefore draw $\delta$ uniformly from $[0, 2\pi)$.

Using the $u$ distribution of \citet{Li2020} results in a substantial fraction of sampled orbits with initial orbital energies that lie above the maximum value sampled in the DASH simulations (corresponding to $x_\rmc = 2$). This fraction has a slight dependence on the host concentration. For example, for $c_\mathrm{vir,h}=10$, a total of $25\%$ of subhaloes have $x_\rmc > 2$ at infall and $2\%$ are initially unbound (i.e., $v$ at infall is larger than the escape velocity). Fortunately, the performance of our DASH-calibrated evolved subhalo density profile model and mass-loss prescription both exhibit minimal dependence on the orbital parameters. We emphasize that the combined impact of dynamical friction and the growth of the host potential results in continuous reduction of the subhalo orbital energy, lowering $x_\rmc$ over time. These effects also drive subhaloes that are initially unbound to eventually become bound after infall; thus, we include these initially unbound orbits in the \texttt{SatGen} subhalo population.

\bsp
\label{lastpage}
\end{document}